\documentclass{article}

\usepackage{microtype}
\usepackage{graphicx}
\usepackage{subcaption}
\usepackage{booktabs} % for professional tables
\usepackage{setspace}

\newcommand{\dl}{\,\mathrm{d}}

\usepackage{hyperref}

\usepackage[preprint]{icml2026}

\usepackage{amsmath}
\usepackage{amssymb}
\usepackage{mathtools}
\usepackage{amsthm}
\usepackage{adjustbox}
\usepackage{hyperref}
\usepackage{url}
\usepackage{graphicx}
\usepackage{comment}
\usepackage{subcaption}
\usepackage{multirow}
\usepackage{booktabs}
\usepackage{wrapfig} 

\usepackage[capitalize,noabbrev]{cleveref}

\theoremstyle{plain}

\theoremstyle{definition}

\theoremstyle{remark}

\usepackage[textsize=tiny]{todonotes}

\icmltitlerunning{Preprint. Under review.}

\begin{document}

\twocolumn[
  \icmltitle{VoiceBridge: General Speech Restoration with One-step Latent Bridge Models}



  \icmlsetsymbol{equal}{*}

  \begin{icmlauthorlist}
    \icmlauthor{Chi Zhang$^*$}{thu,shengshu}
    \icmlauthor{Zehua Chen$^{*\dagger}$}{thu,shengshu}
    \icmlauthor{Kaiwen Zheng}{thu}
    \icmlauthor{Jun Zhu$^\dagger$}{thu,shengshu}
  \end{icmlauthorlist}

  \icmlaffiliation{thu}{Tsinghua University}
  \icmlaffiliation{shengshu}{Shengshu AI}
  \icmlcorrespondingauthor{Zehua Chen}{zhc23thuml@tsinghua.edu.cn}
  \icmlcorrespondingauthor{Jun Zhu}{dcszj@mail.tsinghua.edu.cn}

  \icmlkeywords{Machine Learning, ICML, VoiceBridge, Latent Bridge Models, General Speech Restoration, Speech Enhancement, Speech Restoration, Speech Signal Processing, Diffusion Models, Generative Models, Schrodinger Bridge}

  \vskip 0.3in
]



\printAffiliationsAndNotice{}  % no special notice (required even if empty)

\begin{abstract}
Bridge models have been investigated in speech enhancement but are mostly single-task, with constrained general speech restoration (GSR) capability.
In this work, we propose VoiceBridge, a one-step latent bridge model (LBM) for GSR, capable of efficiently reconstructing 48 kHz fullband speech from diverse distortions.
To inherit the advantages of data-domain bridge models, we design an energy-preserving variational autoencoder, enhancing the waveform-latent space alignment over varying energy levels.
By compressing waveform into continuous latent representations, VoiceBridge models~\textit{various} GSR tasks with a~\textit{single} latent-to-latent generative process backed by a scalable transformer. To alleviate the challenge of reconstructing the high-quality target from distinctively different low-quality priors, we propose a joint neural prior for GSR, uniformly reducing the burden of the LBM in diverse tasks. 
Building upon these designs, we further investigate bridge training objective by jointly tuning LBM, decoder and discriminator together, transforming the model from a denoiser to generator and enabling \textit{one-step GSR without distillation}. 
Extensive validation across in-domain (\textit{e.g.}, denoising and super-resolution) and out-of-domain tasks (\textit{e.g.}, refining synthesized speech) and datasets demonstrates the superior performance of VoiceBridge.
Demos:~\url{https://VoiceBridgedemo.github.io/}.
% Bridge models have been applied to speech enhancement but are mostly single-task, with constrained general speech restoration (GSR) capability. We propose VoiceBridge, a latent bridge model for GSR, reconstructing 48 kHz fullband speech from diverse distortions. By compressing speech waveform into continuous latent representations, VoiceBridge models the GSR process via unified latent-to-latent single step generation. VoiceBridge includes an energy-preserving VAE for improved waveform–latent alignment, a joint neural prior to handle diverse low-quality conditions, and an DGA stage that aligns with human perception and enables single-step generation. Experiments on in- and out-of-domain datasets show strong performance. Demos: \url{https://VoiceBridgedemo.github.io/}
\end{abstract}

\section{Introduction}
Tractable Schrödinger Bridge (SB) models have proven to faithfully reconstruct the target distribution from an informative prior, overcoming the limitation of the noise prior in diffusion models~\citep{chen2021likelihood,bridgetts,audiolbm}.
Recently, their advantages have been extended to speech enhancement tasks~\citep{sbsen, sbsel, han2025few}, such as denoising~\citep{sbsen}, dereverberation, and super-resolution~\citep{bridgesr}, where the low-quality (LQ) observation naturally provides indicative information for the high-quality (HQ) target.
Although these efforts have proposed innovations spanning forward process design~\citep{sbsen}, model parameterization~\citep{richter2025investigating}, training objectives~\citep{zhang2025sb}, and cascaded architectures~\citep{han2025few,audiolbm}, their designs are usually confined to a single task, not yet achieving general speech restoration (GSR) at scale to address various real-world speech degradations.
%which may limit their applicability to address various real-world speech degradations.

% a single latent to latent for diverse LQ to HQ
In this work, we propose \textit{VoiceBridge}, a one-step general speech restoration (GSR) system~\citep{benesty2006speech, kolbaek2016speech, das2021fundamentals} rooted in a latent bridge model (LBM), where the~\textit{diverse LQ-to-HQ tasks} in GSR are modeled with~\textit{a single latent-to-latent generative process} backed by a transformer architecture.
%Latent space
By encoding both the LQ and HQ speech signals into continuous representations in a small latent space, VoiceBridge aims to preserve the advantages of the~\textit{data-to-data} sampling nature of bridge models from data space to latent space while achieving efficient training.
%scalable modeling
% Towards generalizable modeling, VoiceBridge explores combining the advantages of transformer architecture, namely good scalability that has been verified in both image~\citep{dit,u-vit} and audio~\citep{stableaudio} generation, with the bridge generative framework, further strengthening GSR performance at scale.

To better inherit the advantages of bridge generative models on speech enhancement tasks in the data domain within LBMs, a natural idea is to preserve the characteristics in data space into latent representations. 
Motivated by this, we develop an \textit{energy-preserving variational autoencoder (EP-VAE)} to achieve stronger consistency between the waveform and latent space than vanilla VAE~\citep{vae}.
Specifically, we modify the VAE training objective, requiring a linear scaling operation in the latent space to be manifested in the decoded waveform space.
In comparison with vanilla VAE that regularizes data reconstruction in a single scale, EP-VAE introduces reconstruction regularization over varying energy levels, naturally improving the waveform-latent consistency and resulting in a more structural latent space~\citep{kouzelis2025eq, yao2025reconstruction, skorokhodov2025improving, yu2024representation} for LBM modeling.

Moreover, considering the difficulty of reconstructing the HQ target from distinctively different LQ priors caused by flexible degradation methods, such as noise, down-sampling, clipping, reverberation, vocal effects, and their random combinations~\citep{anyenhance, universe, scheibler_universalpp_2024}, we propose~\textit{a joint neural prior}, further enhancing the prior distribution of LBM.
In detail, given the waveform encoder pre-trained by EP-VAE, we replace the reference signal from the LQ prior to the HQ target, enabling the encoder to uniformly reduce the distance between each LQ prior and the HQ target in the latent space.
Similar to bridge models outperforming diffusion models when an informative prior is provided~\citep{bridgetts,wang2024framebridge,bridgesr}, a joint neural prior enhances LBM-based GSR by further alleviating the~\textit{LQ-to-HQ} generation burden.

% Given the joint neural prior and the HQ target encoded in a structural latent space, we have been able to establish a bridge transformer to tackle the various restoration tasks at scale.
To mitigate cascading mismatches between the LBM and the decoder, we perform a post-training stage that jointly fine-tunes the LBM and the VAE decoder with the encoder fixed. Direct supervision in the data domain improves the realism of LBM predictions and calibrates the decoder to the LBM-induced latent distribution, while preserving the encoder’s learned latent geometry. Furthermore, We novelly incorporate adversarial and perceptual losses to align the one-step prediction with the full conditional distribution, converting the LBM from a multi-step denoiser into a \textit{one-step generator} without distillation, achieving streaming-rate synthesis with state-of-the-art quality.

% As the joint neural prior is already pulled toward the HQ target, the small remaining distance may be covered by a single forward pass, rather than a regressive sampling process, enabling inference acceleration. Novelly, we introduce a~\textit{denoiser-to-generator alignment stage}, which jointly finetunes the LBM and the decoder with adversarial and perceptual training objectives. This step transforms the multi-step bridge model into a \textit{one-step generator} without distillation, meanwhile minimizing the cascading mismatches in inference and aligning the generation process with human perceptual quality.

% This design complements the LBM which is typically trained to fit the target distribution from prior distribution without explicitly considering human perception quality.
% Concretely, we introduce two popular speech metrics, PESQ~\citep{pesq} and UTMOS~\citep{utmos}, into the post-training stage of VoiceBridge, which optimizes both the LBM sampling and the EP-VAE decoding, minimizing their cascading mismatches in inference and aligning the entire generation process with human perceptual quality.

In summary, we make the following contributions.
\begin{itemize}
    \item We design VoiceBridge, a LBM-based GSR system tackling~\textit{diverse LQ-to-HQ tasks} with~\textit{a single latent-to-latent generative process} backed a transformer architecture, and integrating multiple innovations to the LBM system.
    \item We propose~\textit{EP-VAE} to inherit the data-space advantages of bridge models by strengthening consistency between the waveform and latent across energy levels, and a~\textit{joint neural prior} to alleviate LQ-to-HQ generation burden by enhancing the prior distribution.
    \item We propose a novel \textit{denoiser-to-generator} post-training process, mitigating bridge-decoder mismatches, improving generation quality, and enabling \textit{one-step inference} without distillation.
    \item Comprehensive validation across in-domain and out-of-domain tasks and datasets, including refining recent zero-shot speech and podcast generation results, shows the superiority of VoiceBridge in efficiently reconstructing high-fidelity 48~kHz speech from various distortions.
\end{itemize}

\section{Related Work}

% In general speech restoration, the goal is to recover high-quality speech from audio degraded by real-world distortions such as additive noise, room reverberation, amplitude clipping, bandwidth reduction, and dynamic equalization. These degradations often occur in combination, forming complex and unpredictable corruption patterns.

% Our degradation modeling follows a probabilistic composite process similar to that used in AnyEnhance~\citep{anyenhance}, where each distortion type is applied independently with a predefined probability. This setup ensures a diverse training distribution that reflects real-world acoustic conditions. The restoration task is then to reconstruct clean speech from these corrupted observations as faithfully as possible.

%\subsection{Motivation}
%We introduce our motivation for this work from both the task (\textit{i.e.}, GSR) and the method (\textit{i.e.}, LBMs) perspectives. 
%Why GSR
\textbf{Bridge-based speech enhancement.} Considering the advantages of the data-to-data generative framework on the tasks with an indicative prior,~\textit{e.g.}, speech denoising, dereverberation, and upsampling, several recent works have explored the parameterization method~\citep{richter2025investigating}, the noise schedule~\citep{bridgesr, sbsen}, and the training objectives~\citep{zhang2025sb} for bridge-based speech enhancement models. % which training objective?
However, their designs are usually proposed for a single task, including a recent work, AudioLBM~\citep{audiolbm}, specialized in audio super-resolution. 
Moreover, a part of recent efforts are verified with narrow-band~\textit{e.g.}, 16~kHz or 24~kHz speech samples~\citep{metis} or small-scale benchmark datasets~\citep{sbsen}. 
These restrictions may limit the applicability of these designs to reconstruct 48~kHz high-fidelity speech samples from diverse real-world degradations (\textit{i.e.}, GSR). 

\textbf{General speech restoration.} Previous GSR systems have exploited mapping-based methods~\citep{voicefixer}, adversarial training~\citep{finally}, masked generative models~\citep{li2024masksr,anyenhance} or pretraining~\citep{metis}, conditional diffusion models~\citep{universe, scheibler_universalpp_2024, hiresldm}, and flow-matching methods~\citep{resembleenhance, speechflow, ku2025generative}.
In VoiceBridge, we design an LBM that extends the advantages of the~\textit{data-to-data} generative framework from specific SE tasks to GSR at scale and introduce three novel techniques that holistically improve generation results.
%featuring two advantages: (1) a full exploitation of informative prior for HQ target reconstruction with a unified probabilistic generative framework, and (2) the iterative refinement nature along the sampling trajectory backed by a scalable network architecture, aiming at achieving high-fidelity GSR at scale.
We provide a more detailed introduction to related work in Appendix~\ref{appendix:related}.

\begin{figure*}[t]
    \centering
    \vspace{-3mm}
    \includegraphics[width=0.95\linewidth]{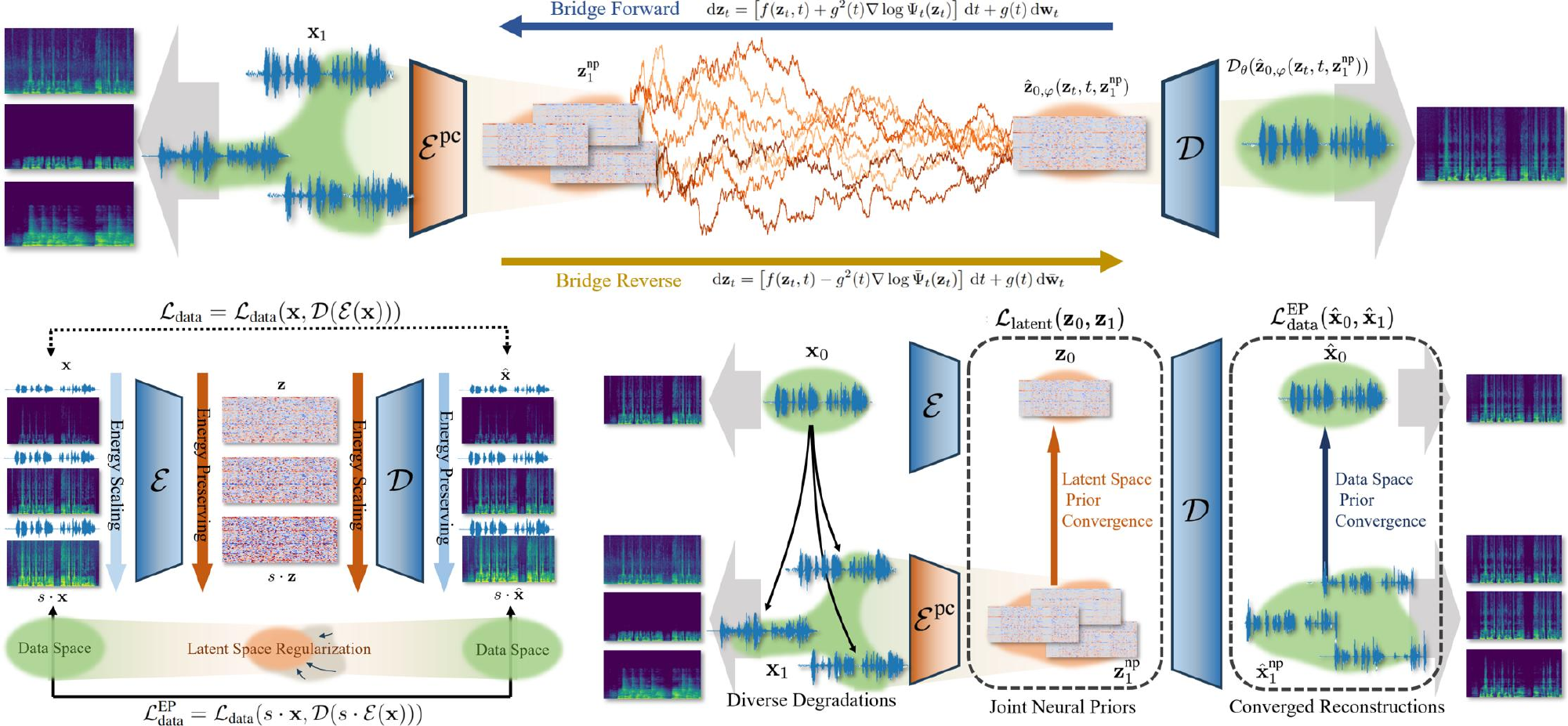}
    \vspace{-3mm}
    \caption{\small Overview of VoiceBridge. The upper part demonstrates the designed LBM-based GSR system. The lower part shows our approaches to building a structural latent space and converging a joint neural prior. On the left,~\textit{EP-VAE} requires alignment between the latent and data space at varying energy levels. On the right,~\textit{a joint neural} is encoded to reduce the distance between LQ priors and the HQ target, facilitating LBM reconstruction. }
    \vspace{-3mm}
    \label{fig:main}
\end{figure*}

\section{VoiceBridge}
In this section, we first present our designed LBM-based GSR system, and then introduce the three innovations leading to holistic improvement, namely~\textit{EP-VAE},~\textit{joint neural prior}, and~\textit{denoiser-to-generator alignment}, respectively.

\subsection{Latent Bridge Transformer for GSR}
\label{LBMGSR}
%Bridge models~\citep{sbtheory,  wang2021deep, chen2021likelihood, bridgetts} can capture the target distribution from an informative prior distribution, overcoming the restrictions of the uninformative noisy prior in diffusion models~\citep{ddpm, song2020score} and therefore allowing full exploitation of the provided instructive informative, \textit{e.g.}, low-quality observation in GSR, at the inference stage.
In VoiceBridge, we construct the boundary distributions of bridge models, namely the high-quality target and the low-quality prior, as follows.
Given a clean and full-band speech signal $\mathbf{x}_0 \in \mathbb{R}^L$, to simulate real-world speech distortions, we construct the degraded version $\mathbf{x}_1 \in \mathbb{R}^L$ with a degradation operator $\mathcal{T}$:
\begin{align}
    \mathbf{x}_1 = \mathcal{T}(\mathbf{x}_0),
\end{align}
where $\mathcal{T}$ is sampled from a range of speech degradation methods, including additive noise, reverberation, bandwidth limitation, clipping, vocal effects, and their random mixtures.
Namely, for each ground-truth $\mathbf{x}_0$, VoiceBridge considers numerous different LQ prior $\mathbf{x}_1$, rather than considering a fixed degradation method as a single speech enhancement task~\citep{sbsen,bridgesr,audiolbm}. 
A more detailed data construction process at the training stage is introduced in Appendix~\ref{appendix:augmentation}.

Given constructed LQ and HQ waveform pairs $(\mathbf{x}_0,\mathbf{x}_1)$, we directly compress the speech waveform into latent representations, obtaining the latent target and prior, $\mathbf{z}_0 \in \mathbb{R}^{c\times l}$ and $\mathbf{z}_1 \in \mathbb{R}^{c\times l}$.
Prior works~\citep{bridgesr, sbsen} mainly conduct bridge modeling in the data space, either in the waveform or spectrogram domain, to handle single tasks such as speech super-resolution or speech enhancement. However, we find that models in these domains struggle in the complex GSR scenario and 48kHz settings. Directly performing SB in the data space takes great modeling and computational effort. Meanwhile, raw speech signals include massive redundant information at high frequencies, which can be effectively compressed by VAEs, resulting in a more compact latent space. Latent compression reduces sequence length by more than an order of magnitude, enabling a single 544M transformer LBM to handle all tasks. Details experiments proving the necessity of the~\textit{latent SB} design is shown in Appendix~\ref{appendix:ablation}

In comparison with alternative speech compression representations, the waveform latent naturally preserves the~\textit{data-to-data} generative process for the~\textit{LQ-to-HQ} speech tasks without suffering from area removal~\citep{kong2025a2sb}, which is important to a range of tasks in GSR.
For example, for the down-sampling degradation and its mixture with other degradation methods, GSR systems are usually required to generate the 48~kHz waveform from its 2~kHz or even 1~kHz version~\citep{andreev2023hifi++, ku2025generative}. 
In these settings, the prior information contained in the spectrograms~\citep{mandel2023aero,kong2025a2sb} and the latent space of the mel-spectrogram~\citep{voicefixer} will be very limited because of their large-scale area removal, which may hinder the final restoration performance as mentioned by~\citep{lim2018time, kong2025a2sb}.

Given these paired latents $(\mathbf{z}_0,\mathbf{z}_1)$, we model the generative process between them as a Tractable Schrödinger Bridge, a probabilistic interpolation between two marginal distributions over time. The SB problem seeks the stochastic trajectory $p_t$ connecting two distributions $p_0, p_1$ which minimizes the KL divergence to a reference diffusion process. When we take the marginal distributions $p_0, p_1$ to be Gaussian distributions around the latent pairs $(\mathbf{z}_0,\mathbf{z}_1)$, the generally intractable SB problem will have a closed-form solution~\citep{bunne2023schrodinger,bridgetts}, where the interpolated distribution $p_t$ at time $t$ is a Gaussian:
\begin{align}
\label{forward}
p_t= \mathcal{N}(
    \frac{\alpha_t\bar{\sigma}_t^2}{\sigma_1^2}\mathbf{z}_0 + \frac{\bar{\alpha_t}\sigma_t^2}{\sigma_1^2} \mathbf{z}_1,
    \frac{\alpha_t^2\bar{\sigma}_t^2\sigma_t^2}{\sigma_1^2}\mathbf{I}
)
\end{align}
where $\alpha_t, \sigma_t$ are defined by the reference diffusion process.
A neural network is then applied to solve the bridge trajectory from any middle distribution $p_t$, eventually generating $\mathbf{z}_0\sim p_0$ from given $\mathbf{z}_1\sim p_1$ by sampling through the solved trajectory~\citep{bridgetts}. Following~\citep{ bridgetts,bridgesr}, we re-parameterize the neural network to directly predict $\mathbf{z}_0$, optimizing:
\begin{align}
\label{eq:bridge}
\mathcal{L}_\text{bridge}(\varphi)
=
\mathbb{E}_{\substack{
    \mathbf{x}_0 \sim p_{\text{data}},\;\mathbf{x}_1=\mathcal{T}(\mathbf{x}_0)\\
    \mathbf{z}_0=\mathcal{E}(\mathbf{x}_0),\;\mathbf{z}_1=\mathcal{E}(\mathbf{x}_1)
}}
\bigl\|\,\hat{\mathbf{z}}_{0,\varphi}(\mathbf{z}_t,t,\mathbf{z}_1) - \mathbf{z}_0 \bigr\|_2^2
\end{align}
where $\hat{\mathbf{z}}_{0,\varphi}$ denotes the latent predicted by the neural network with trainable parameters $\varphi$; $ \mathbf{z}_0, \mathbf{z}_1 $ are the paired latent representations encoded from the HQ signal and its degraded LQ versions, respectively; $\mathbf{z}_t$ is constructed with Equation~\ref{forward}.
For generalizable training, we explore combining the advantage of transformer architecture, namely good scalability as shown in diffusion-based both image~\citep{dit,u-vit} and audio generation~\citep{evans2024long}, with the bridge generative framework. 
More discussion of tractable Schrödinger bridge models is provided in Appendix~\ref{appendix:sb}.
The details of network architecture are introduced in Appendix~\ref{appendix:training}.

\subsection{Energy-Preserving VAE}
\label{sectioneq}
As shown in Equation~\ref{eq:bridge}, the prior and target in VoiceBridge, $(\mathbf{z}_0, \mathbf{z}_1)$, are compressed from the corresponding waveform $(\mathbf{x}_0, \mathbf{x}_1)$. 
The compression network, namely VAE~\citep{vae}, is employed to reduce the redundant information in waveform space to allow efficient training, while ensuring the reconstruction quality. 
Generally, the training of audio waveform VAE~\citep{evans2024long} with network parameters $\theta$ optimizes the objectives on data-space reconstruction $\mathcal{L}_{\text{data}}$ and latent-space regularization $\mathcal{L}_{\text{lat.}}$:
\begin{align}
\label{eq:vaeloss}
\mathcal{L}_{\text{vae}}
% &= \mathcal{L}_\text{data} + \mathcal{L}_\text{lat.} \\
&= \mathcal{L}_\text{data}( \mathcal{D}_\theta(\mathcal{E}_\theta(\mathbf{x})),\mathbf{x}) + \mathcal{L}_\text{lat.}(\mathcal{E}_\theta(\mathbf{x}),\mathbf{z}_{\text{ref}}),
% &= \underbrace{\mathcal{L}_{\text{rec}} + \mathcal{L}_{\text{fm}} + \mathcal{L}_{\text{adv}}}_{\mathcal{L}_{\text{data}}} + \underbrace{\mathcal{L}_{\text{KL}}}_{\mathcal{L}_{\text{lat.}}}.
\end{align}
where the term $\mathcal{L}_\text{data}$ is typically composed of reconstruction loss, adversarial loss, and feature matching loss between input signal $\mathbf{x}$ and reconstructed version $\mathcal{D}(\mathcal{E}(\mathbf{x}))$, while the term $\mathcal{L}_\text{lat.}$ represents the KL divergence regularization of the latent $\mathbf{z} = \mathcal{E}(\mathbf{x})$ with a known prior distribution $\mathbf{z}_{\text{ref}}$,~\textit{e.g.}, standard Gaussian.
In recent task-specific bridge-based speech enhancement works, the~\textit{data-to-data} generative framework effectively exploits the instructive information contained in the LQ observation in either waveform space~\citep{bridgesr} or STFT representations~\citep{sbsen, kong2025a2sb, han2025few}.
Hence, when designing a~\textit{latent-to-latent} GSR system, we naturally explore preserving the advantages of bridge models from the data space to the encoded latent space.
In VoiceBridge, to facilitate latent SB modeling, we introduce a novel technique, EP-VAE, to enhance the consistency between data and latent from the perspective of scale equivariance. 
Specifically, we add an EP constraint into the VAE training objective, expanding the data-space loss $\mathcal{L}_\text{data}$ in Equation~\ref{eq:vaeloss} to continuous energy levels with $\mathcal{L}_\text{data}^\text{ep}$:
\begin{align}
\label{eq:epvae}
    \mathcal{L}_\text{ep}
    %&=
    %\mathcal{L}_\text{data}^\text{ep} + \mathcal{L}_\text{lat.} \nonumber\\
    &=\mathcal{L}^{\text{ep}}_\text{data}(\mathcal{D}_\theta(s\cdot \mathcal{E}_\theta(\mathbf{x})), s\cdot \mathbf{x}) + \mathcal{L}_\text{lat.}(\mathcal{E}_\theta(\mathbf{x}),\mathbf{z}_{\text{ref}}),
\end{align}
where $s\sim \mathcal{U}(0.5,2)$ is a random scaling factor sampled from a uniform distribution. The EP-VAE is required to maintain the rescaling of latent $\mathbf{z}$ on the reconstructed signal $\mathbf{x}$: when the latent energy is rescaled by $s$ times to $s\cdot \mathbf{z} = s\cdot \mathcal{E}(\mathbf{x})$, the reconstruction signal should have energy rescaled by $s$ times to $s\cdot \mathbf{x}$ as well.
By strengthening waveform-latent alignment with extra regularization at diverse energy levels, we obtain a more structural latent space, which facilitates LBM to reconstruct the distribution of latent HQ target, namely $\mathbf{z}_0$, from the informative latent prior $\mathbf{z}_1$, boosting the performance for high-quality GSR.
%latent diffusion models, to capture the target distribution~\citep{yu2024representation, yao2025reconstruction, kouzelis2025eq, skorokhodov2025improving}. In VoiceBridge, it boosts the performance of LBM in reconstructing the latent HQ target, namely $\mathbf{z}_0$, from the informative latent prior $\mathbf{z}_1$, for high-quality GSR.

%With a well-structured and smooth latent space~\citep{yao2025reconstruction, kouzelis2025eq}, VoiceBridge can be facilitated from two perspectives: generative process and target modelling. 
%Firstly, by strengthening the alignment between the waveform and the latent spaces, EP-VAE can better preserve the characteristics of the LQ prior and the HQ target from data to latent, respectively, and therefore helps VoiceBridge exploit the advantages of~\textit{data-to-data} generative framework in a~\textit{latent-to-latent} manner for~\textit{general LQ-to-HQ} restoration.
%Secondly, by improving the smoothness of the latent space, it facilitates latent generative models to capture the target distribution~\citep{yu2024representation, yao2025reconstruction, kouzelis2025eq, skorokhodov2025improving}, and therefore boosts the performance of VoiceBridge in reconstructing the latent HQ target, namely $\mathbf{z}_0$, from the informative latent prior $\mathbf{z}_1$, for high-quality GSR.

\subsection{Joint Neural Prior}
\label{sectionla}
\begin{figure}
    \centering
    \vspace{-3mm}
    \includegraphics[width=\linewidth]{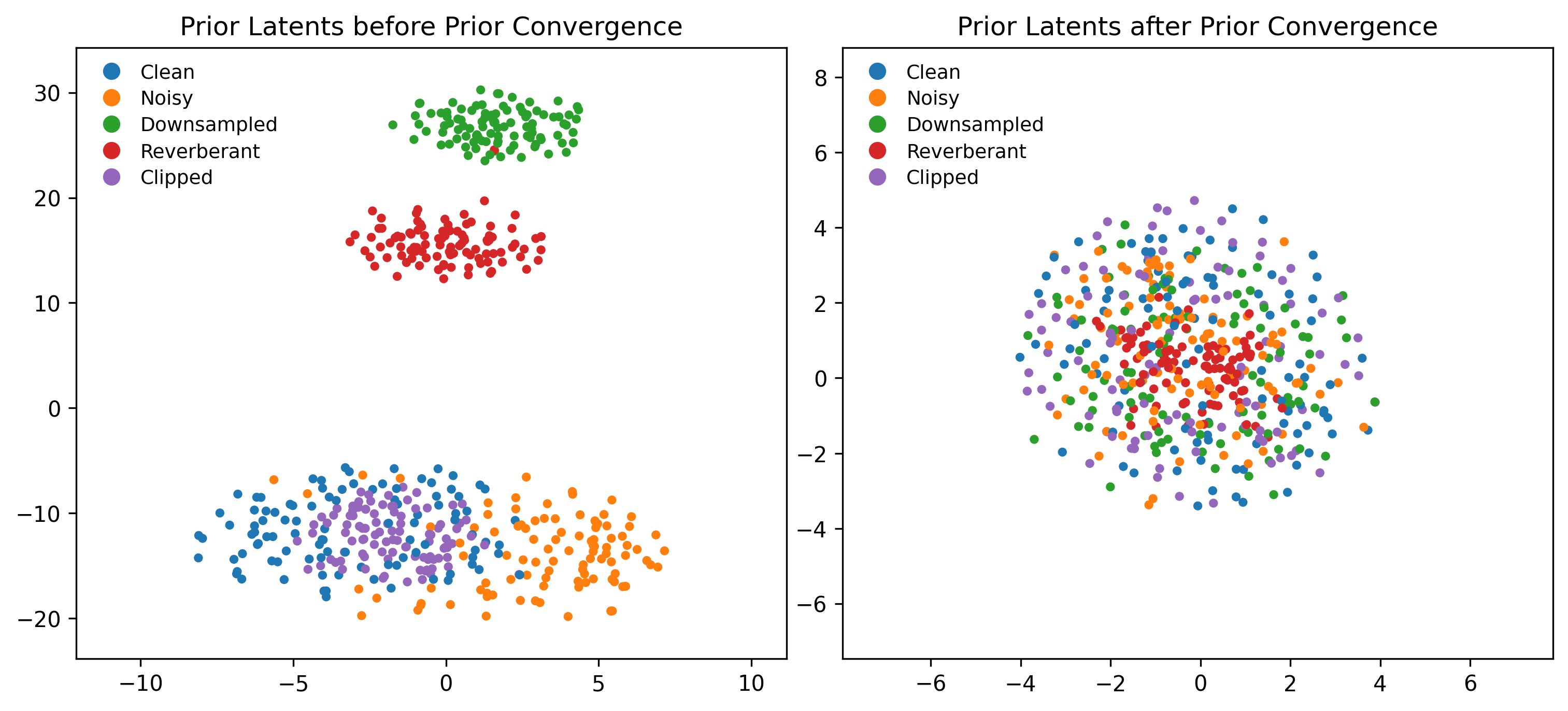}
    \caption{\small The tSNE visualization~\citep{maaten2008visualizing} of the prior latent before and after~\textit{prior convergence} from augmented VCTK~\citep{vctk}. Note the difference between the scale of axes of the two figures.}
    \label{fig:pc}
    \vspace{-4mm}
\end{figure}

In GSR systems, task diversity is one of the key features. 
Namely, one target $\mathbf{x}_0$ can suffer from diverse degradations as the generation prior $\mathbf{x}_1$, and therefore a GSR system should be able to reconstruct the HQ target from a flexible LQ prior rather than a fixed version. 
However, in the latent space, when the LQ priors $\mathbf{z}_1 = \mathcal{E}(\mathbf{x}_1)$ are distinctly different from each other as shown in Figure~\ref{fig:pc}, it inevitably increases the difficulty for an LBM to model~\textit{diverse LQ-to-HQ} tasks with a~\textit{single latent-to-latent} generative process.

\begin{table*}[t]
\small
\centering
\caption{\small  Evaluation results on in-domain tasks, including GSR and denoising benchmarks. The test sets covers the test splits of used datasets, simulated unseen datasets, and in-the-wild data. Best is bolded, and second best is underlined. * denotes results taken from the original paper. VoiceBridge outperforms most baselines in nearly all metrics with a single inference step, demonstrating strong restoration ability in complex environments.}
\vspace{-3mm}
\label{tab:GSR}
\begin{adjustbox}{width=\linewidth,
center}
\begin{tabular}{llclccccccc}
\toprule
% We define the widest header (Restoration) here as the default
\textbf{Task} & \textbf{Dataset} & \textbf{Data Domain} & \textbf{Model} & \textbf{PESQ($\uparrow$)} & \textbf{SIG($\uparrow$)} & \textbf{BAK($\uparrow$)} & \textbf{OVRL($\uparrow$)} & \textbf{UTMOS($\uparrow$)} & \textbf{WVMOS($\uparrow$)} & \textbf{NISQA($\uparrow$)} \\ 
\midrule

% =================================================================================
% RESTORATION & ENHANCEMENT BLOCK (Uses all 7 columns)
% =================================================================================
\multirow{21}{*}{\textbf{\shortstack{General Speech \\ Restoration}}} 
% --- Voicefixer-GSR ---
& \multirow{7}{*}{\textbf{Voicefixer-GSR}} & \multirow{7}{*}{In-Domain Data} 
  & Input & 1.943 & 2.962 & 2.858 & 2.395 & 2.633 & 1.965 & 2.674 \\
& & & Miipher & 1.491 & 3.335 & 3.992 & 3.051 & \underline{3.617} & 4.171 & 4.136 \\
& & & VoiceFixer & 2.043 & 3.302 & 3.971 & 3.005 & 3.439 & 3.800 & 4.159 \\
& & & Resemble-Enhance & 2.000 & \underline{3.412} & 4.045 & \underline{3.137} & 3.608 & \underline{4.224} & \textbf{4.507} \\
& & & UniverSE++ & \underline{2.346} & 3.275 & 3.961 & 2.976 & 3.376 & 3.648 & 4.009 \\
& & & AnyEnhance & - & 3.406* & \underline{4.073}* & 3.136* & - & - & 4.308* \\
& & & \textbf{VoiceBridge} & \textbf{2.471} & \textbf{3.494} & \textbf{4.113} & \textbf{3.285} & \textbf{4.339} & \textbf{4.295} & 4.203 \\
\cmidrule{2-11}

% --- DNS-with-Reverb ---
& \multirow{7}{*}{\textbf{\shortstack{DNS-with-\\Reverb}}} & \multirow{7}{*}{Out-Domain Data} 
  & Input & 1.176 & 1.767 & 1.481 & 1.377 & 1.323 & -0.019 & 1.689 \\
& & & Miipher & 1.324 & 3.401 & 3.975 & 3.128 & \underline{2.973} & 3.195 & 3.629 \\
& & & VoiceFixer & \underline{1.407} & 3.355 & 3.982 & 3.052 & 2.526 & 2.961 & 3.704 \\
& & & Resemble-Enhance & 1.154 & \underline{3.502} & 4.021 & \underline{3.205} & 2.756 & \underline{3.248} & \underline{4.583} \\
& & & UniverSE++ & 1.098 & 2.548 & 3.649 & 2.244 & 1.669 & 2.463 & 3.006 \\
& & & AnyEnhance & - & 3.500* & \underline{4.040}* & 3.204* & - & - & 3.722* \\
& & & \textbf{VoiceBridge} & \textbf{1.583} & \textbf{3.581} & \textbf{4.127} & \textbf{3.318} & \textbf{3.382} & \textbf{4.195} & \textbf{4.588} \\

% --- DNS-Real ---
\cmidrule{2-11}
& \multirow{7}{*}{\textbf{DNS-Real}} & \multirow{7}{*}{\shortstack{Out-Domain Data\\(Real Data)}}
& Input & - & 2.985 & 2.510 & 2.212 & 1.940 & 1.483 & 2.160 \\
& & &Miipher & - & 3.325 & 3.976 & \underline{3.171} & \underline{3.911} & 3.171 & 4.124 \\
& & &VoiceFixer & - & 3.174 & 3.919 & 2.875 & 2.351 & 2.938 & 3.529 \\
& & &Resemble-Enhance & - & 3.395 & \underline{3.993} & 3.100 & 2.767 & \underline{3.298} & \underline{4.329} \\
& & &UniverSE++ & - & 2.999 & 3.660 & 2.641 & 2.306 & 2.603 & 3.317 \\
& & &AnyEnhance & - & \textbf{3.488}* & 3.977* & 3.161* & - & - & - \\
& & &\textbf{VoiceBridge} & - & \underline{3.473} & \textbf{4.025} & \textbf{3.185} & \textbf{4.223} & \textbf{4.045} & \textbf{4.501} \\

\midrule

\multirow{13}{*}{\textbf{\shortstack{Speech \\ Enhancement}}} 

% --- VB-Demand ---
& \multirow{6}{*}{\textbf{VB-Demand}} & \multirow{6}{*}{In-Domain Data} 
  & Input & 1.972 & 3.343 & 3.124 & 2.694 & 3.063 & 2.985 & 2.998 \\
& & & SBSE & 2.140 & 2.571 & 3.658 & 2.946 & 3.407 & 3.819 & 3.976 \\
& & & VoiceFixer & 2.454 & 3.454 & \underline{4.047} & 3.174 & 3.645 & 4.129 & 4.433 \\
& & & Resemble-Enhance & 2.352 & 3.454 & 3.986 & 3.141 & 3.678 & 4.270 & \underline{4.503} \\
& & & UniverSE++ & \textbf{3.020} & \textbf{3.486} & 4.042 & \underline{3.198} & \underline{3.968} & \underline{4.385} & \underline{4.503} \\
& & & \textbf{VoiceBridge} & \underline{2.831} & \underline{3.483} & \textbf{4.062} & \textbf{3.216} & \textbf{4.296} & \textbf{4.392} & \textbf{4.536} \\
\cmidrule{2-11}

% --- WSJ0-CHiME3 ---
& \multirow{7}{*}{\textbf{WSJ0-CHiME3}} & \multirow{7}{*}{Out-Domain Data} 
  & Input & 1.248 & 2.486 & 1.868 & 1.795 & 1.823 & 0.128 & 1.520 \\
& & & SGMSE+ & 1.723 & 3.517 & 3.946 & 3.191 & 3.282 & 3.547 & 4.294 \\
& & & StoRM & \textbf{1.826} & 3.452 & \underline{4.115} & 3.214 & \underline{3.468} & \underline{3.618} & \underline{4.611} \\
& & & VoiceFixer & 1.493 & 3.293 & 4.059 & 3.052 & 2.863 & 3.255 & 3.921 \\
& & & Resemble-Enhance & 1.171 & \underline{3.524} & 4.113 & \underline{3.263} & 2.749 & 3.474 & 4.610 \\
& & & UniverSE++ & 1.320 & 3.173 & 3.997 & 2.914 & 2.984 & 3.248 & 3.918 \\
& & & \textbf{VoiceBridge} & \underline{1.742} & \textbf{3.556} & \textbf{4.118} & \textbf{3.297} & \textbf{4.445} & \textbf{4.374} & \textbf{4.649} \\

\bottomrule
\end{tabular}%
\end{adjustbox}
\vspace{-3mm}
\end{table*}

Hence, we present~\textit{joint neural prior}, aiming at uniformly reducing the distance between each LQ prior $\mathbf{z}_1$ and the ground-truth target $\mathbf{z}_0$, thereby facilitating the bridge generation process. 
Specifically, given the pre-trained EP-VAE encoder $\mathcal{E}$ used for $\mathbf{x}_1$, we fine-tune it to obtain a new version, $\mathcal{E}^\text{np}$, which is capable of converging LQ priors $\mathbf{z}_1$ to a joint neural prior, $\mathbf{z}^\text{np}_1$.
Given the waveform pair $(\mathbf{x}_0,\mathbf{x}_1)$, we encode $\mathbf{x}_0$ with the pre-trained $\mathcal{E}$, and $\mathbf{x}_1$ with a trainable encoder $\mathcal{E}^\text{np}$ initialized from $\mathcal{E}$. Their latent representations $ \mathbf{z}_0 = \mathcal{E}(\mathbf{x}_0)$ and $ \mathbf{z}_1^\text{np} = \mathcal{E}^\text{np}(\mathbf{x}_1)$ are then decoded via the shared pre-trained EP-VAE decoder $\mathcal{D}$ into waveform $ \hat{\mathbf{x}}_0 = \mathcal{D}(\mathbf{z}_0)$ and $ \hat{\mathbf{x}}_1^\text{np} = \mathcal{D}(\mathbf{z}_1^\text{np})$. 
%We hope to reduce the latent diversity and prior-target gap by enclosing the distance between the latent distribution of $\mathbf{z}_0$ and $\mathbf{z}_1$ through finetuning the new Encoder $\mathcal{E}^\text{np}$, minimizing $\mathcal{L}(\mathbf{z}_0,\mathbf{z}_1^\text{np})$. Further, we hope that the enclosure happens in all measures between the two spaces, not only the Euclidean distance. 
Then, we reduce the distance between each LQ prior and the ground-truth target from both data and latent spaces with:
%align the pair in latent space as well as reconstructed data space to ensure both geometrical and perceptual similarity, further enhancing the semantic proximity between prior and target latents, via objective
\begin{align}
\label{eq:pc-encoder}
\mathcal{L}_\text{np-enc} &= \mathcal{L}^\text{ep}_\text{data}(\mathcal{D}(s\cdot \mathcal{E}^\text{np}_{\theta}(\mathbf{x}_1)), s\cdot \hat{\mathbf{x}}_0)+\mathcal{L}_{\text{lat.}}(\mathcal{E}^\text{np}_{\theta}(\mathbf{x}_1), \mathbf{z}_0),
%\mathcal{L}_\text{np} &= \mathcal{L}_\text{data}(s\cdot \mathcal{D}(\mathcal{E}(\mathbf{x}_0)), \mathcal{D}(s\cdot \mathcal{E}^\text{np}(\mathbf{x}_1))) \\
%&+\mathcal{L}_{\text{lat.}}(\mathcal{E}(\mathbf{x}_0), \mathcal{E}^\text{np}(\mathbf{x}_1)) \nonumber
%\\&+ \mathcal{L}_{\text{data}}^\text{EP}( \mathcal{D}(\mathcal{E}(\mathbf{x}_0)), \mathcal{D}(\mathcal{E}^\text{np}(\mathbf{x}_1))) \nonumber\\
%&=\mathcal{L}_{\text{lat.}}(\mathcal{E}(\mathbf{x}_0), \mathcal{E}^\text{np}(\mathbf{x}_1))\nonumber\\
%+ \mathcal{L}_\text{data}(s\cdot \mathcal{D}(\mathcal{E}(\mathbf{x}_0)), \mathcal{D}(s\cdot \mathcal{E}^\text{np}(\mathbf{x}_1)))
\end{align}
\begin{comment}
\begin{align}
\mathcal{L}_\text{np} &= \mathcal{L}_{\text{lat.}}(\mathbf{z}_0, \mathbf{z}_1) + \mathcal{L}_{\text{data}}^\text{EP}( \hat{\mathbf{x}}_0,\hat{\mathbf{x}}_1) \\
&=\mathcal{L}_{\text{lat.}}(\mathbf{z}_0, \mathbf{z}_1) + \mathcal{L}_\text{data}(s\cdot \hat{\mathbf{x}}_0, \mathcal{D}(s\cdot \mathbf{z}_1))
\end{align}
\end{comment}
where $\mathcal{L}_{\text{lat.}}$ has been the joint neural prior loss consisting of MSE loss and cosine similarity loss as introduced in Appendix~\ref{appendix:training}, and $\mathcal{L}_{\text{data}}^\text{ep}$ is the data-space loss in EP-VAE training modified for prior convergence. Prior works~\citep{fang2021variational} applies a KL convergence loss to a VAE for speech denoising. In contrast, we add objectives including cosine similarity in latent space and multiple supervision signals in the reconstructed data space, in order to reduce the distance between the latents in all geometrical measures. We prove our method achieves better results in Appendix~\ref{appendix:ablation}.

Namely, as shown in Equation~\ref{eq:pc-encoder}, the encoder $\mathcal{E}^\text{np}$ is fine-tuned with changed objectives to converge different LQ priors to HQ target, rather than still learning the latent presentation for reconstruction as Equation~\ref{eq:epvae}.
Moreover, the EP constraint is preserved. When reducing the distance between LQ priors $\mathbf{z}_1$ and the HQ target $\mathbf{z}_0$ to achieve $\mathbf{z}_1^{\text{np}}$, scale equivariance is simultaneously required, which inherits the EP property to measure the consistency at varying energy levels. %Since the HQ latent representation is our generation target, the latent regularization term with $z_\text{ref}$ is omitted.
As shown in Figure~\ref{fig:pc}, this technique significantly closes the distance between different LQ priors and the HQ target in the latent space, and naturally reduces the burden of LBM modeling. A further quantitative analysis on the prior distance is shown in Appendix~\ref{appendix:latents}

\subsection{From Denoiser to Generator}
\label{sectionpagen}
After training the EP-VAE encoder $\mathcal{E}$ with Equation~\ref{eq:epvae} to encode HQ target, its fine-tuned version $\mathcal{E}^\text{np}$ with Equation~\ref{eq:pc-encoder} for diverse LQ priors, and the shared decoder $\mathcal{D}$, we can already create a structured latent space where the different LQ priors have converged.
Then, we train the bridge transformer model that maps the joint neural prior and the HQ target using Equation~\ref{eq:bridge}, while the LQ prior and condition $\mathbf{z}_1$ have been changed to $\mathbf{z}_1^{\text{np}}$ encoded by $\mathcal{E}^\text{np}$, and the HQ target $\mathbf{z}_0$ is compressed from $\mathbf{x}_0$ with EP-VAE encoder $\mathcal{E}$.

Although the latent bridge model (LBM) and the VAE decoder are trained sequentially, their objectives are not perfectly aligned: the LBM is optimized to predict target latents under the bridge loss, while the decoder is optimized to reconstruct waveforms from encoder latents. At inference, this discrepancy can lead to \emph{cascading mismatch}: small deviations in LBM-predicted latents may be amplified by the decoder, degrading perceptual quality. 
To mitigate this issue, we introduce a post-training stage that \emph{jointly} fine-tunes the LBM and the decoder, while keeping the encoder fixed. This preserves the latent geometry learned by the encoder, and simultaneously calibrates the decoder to the LBM-induced latent distribution.

To ensure fine supervision from the data domain, we introduce the data-level reconstruction loss $\mathcal{L}_\text{data}$ in VAE training to the objectives of the post-training process. Meanwhile, as GSR systems are required to generate speech samples with high perceptual quality~\citep{manjunath2009limitations, manocha2022audio,finally}, we incorporate a perceptual metric aligned loss $\mathcal{L}_\text{perc}$ utilizing the PESQ~\citep{pesq} and UTMOS~\citep{utmos} metrics. However, including only this term causes severe overfitting problems~\citep{oliveira2024pesqetarian}, as the decoder falls into a local minima of these metrics and forget faithful reconstruction.
To tackle this issue, we activate the discriminator $\omega$ used in VAE training, and incorporate an adversarial objective $\mathcal{L}_\text{GAN}$ to the joint post-training process. The perceptual objective aligns the generation process with human perceptual quality, and the discriminator is reponsible to spot artifacts caused by optimizing the perceptual indicators and prevent overfitting. 

Additionally, the adversarial objective surprisingly can transform the LBM from a multi-step denoiser into a single-step generator, enabling one-step inference without distillation. In the LBM training stage guided by the MSE loss~\ref{eq:bridge}, the model learns an optimal $\hat{\mathbf{z}}_{0,\varphi}$ estimating
\begin{align}
    \arg\min_{\hat{\mathbf{z}}_{0,\varphi}}\ \mathbb{E}\left[\|\mathbf{z}_0 - \hat{\mathbf{z}}_{0,\varphi}(\mathbf{z}_t, \mathbf{z}_1^{\text{np}})\|_2^2\right]=
\mathbb{E}[\mathbf{z}_0 \mid \mathbf{z}_t,\mathbf{z}_1^{\text{np}}],
\label{eq:regression_mean}
\end{align}
The one-step prediction of $\hat{\mathbf{z}}_{0,\varphi}$ is not a sample from the target distribution, but a conditional expectation symbolizing the denoising target. However, the adversarial loss aims to
$\min_{\varphi}\ \max_{\omega}\ \mathcal{L}_{\text{GAN}}(\varphi,\omega)$, 
leading $\hat{\mathbf{z}}_{0,\varphi}$ from the conditional expectation toward the conditional distribution 
\begin{align}
p_{\varphi}(\mathbf{z}_0\mid \mathbf{z}_t,\mathbf{z}_1^{\text{np}})
=
p_{\text{data}}(\mathbf{z}_0\mid \mathbf{z}_t,\mathbf{z}_1^{\text{np}}),
\label{eq:gan_match_dist}
\end{align}
at the global optimum, which gives the model one-step restoration ability.
Detailed derivations are in Appendix~\ref{appendix:math}

%Specifically, after training EP-VAE with Equation~\ref{eq:epvae}, fine-tuned encoder $\mathcal{E}^\text{np}$ with Equation~\ref{eq:pc-encoder}, and LBM with Equation~\ref{eq:bridge} in turn, 
Technically, we perform a post-training stage for both the LBM and the EP-VAE decoder, mitigating their cascading mismatches in inference as well as aligning both the bridge sampling and VAE decoding with human perceptual quality.
The post-training objectives for LBMs and the decoder include:
\begin{align}
\label{eq:ap}
    \mathcal{L}_\text{pt} 
    =& \mathcal{L}_\text{bridge}(\varphi)
    + \mathcal{L}_\text{data}(\theta, \varphi) \nonumber\\
    +& \mathcal{L}_\text{perc}(\theta, \varphi) + \mathcal{L}_\text{GAN}(\theta, \varphi, \omega),
    % (\mathcal{D}_{\theta}(\hat{\mathbf{z}}_{0,\varphi}(\mathbf{z}_t,t,\mathbf{z}^{\text{np}}_1)), \mathbf{x}_0) \nonumber\\
    % +& \mathcal{L}_\text{GAN}(\mathcal{D}_{\theta}(\hat{\mathbf{z}}_{0,\varphi}(\mathbf{z}_t,t,\mathbf{z}^{\text{np}}_1)), \mathbf{x}_0) \nonumber\\
    % +& \mathcal{L}_\text{perc}(\mathcal{D}_{\theta}(\hat{\mathbf{z}}_{0,\varphi}(\mathbf{z}_t,t,\mathbf{z}^{\text{np}}_1)), \mathbf{x}_0),
\end{align}

where $\mathcal{L}_\text{perc}$ denotes a PESQ loss and a UTMOS loss, and $\mathcal{L}_\text{data}$ is the data level MSE loss and $\mathcal{L}_\text{GAN}$ the adversarial loss. These three losses are all calculated between decoded signal $\hat{\mathbf{z}}_{0,\varphi}(\mathbf{z}_t,t,\mathbf{z}^{\text{np}}_1)$ and ground truth $\mathbf{x}_0$, with weights updated on LBM $\varphi$, decoder $\theta$, and discriminator $\omega$ differently.

% Intuitively, such a post-training procedure helps the bridge model to generate the speech samples on the sub-manifold of the latent space corresponding to the waveform with high perceptual quality, as well as encourages the decoder to generate HQ audio. 
% As the encoder is frozen in this training section, this technique does not affect the good latent properties established by EP-VAE and prior convergence, allowing the utilization of each proposed innovation in VoiceBridge.

%\vspace{-0.15cm}

% \begin{wraptable}{0.55\textwidth}
%   \centering
%   \caption{\small Bandwidth extension results on VCTK-BWE. VoiceBridge beats other GSR models and audio super-resolution models on both subjective and objective metrics.}
%   \label{tab:SR}
%   \begin{adjustbox}{width=\linewidth,
% center}
%     \input{assets/SR_table.tex}
%     \end{adjustbox}
% \end{wraptable}

\subsection{Training Pipeline Overview}
\label{pipeline}
In actual training, VoiceBridge integrates EP-VAE and the joint neural prior, and achieves denoiser-to-generator alignment through a four-stage training pipeline: (1) pre-training an EP-VAE on clean speech with Equation~\ref{eq:vaeloss}; (2) fine-tuning the EP-VAE encoder to learn a joint neural prior on distorted inputs with Equation~\ref{eq:pc-encoder}; (3) training the transformer model to solve the SB in latent space with Equation~\ref{eq:bridge} and (4) jointly fine-tuning the LBM and the decoder to enhance perceptual alignment with Equation~\ref{eq:ap}.

\begin{table*}[t]
\small
\centering
\caption{\small  Enhancement results on OOD tasks, including codec artifact removal and TTS quality improvement. VoiceBridge exhibits consistent improvement on generated speech quality.}
\label{tab:table2}
\vspace{-3mm}
\begin{adjustbox}{width=1\linewidth, center}

\begin{tabular}{llclcccccc}
\toprule
\textbf{Task} & \textbf{Dataset} & \textbf{Data Domain} & \textbf{Model($\uparrow$)} & \textbf{WVMOS($\uparrow$)} & \textbf{UTMOS($\uparrow$)} & \textbf{DNSMOS($\uparrow$)} & \textbf{NISQA($\uparrow$)} & \textbf{PESQ($\uparrow$)} & \textbf{WER($\downarrow$)} \\ 
\midrule

% =================================================================================
% CODEC BLOCK
% Columns: WVMOS, UTMOS, DNSMOS, NISQA, PESQ. (WER is empty)
% =================================================================================
\multirow{5}{*}{\textbf{\shortstack{Codec Artifact \\ Removal}}} & \multirow{5}{*}{\shortstack{\textbf{VCTK}\\\textbf{(Encodec)}}} & \multirow{5}{*}{In-Domain Data} 
  & Input & 4.036 & 2.889 & 2.822 & 3.661 & 2.210 & - \\
& & & VoiceFixer & 3.956 & 3.229 & 3.003 & 4.436 & 1.984 & - \\
& & & Resemble-Enhance & \underline{4.054} & \underline{3.294} & \underline{3.063} & \underline{4.665} & 2.059 & - \\
& & & UniverSE++ & 4.002 & 2.954 & 2.868 & 3.861 & \underline{2.320} & - \\
& & & \textbf{VoiceBridge} & \textbf{4.131} & \textbf{3.683} & \textbf{3.132} & \textbf{4.469} & \textbf{2.361} & - \\
\midrule

% =================================================================================
% TTS EVAL BLOCK
% Columns: WVMOS, UTMOS, DNSMOS, NISQA, WER. (PESQ is empty)
% =================================================================================
\multirow{10}{*}{\shortstack{\textbf{TTS Quality}\\\textbf{Improvement}}} & \multirow{5}{*}{\shortstack{\textbf{Seed-TTS}\\\textbf{(MoonCast)}}} & \multirow{5}{*}{Out-Domain Data} 
  & Input & \underline{3.971} & \underline{3.623} & 3.120 & 3.922 & - & 5.93\% \\
& & & VoiceFixer & 3.832 & 3.481 & 3.145 & 4.004 & - & \underline{5.51}\% \\
& & & Resemble-Enhance & 3.612 & 3.291 & \underline{3.176} & 4.121 & - & 12.88\% \\
& & & UniverSE++ & 3.747 & 3.446 & 3.143 & \underline{4.217} & - & 8.59\% \\
& & & \textbf{VoiceBridge} & \textbf{4.098} & \textbf{3.821} & \textbf{3.227} & \textbf{4.464} & - & \textbf{5.40}\% \\
\cmidrule{2-10}

& \multirow{5}{*}{\shortstack{\textbf{Seed-TTS}\\\textbf{(MaskGCT)}}} & \multirow{5}{*}{Out-Domain Data} 
  & Input & 3.673 & 3.427 & 3.122 & 3.964 & - & \textbf{5.63}\% \\
& & & VoiceFixer & 3.483 & 3.288 & 3.186 & 4.349 & - & 6.02\% \\
& & & Resemble-Enhance & \underline{3.818} & \underline{3.594} & \textbf{3.312} & \underline{4.691} & - & 7.78\% \\
& & & UniverSE++ & 3.324 & 3.277 & 3.180 & 4.308 & - & 7.21\% \\
& & & \textbf{VoiceBridge} & \textbf{4.172} & \textbf{4.051} & \underline{3.326} & \textbf{4.769} & - & \underline{5.66}\% \\
% \midrule

% % =================================================================================
% % ENHANCEMENT BLOCK
% % Columns: WVMOS, UTMOS, DNSMOS, NISQA, PESQ. (WER is empty)
% % Note: DNSMOS comes from OVRL column in source.
% % =================================================================================
% \multirow{7}{*}{\shortstack{\textbf{Real-World}\\\textbf{Speech}\\\textbf{Restoration}}} & \multirow{7}{*}{\textbf{DNS-Real}} & \multirow{7}{*}{\shortstack{Out-Domain Data\\(Real Data)}}
%   & Input & 1.483 & 1.940 & 2.212 & 2.160 & - & - \\
% & & & Miipher & 3.171 & \underline{3.911} & \textbf{3.171} & 4.124 & - & - \\
% & & & VoiceFixer & 2.938 & 2.351 & 2.875 & 3.529 & - & - \\
% & & & Resemble-Enhance & \underline{3.298} & 2.767 & 3.100 & \underline{4.329} & - & - \\
% & & & UniverSE++ & 2.603 & 2.306 & 2.641 & 3.317 & - & - \\
% & & & AnyEnhance & - & - & 3.161* & - & - & - \\
% & & & \textbf{VoiceBridge} & \textbf{4.034} & \textbf{4.197} & \underline{3.166} & \textbf{4.497} & - & - \\

\bottomrule
\end{tabular}%
\end{adjustbox}
\vspace{-3mm}
\end{table*}

\section{Experiments}

% Due to page limitations, all detailed composition of experimental setup are listed in Appendix~\ref{appendix:}

\subsection{Experimental Setup}
\begin{table}
\small
\centering
\vspace{-2mm}
\caption{\small  MOS results on VoiceFixer-GSR and DNS-Real datasets, covering both simulated and real data. VoiceBridge outperforms other baselines on subjective evaluation results collected in human listening experiments.}
\label{tab:Mos}
\vspace{-1mm}
\begin{adjustbox}{width=0.8\linewidth, center}
\begin{tabular}{l|cccc|c}
\toprule
 & VF & RE & USE & VB & GT \\
\midrule
Voicefixer-GSR  & 3.77 & \underline{3.97} & 2.91 & \textbf{4.32} & 4.43 \\
DNS-Real & 2.90 & \underline{3.02} & 2.04 & \textbf{4.28} & - \\
\bottomrule
\end{tabular}
\vspace{-3mm}
\end{adjustbox}%
\vspace{-11mm}
\end{table}

% \begin{table*}[t]
% \small
% \centering
% \caption{\small Evaluation results on three denoising benchmarks. Best is bolded and second best is underlined. VoiceBridge achieves strong performance on the traditional speech enhancement task compared with both other GSR models and specific SE models.}
% \label{tab:SE}
% \vspace{-3mm}
% \begin{adjustbox}{width=\linewidth,
% center}
% \input{assets/SE_table.tex}
% \end{adjustbox}
% \vspace{-3mm}
% \end{table*}
% \begin{wraptable}{0.55\textwidth}
%   \centering
%   \vspace{-19mm}
%   \caption{\small  Bandwidth extension results on VCTK-BWE. VoiceBridge beats other GSR models and audio super-resolution models on both subjective and objective metrics.}
%   \label{tab:SR}
%   \vspace{-3mm}
%   \begin{adjustbox}{width=\linewidth,
% center}
%     \input{assets/SR_table.tex}
%     \end{adjustbox}
%     \vspace{-3mm}
% \end{wraptable}

We train and evaluate VoiceBridge on GSR task, and also test its zero-shot generalization to out-of-domain (OOD) data and tasks. Our full training corpus contains approximately 1138 hours of clean speech audio at 48 kHz, constructed only by combining \textbf{public datasets}. 
Distortions are synthesized using publicly available noise datasets and room impulse responses, following previous works~\citep{voicefixer, anyenhance}. 
The compression network adopts the Oobleck~\citep{evans2024long} arthitecture. The LBM is built with a 544M-parameter Transformer backbone~\citep{stableaudio}. 
The training process is described in Section~\ref{pipeline}. Only \textbf{one inference step} is used.
We provide detailed information on the VAE compression network, datasets, and evaluation methods in Appendix~\ref{appendix:training},~\ref{appendix:dataset}, and~\ref{appendix:evaluation}, respectively.

\subsection{General Speech Restoration}
We first evaluate VoiceBridge's performance on in-domain restoration tasks, including simulated GSR task and downstream single-task speech enhancement. Following previous works, we use VoiceFixer-GSR~\citep{voicefixer}, DNS-with-Reverb~\citep{reddy2021interspeech} as the simulated GSR benchmarks. Notably, data from the DNS-Challenge~\citep{reddy2021interspeech} are excluded in our training data, providing a mismatch in training test conditions for VoiceBridge.
We compare VoiceBridge against other GSR models including Miipher~\citep{koizumi_miipher_2023}, VoiceFixer~\citep{voicefixer}, Resemble-Enhance~\citep{resembleenhance}, UniverSE++~\citep{scheibler_universalpp_2024} and AnyEnhance~\citep{anyenhance}. Despite these models, we further compare VoiceBridge to close-sourced models with there reported results or using there demo samples in Appendix~\ref{appendix:evaluation}.

\begin{figure}
  \centering
  \vspace{-3mm}
  % \hfill
  \begin{subfigure}[b]{0.17\linewidth}
  \centering
    \includegraphics[width=\linewidth]{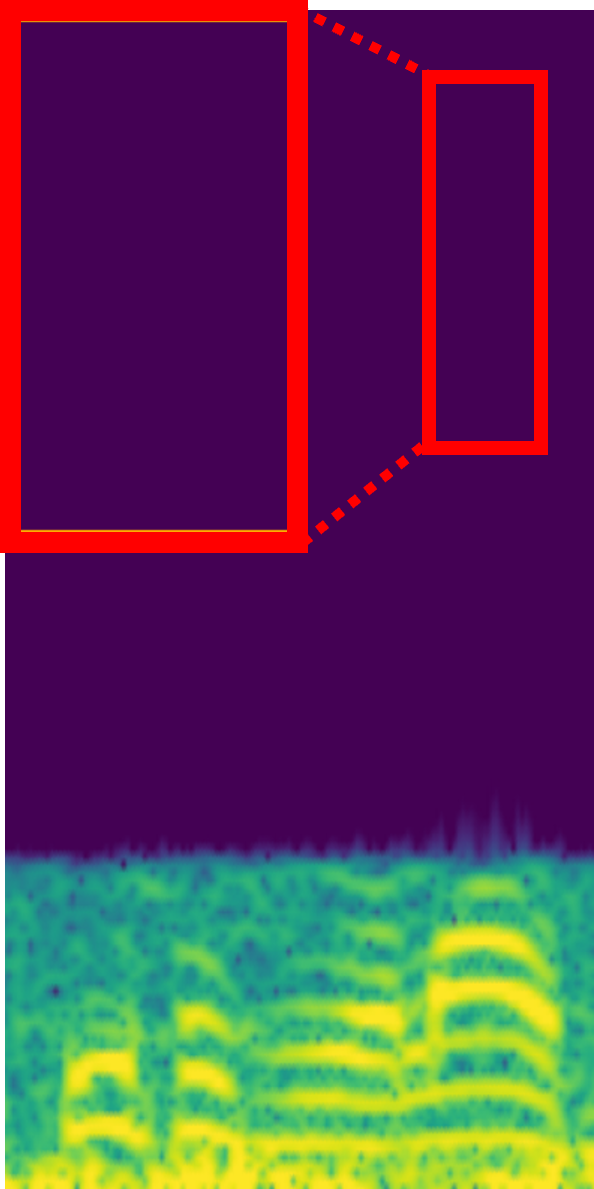}
    \caption{\small  LQ}
  \end{subfigure}
 \hspace{0.02cm}
  \begin{subfigure}[b]{0.17\linewidth}
  \centering
    \includegraphics[width=\linewidth]{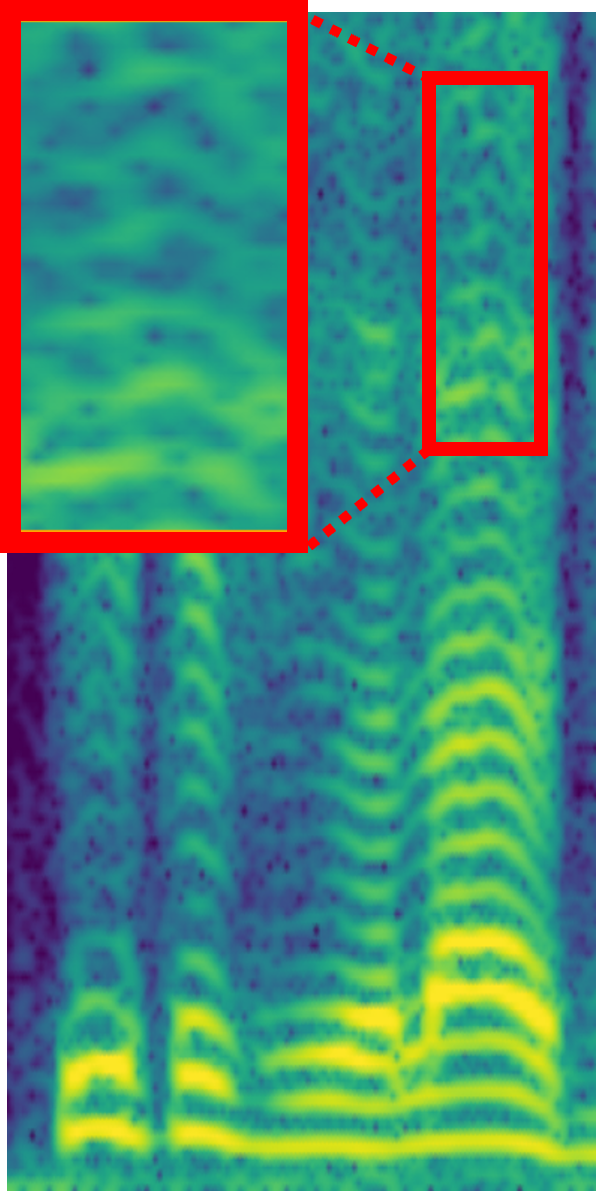}
    \caption{\small  VF}
  \end{subfigure}
\hspace{0.02cm}
  \begin{subfigure}[b]{0.17\linewidth}
  \centering
    \includegraphics[width=\linewidth]{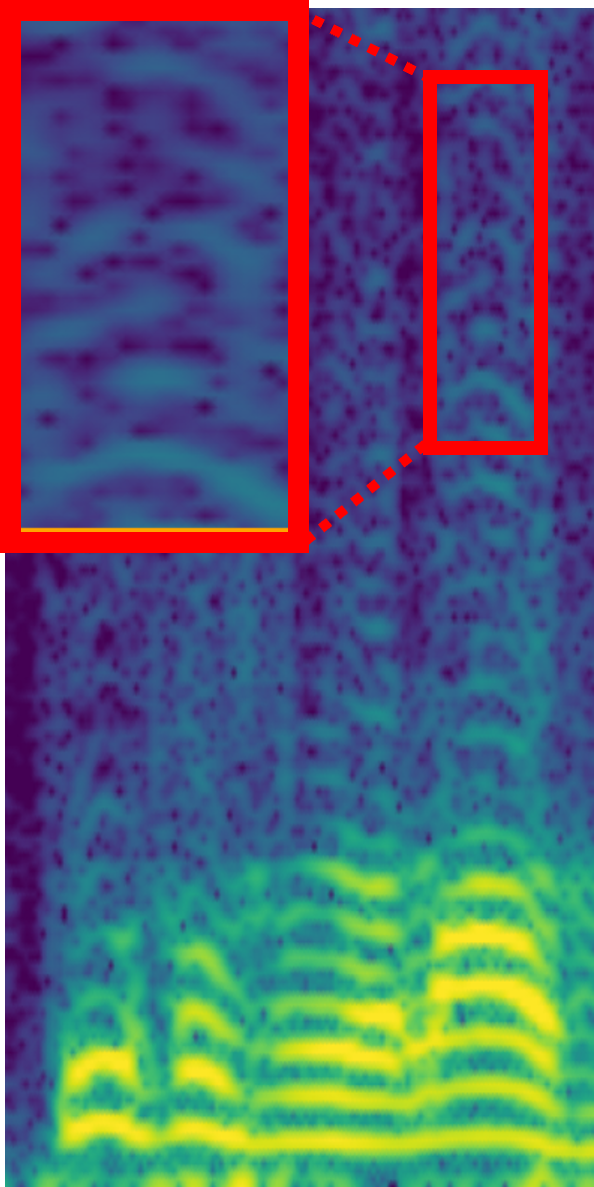}
    \caption{\small  RE}
  \end{subfigure}
\hspace{0.02cm}
  \begin{subfigure}[b]{0.17\linewidth}
  \centering
    \includegraphics[width=\linewidth]{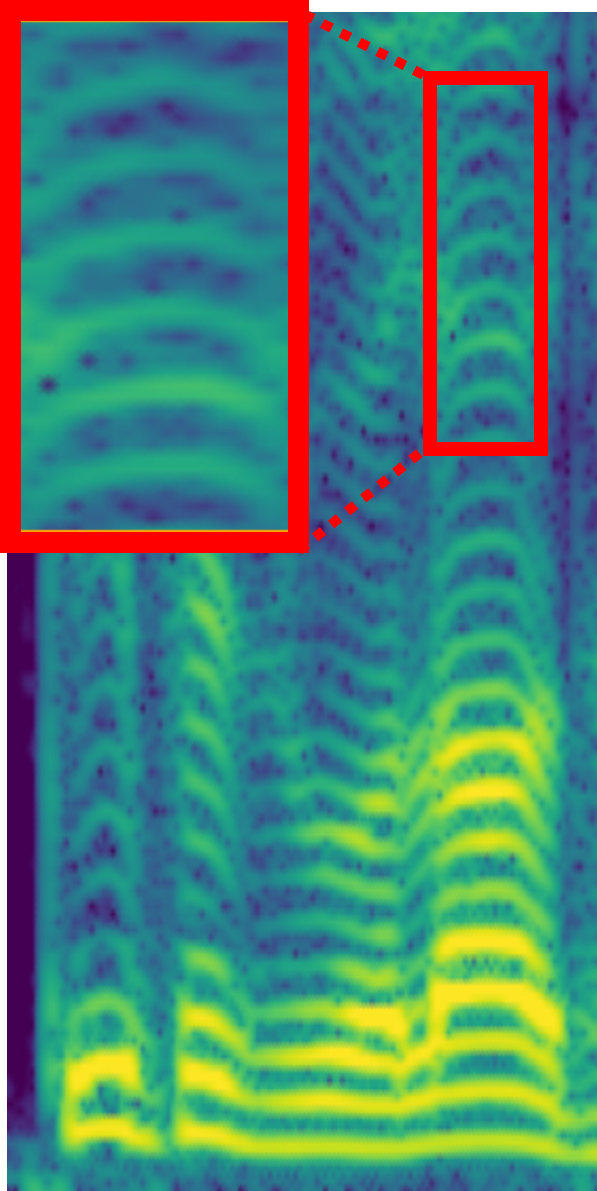}
    \caption{\small  VB}
  \end{subfigure}
\hspace{0.02cm}
  \begin{subfigure}[b]{0.17\linewidth}
  \centering
    \includegraphics[width=\linewidth]{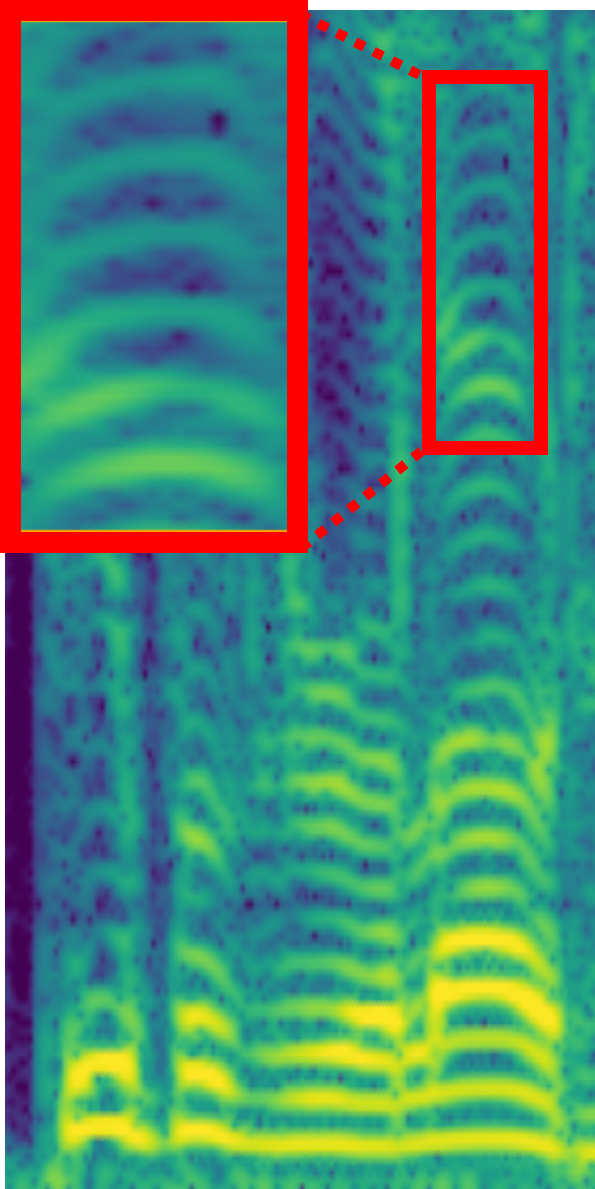}
    \caption{\small  GT}
  \end{subfigure}

  \vspace{-1mm}
  \caption{\small  STFT spectrograms of the same piece of speech restored by different models (a) Low-Quality Signal. (b) VoiceFixer (b) Resemble-Enhance  (d) VoiceBridge (Ours) (e) Ground Truth.}
  \label{casestudy}
  \vspace{-9mm}
\end{figure}
 
VoiceBridge and the four baselines have included five classes of generative models: bridge, mapping-based network, flow-matching, diffusion, and masked generative models. Such diversity enables a broad comparison across generative modeling strategies. Recently, non-intrusive metrics which measure perceptual quality without reference signals are believed to be more relevant to generative speech restoration~\citep{manjunath2009limitations, manocha2022audio, finally}. In our evaluation, we report both intrusive and non-intrusive metrics, incorporating PESQ~\citep{pesq}, DNSMOS~\citep{dnsmos} (SIG, BAK, OVRL), UTMOS~\citep{utmos}, WV-MOS~\citep{wvmos}, and NISQA~\citep{nisqa}.

The GSR evaluation results are shown in Table~\ref{tab:GSR}. VoiceBridge achieves the best or the second-best results on almost every metric across all datasets, demonstrating strong multi-degradation restoration capability on both simulated and real-world audios. 
We further conduct a user study with VoiceFixer-GSR and DNS-Real to investigate human preference for each GSR system. 
The result, as shown in Table~\ref{tab:Mos}, demonstrates VoiceBridge's superior performance, achieving significantly higher Mean Opinion Score (MOS) on both simulated and real audio samples.
A case study is shown in Figure~\ref{casestudy}, where VoiceBridge faithfully generates different components of the ground truth with four steps in comparison with other methods. 

Further, as a GSR model, VoiceBridge naturally handles all kinds of restoration tasks within the data distribution. We test VoiceBridge on the task of traditional Speech Enhancement (SE), comparing it to task-specific models including SGMSE+~\citep{richter2023speech, sgmse}, StoRM~\citep{storm} and SBSE~\citep{sbsen} on benchmarks involving VoiceBank-DEMAND~\citep{valentini2017noisy} and WSJ-CHiME~\citep{chime3}. The results are shown in Table~\ref{tab:GSR}, where VoiceBridge outperforms data-space SB models SBSE~\citep{sbsen}, diffusion models, and flow-matching models on various metrics. Note that the WSJ0 dataset, which the SGMSE+ and StoRM models are trained on, is excluded from our training dataset, exceptionally demonstrating the superiority of VoiceBridge. 
Apart from SE, evaluation for other tasks has also been conducted, with details in Appendix~\ref{appendix:evaluation}.

\subsection{Out of Domain Tasks}

Moreover, as a probabilistic generative model, VoiceBridge captures target distributional information rather than establishing a point-to-point mapping function. Therefore, it shows strong zero-shot ability on OOD restoration tasks that are unseen during the training stage. We evaluate VoiceBridge and other GSR models on codec artifact removal, where all models are tested to repair reconstructed audio from VCTK by the Encodec~\citep{defossez2022high} model operating at 3~kbps bitrate. 
Beyond that, We originally propose another application scenario of GSR models: further refining the perceptual quality of generation results of text-to-speech (TTS) models.
We take two recent methods, MaskGCT~\citep{wang2024maskgct} and MoonCast~\citep{ju2025mooncast}, as the TTS baseline methods to synthesize speech on the Seed-TTS~\citep{higgsaudio2025, peng2025vibevoice} benchmark, and then apply VoiceBridge and other GSR baselines to refine the generated speech samples. We evaluate using both GSR and TTS metrics.

As Shown in Table~\ref{tab:table2}, VoiceBridge consistently outperforms baseline models on various metrics, exhibiting the zero-shot performance of VoiceBridge on OOD tasks.
Importantly, in Table~\ref{tab:table2}, other GSR models may fail to further enhance the model-generated speech, while VoiceBridge reliably improves the quality. This demonstrates VoiceBridge's strong capability of comprehensively refining audio quality.

\subsection{Ablation Studies}

\begin{figure*}[t]
    \centering
    \vspace{-1mm}
     \caption{\small  Ablation study with different GSR metrics. The horizontal axis shows training steps, and the vertical axis displays performance. The models differ in whether \textit{EP-VAE} and \textit{joint neural prior} are employed.}
     
    \begin{subfigure}[b]{0.24\linewidth}
        \includegraphics[width=\linewidth]{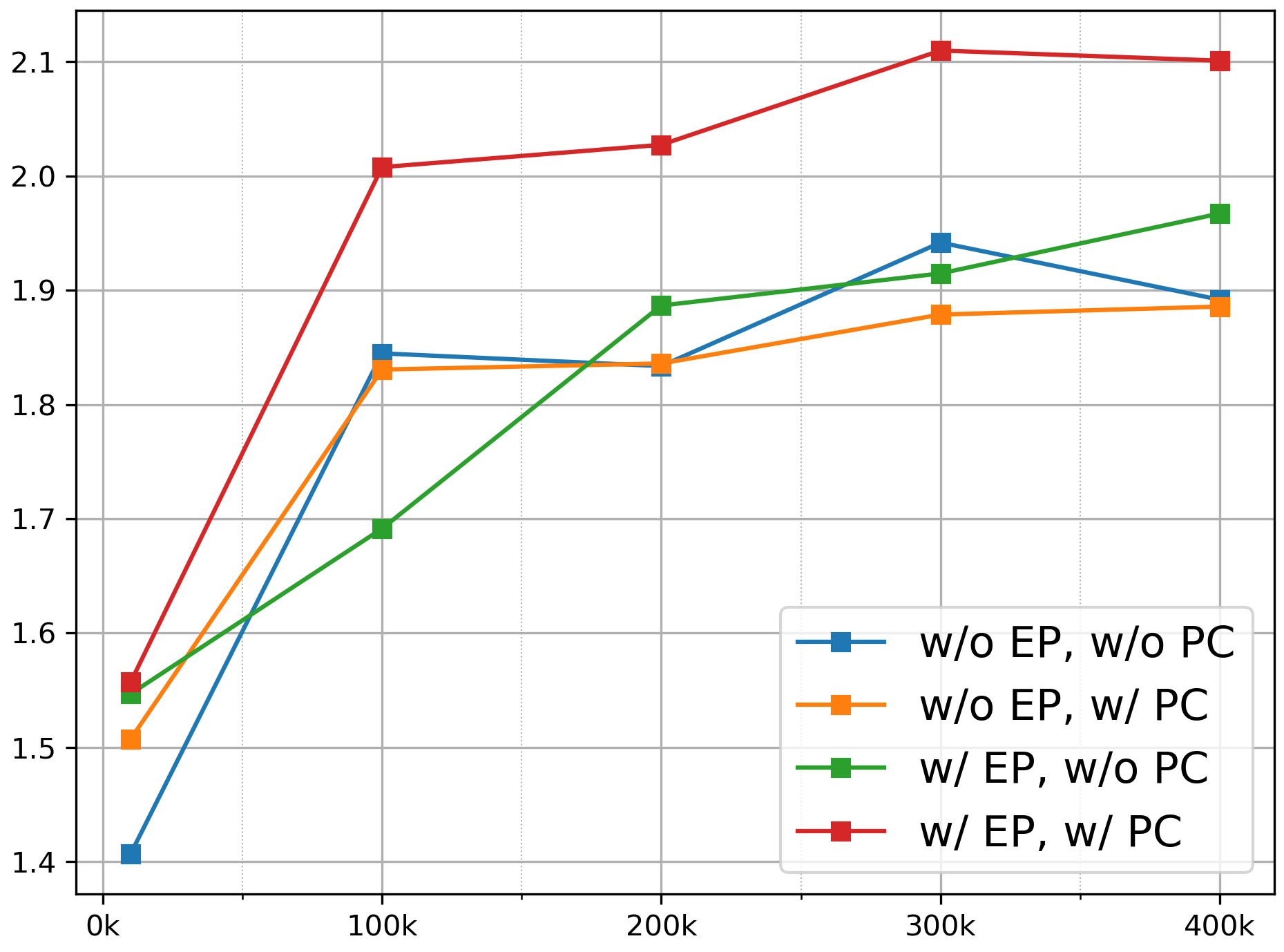}
        \caption{\small  PESQ}
        \label{fig:ablation_1}
    \end{subfigure}
    \hfill
    \begin{subfigure}[b]{0.24\linewidth}
        \includegraphics[width=\linewidth]{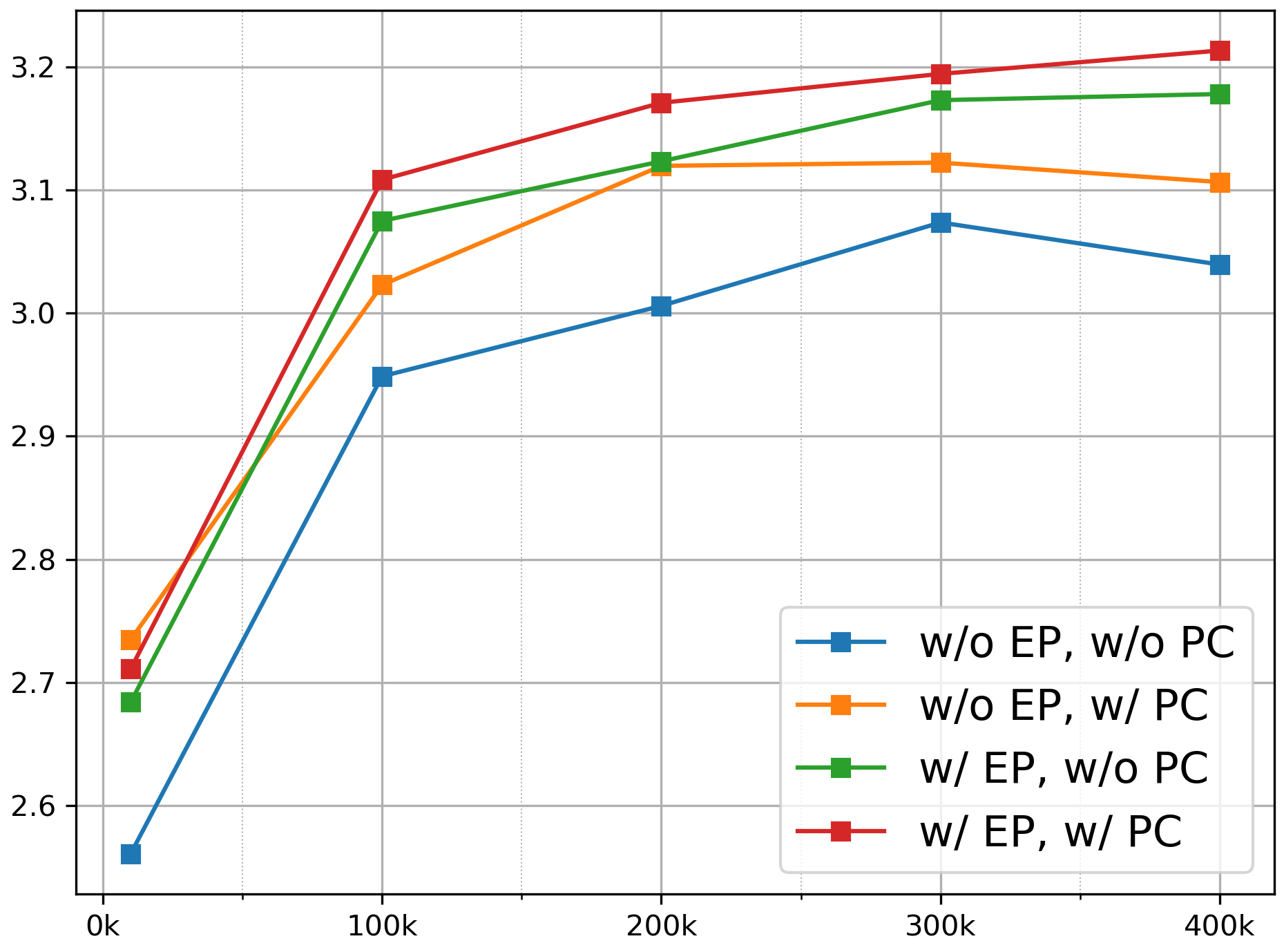}
        \caption{\small  DNSMOS-OVRL}
        \label{fig:ablation_2}
    \end{subfigure}
    \hfill
    \begin{subfigure}[b]{0.24\linewidth}
        \includegraphics[width=\linewidth]{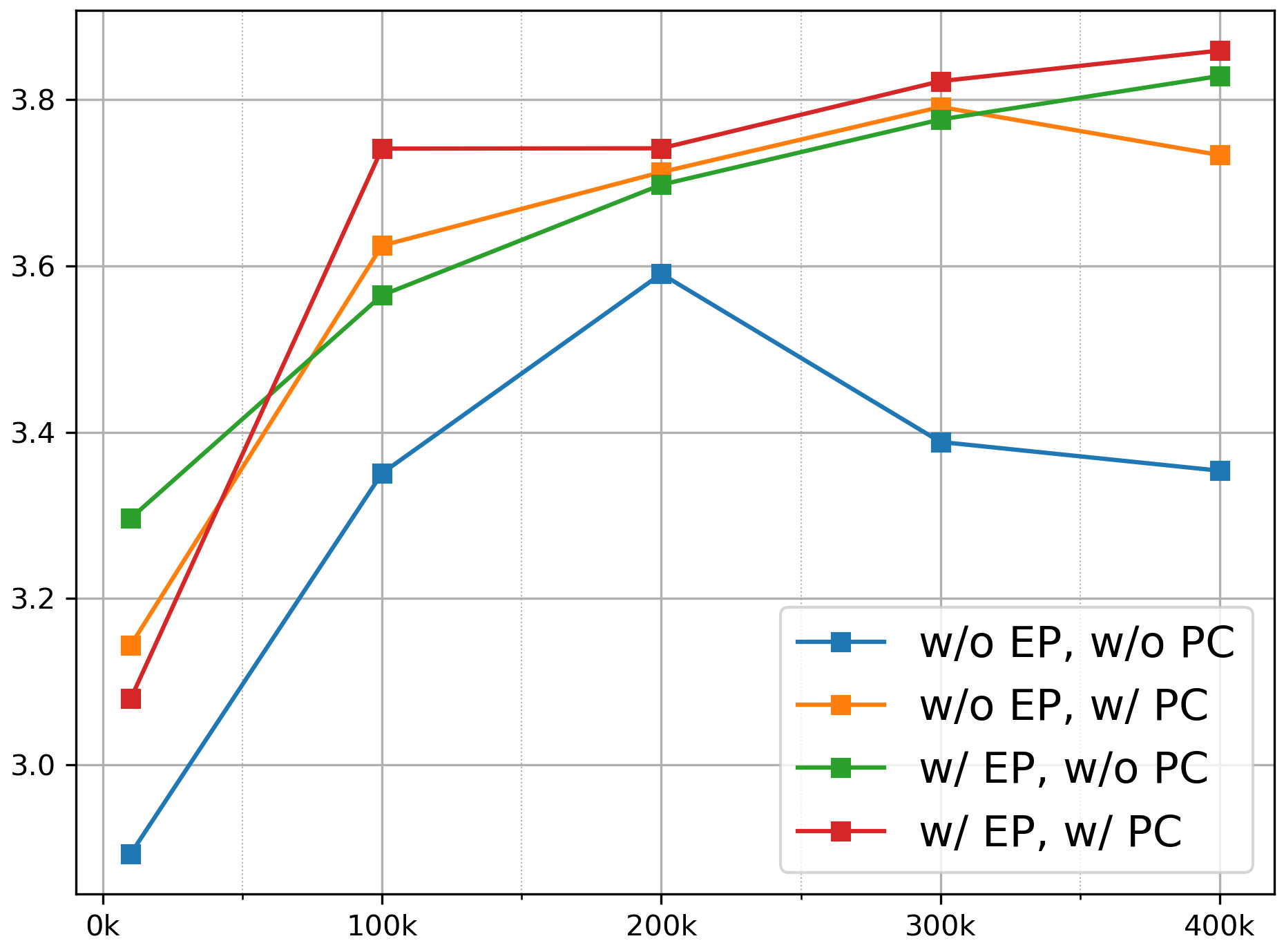}
        \caption{\small  WV-MOS}
        \label{fig:ablation_3}
    \end{subfigure}
    \hfill
    \begin{subfigure}[b]{0.24\linewidth}
        \includegraphics[width=\linewidth]{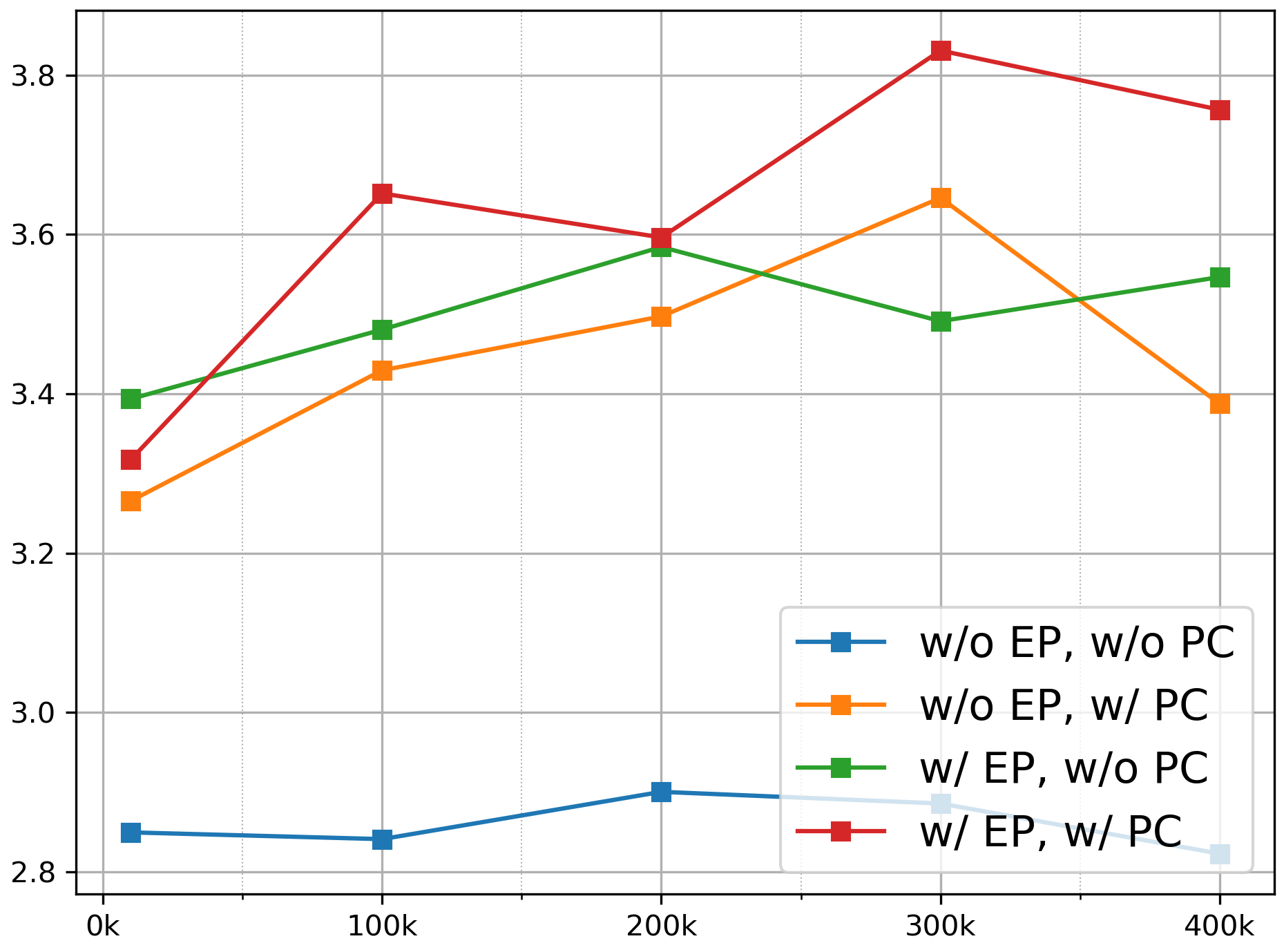}
        \caption{\small  NISQA}
        \label{fig:ablation_4}
    \end{subfigure}
    \vspace{-1mm}
   
    \label{fig:ablation}
\end{figure*}

We conduct an ablation study on our designs by training four model variants: with or without EP, and with or without the joint neural prior. 
%This results in four parallel experiments: baseline (no EP, no PC), EP only, PC only, and the full model with both. 
All variants share the same latent bridge and decoder training pipeline for fair comparison.
As shown in Figure~\ref{fig:ablation}, introducing the joint neural prior or adding the EP constraint leads to a noticeable improvement, while adding both achieves the best overall performance, demonstrating the complementary benefits of these two mechanisms. 

\begin{table*}[t]
\centering
\small
\vspace{-1mm}
\caption{\small Evaluation results on VoiceFixer-GSR~\citep{voicefixer} for different post-training processes of VoiceBridge. }
\vspace{-1mm}
\label{tab:abla}
\begin{adjustbox}{width=\linewidth,
center}

\begin{tabular}{llccccc}

\toprule
\multicolumn{7}{c}{Voicefixer-GSR~\cite{voicefixer}}\\
\midrule
Finetuned Module & Finetuning Objective & PESQ ($\uparrow$) & DNSMOS ($\uparrow$) & UTMOS ($\uparrow$) & WV-MOS ($\uparrow$) & NISQA ($\uparrow$) \\
\midrule
\multirow{5}{*}{Bridge} 
& $\mathcal{L}_\text{lat.}$ & 2.067 & 3.031 & 3.645 & 4.153 & \underline{4.067} \\
\cmidrule(lr){2-7}
& $\mathcal{L}_\text{lat.}+\mathcal{L}_\text{rec}$ & 2.125 & 3.049 & 3.649 & 4.207 & 3.524 \\
& $\mathcal{L}_\text{lat.}+\mathcal{L}_\text{rec}+\mathcal{L}_\text{gan}$ & 2.058 & \underline{3.069} & 3.533 & 4.055 & 3.362 \\
& $\mathcal{L}_\text{lat.}+\mathcal{L}_\text{rec}+\mathcal{L}_\text{perc}$ & 2.171 & 3.038 & 3.625 & 4.122 & 3.466 \\
& $\mathcal{L}_\text{lat.}+\mathcal{L}_\text{rec}+\mathcal{L}_\text{gan} + \mathcal{L}_\text{perc}$ & 2.182& 3.049 & 3.599 & 4.158 & 3.487 \\
\midrule
\multirow{4}{*}{Bridge $+$ Decoder} 
& $\mathcal{L}_\text{lat.}+\mathcal{L}_\text{rec}$ & 2.297 & 2.952 & 3.588 & 3.972 & 2.301 \\
& $\mathcal{L}_\text{lat.}+\mathcal{L}_\text{rec}+\mathcal{L}_\text{gan}$ & 2.169 & 2.938 & 3.475 & 3.709 & 3.091 \\
& $\mathcal{L}_\text{lat.}+\mathcal{L}_\text{rec}+\mathcal{L}_\text{perc}$ & \textbf{3.014} & 3.025 & \textbf{4.267} & \underline{4.252} & 3.167 \\
& $\mathcal{L}_\text{lat.}+\mathcal{L}_\text{rec}+\mathcal{L}_\text{gan} + \mathcal{L}_\text{perc}$ & \underline{2.722}& \textbf{3.091} & \underline{4.228} & \textbf{4.366} & \textbf{4.172} \\
\bottomrule
\end{tabular}
\end{adjustbox}
\vspace{-3mm}
\end{table*}

Further, we ablate the training objectives and update strategy in the post-training stage. Starting from the pre-trained LBM, we run 9 post-training variants that differ in (i) the loss objective and (ii) whether the decoder is updated. We consider two groups: (1) decoder frozen, where we fine-tune only the bridge using a data-space loss between decoded audio and clean audio; and (2) joint post-training, where the bridge and decoder are fine-tuned together.

We evaluate the loss combinations in Table~\ref{tab:abla}. As a reference baseline, we continue training with the original latent-space MSE objective $\mathcal{L}_\text{lat.}$ (Line~1). For post-training in the data domain, we use MRSTFT reconstruction loss $\mathcal{L}_\text{rec}$ (the same loss used in VAE training) as the base objective (Lines~2 and~6). We then add the GAN loss $\mathcal{L}_\text{gan}$ (adversarial + feature matching) and the perceptual term $\mathcal{L}_\text{perc}$, first individually (Lines~3,4,7,8) and then together (Lines~5 and~9). Each objective variant is tested in both update strategies (decoder frozen vs. joint). All hyperparameters follow the VAE training setup in Appendix~\ref{appendix:training}. All results use 4 inference steps.

As shown in Table~\ref{tab:abla}, freezing the decoder yields little improvement regardless of the objective. Under joint post-training, using $\mathcal{L}_\text{rec}$ alone, or $\mathcal{L}_\text{rec}$ with $\mathcal{L}_\text{gan}$, produces negligible or even negative gains. This may leads to the change in the MSE objective, causing approximation error in the bridge transport trajectory. In contrast, adding only $\mathcal{L}_\text{perc}$ substantially increases PESQ and UTMOS but degrades other metrics, consistent with metric-overfitting behavior reported in PESQetarian~\citep{oliveira2024pesqetarian}. The generated audio shares a common audible artifact, degrading listening experience. Intuitively, since the evaluator is fixed, the decoder can learn “hacks” that exploit the perceptual scorer rather than improving true audio quality.

Combining $\mathcal{L}_\text{gan}$ and $\mathcal{L}_\text{perc}$ prevents such metric hacking. The final setting (Line~9) yields a consistent and significant improvement across all metrics, indicating a genuine gain in perceptual quality. A plausible explanation is that overfitting-induced artifacts are readily detected by the discriminator, forcing the model to raise perceptual scores by improving generation quality.

We also conduct further ablation experiments regarding the design space of bridge modeling and EP-VAE training objectives. Results are shown in Appendix~\ref{appendix:ablation} due to page limit.

\section{Conclusion}
Our work presents VoiceBridge, modeling~\textit{diverse LQ-to-HQ tasks} in GSR with~\textit{a single latent-to-latent generative framework} backed by a transformer architecture. 
By introducing three novel techniques, including scale-equivariant regularization, joint neural prior, and the denoiser to generator post-training process, we allow VoiceBridge to consistently outperform strong GSR baselines on 48 kHz benchmarks with single step inference and demonstrate strong performance on unseen tasks. Ablations confirm the holistic improvements made by our innovations.

\section{Impact Statement}
This paper presents work whose goal is to advance the field of Generative Bridge Models, and General Speech Restoration. There are many potential societal consequences of our work, none which we feel must be specifically highlighted here.

\bibliography{VoiceBridge}

@INPROCEEDINGS{koizumi_miipher_2023,
  author={Koizumi, Y and Zen, H and Karita, S and others},
  booktitle={2023 IEEE Workshop on Applications of Signal Processing to Audio and Acoustics (WASPAA)}, 
  title={Miipher: A Robust Speech Restoration Model Integrating Self-Supervised Speech and Text Representations}, 
  year={2023},
  volume={},
  number={},
  pages={1-5},
  doi={10.1109/WASPAA58266.2023.10248089}}

@article{fang2021variational,
    title={Variational Autoencoder for Speech Enhancement with a Noise-Aware Encoder},
    author={ Fang, Huajian and Carbajal, Guillaume and Wermter, Stefan and Gerkmann, Timo },
    booktitle={ICASSP 2021-2021 IEEE International Conference on Acoustics, Speech and Signal Processing (ICASSP)},
    year={2021},
}

@article{oliveira2024pesqetarian,
  title={The PESQetarian: On the Relevance of Goodhart's Law for Speech Enhancement},
  author={de Oliveira, Danilo and  Welker, Simon and Richter,  Julius and Gerkmann, Timo}
}

@article{kolbaek2016speech,
  title={Speech intelligibility potential of general and specialized deep neural network based speech enhancement systems},
  author={Kolb{\ae}k, Morten and Tan, Zheng-Hua and Jensen, Jesper},
  journal={IEEE/ACM Transactions on Audio, Speech, and Language Processing},
  volume={25},
  number={1},
  pages={153--167},
  year={2016},
  publisher={IEEE}
}

@book{benesty2006speech,
  title={Speech enhancement},
  author={Benesty, Jacob and Makino, Shoji and Chen, Jingdong},
  year={2006},
  publisher={Springer Science \& Business Media}
}

@article{das2021fundamentals,
  title={Fundamentals, present and future perspectives of speech enhancement},
  author={Das, Nabanita and Chakraborty, Sayan and Chaki, Jyotismita and Padhy, Neelamadhab and Dey, Nilanjan},
  journal={International Journal of Speech Technology},
  volume={24},
  number={4},
  pages={883--901},
  year={2021},
  publisher={Springer}
}

@inproceedings{andreev2023hifi++,
  title={Hifi++: a unified framework for bandwidth extension and speech enhancement},
  author={Andreev, Pavel and Alanov, Aibek and Ivanov, Oleg and Vetrov, Dmitry},
  booktitle={ICASSP 2023-2023 IEEE International Conference on Acoustics, Speech and Signal Processing (ICASSP)},
  pages={1--5},
  year={2023},
  organization={IEEE}
}

@article{han2022nu,
  title={NU-Wave 2: A general neural audio upsampling model for various sampling rates},
  author={Han, Seungu and Lee, Junhyeok},
  journal={arXiv preprint arXiv:2206.08545},
  year={2022}
}

@article{birnbaum2019temporal,
  title={Temporal FiLM: Capturing long-range sequence dependencies with feature-wise modulations.},
  author={Birnbaum, Sawyer and Kuleshov, Volodymyr and Enam, Zayd and Koh, Pang Wei W and Ermon, Stefano},
  journal={Advances in Neural Information Processing Systems},
  volume={32},
  year={2019}
}

@inproceedings{liu2024audiosr,
  title={AudioSR: Versatile audio super-resolution at scale},
  author={Liu, Haohe and Chen, Ke and Tian, Qiao and Wang, Wenwu and Plumbley, Mark D},
  booktitle={ICASSP 2024-2024 IEEE International Conference on Acoustics, Speech and Signal Processing (ICASSP)},
  pages={1076--1080},
  year={2024},
  organization={IEEE}
}

@inproceedings{manjunath2009limitations,
  title={Limitations of perceptual evaluation of speech quality on VoIP systems},
  author={Manjunath, T},
  booktitle={2009 IEEE International Symposium on Broadband Multimedia Systems and Broadcasting},
  pages={1--6},
  year={2009},
  organization={IEEE}
}

@article{manocha2022audio,
  title={Audio similarity is unreliable as a proxy for audio quality},
  author={Manocha, Pranay and Jin, Zeyu and Finkelstein, Adam},
  journal={arXiv preprint arXiv:2206.13411},
  year={2022}
}

@article{liu20232,
  title={I2SB: Image-to-Image Schrodinger Bridge},
  author={Liu, Guan-Horng and Vahdat, Arash and Huang, De-An and Theodorou, Evangelos A and Nie, Weili and Anandkumar, Anima},
  journal={arXiv preprint arXiv:2302.05872},
  year={2023}
}

@article{wang2024framebridge,
  title={Framebridge: Improving image-to-video generation with bridge models},
  author={Wang, Yuji and Chen, Zehua and Chen, Xiaoyu and Wei, Yixiang and Zhu, Jun and Chen, Jianfei},
  journal={arXiv preprint arXiv:2410.15371},
  year={2024}
}

@inproceedings{lu2021study,
  title={A study on speech enhancement based on diffusion probabilistic model},
  author={Lu, Yen-Ju and Tsao, Yu and Watanabe, Shinji},
  booktitle={2021 Asia-Pacific Signal and Information Processing Association Annual Summit and Conference (APSIPA ASC)},
  pages={659--666},
  year={2021},
  organization={IEEE}
}

@inproceedings{lu2022conditional,
  title={Conditional diffusion probabilistic model for speech enhancement},
  author={Lu, Y and Wang, Z and Watanabe, S and Richard, A and Yu, C and Tsao, Y},
  booktitle={ICASSP 2022-2022 IEEE International Conference on Acoustics, Speech and Signal Processing (ICASSP)},
  pages={7402--7406},
  year={2022},
  organization={Ieee}
}

@article{sgmse,
  title={Speech enhancement with score-based generative models in the complex STFT domain},
  author={Welker, S and Richter, J and Gerkmann, T},
  journal={arXiv preprint arXiv:2203.17004},
  year={2022}
}

@article{speechflow,
  title={Generative pre-training for speech with flow matching},
  author={Liu, A and Le, M and Vyas, A and Shi, B and Tjandra, A and Hsu, W},
  journal={arXiv preprint arXiv:2310.16338},
  year={2023}
}

@inproceedings{ku2025generative,
  title={Generative speech foundation model pretraining for high-quality speech extraction and restoration},
  author={Ku, Pin-Jui and Liu, Alexander H and Korostik, Roman and Huang, Sung-Feng and Fu, Szu-Wei and Juki{\'c}, Ante},
  booktitle={ICASSP 2025-2025 IEEE International Conference on Acoustics, Speech and Signal Processing (ICASSP)},
  pages={1--5},
  year={2025},
  organization={IEEE}
}

@article{li2024masksr,
  title={Masksr: Masked language model for full-band speech restoration},
  author={Li, Xu and Wang, Qirui and Liu, Xiaoyu},
  journal={arXiv preprint arXiv:2406.02092},
  year={2024}
}

@article{anyenhance,
  title={AnyEnhance: A Unified Generative Model with Prompt-Guidance and Self-Critic for Voice Enhancement},
  author={Zhang, J and Yang, J and Fang, Z and others},
  journal={arXiv preprint arXiv:2501.15417},
  year={2025}
}

@inproceedings{scheibler_universalpp_2024,
	title = {Universal {Score}-based {Speech} {Enhancement} with {High} {Content} {Preservation}},
	url = {https://www.isca-archive.org/interspeech_2024/scheibler24_interspeech.html},
	doi = {10.21437/Interspeech.2024-138},
	language = {en},
	urldate = {2024-09-03},
	booktitle = {Interspeech 2024},
	publisher = {ISCA},
  author={Scheibler, R and Fujita, Y and Shirahata, Y and Komatsu, T},
	month = sep,
	year = {2024},
	pages = {1165--1169},
}

@inproceedings{voicefixer,
	title = {{VoiceFixer}: {A} {Unified} {Framework} for {High}-{Fidelity} {Speech} {Restoration}},
	shorttitle = {{VoiceFixer}},
	url = {https://www.isca-speech.org/archive/interspeech_2022/liu22y_interspeech.html},
	doi = {10.21437/Interspeech.2022-11026},
	language = {en},
	urldate = {2023-05-30},
	booktitle = {Interspeech 2022},
	publisher = {ISCA},
  author={Liu, H and Liu, X and Kong, Q and Tian, Q and Zhao, Y and Wang, D and Huang, C and Wang, Y},
	month = sep,
	year = {2022},
	pages = {4232--4236},
}

@article{richter2023speech,
  title={Speech enhancement and dereverberation with diffusion-based generative models},
  author={Richter, Julius and Welker, Simon and Lemercier, Jean-Marie and Lay, Bunlong and Gerkmann, Timo},
  journal={IEEE/ACM Transactions on Audio, Speech, and Language Processing},
  volume={31},
  pages={2351--2364},
  year={2023},
  publisher={IEEE}
}

@ARTICLE{storm,
  author={Lemercier, J and Richter, J and Welker, S and Gerkmann, T},
  journal={IEEE/ACM Transactions on Audio, Speech, and Language Processing}, 
  title={StoRM: A Diffusion-Based Stochastic Regeneration Model for Speech Enhancement and Dereverberation}, 
  year={2023},
  volume={31},
  number={},
  pages={2724-2737},
  doi={10.1109/TASLP.2023.3294692}}

@ARTICLE{universe,
  title={Universal speech enhancement with score-based diffusion},
  author={Serr{\`a}, J and Pascual, S and Pons, J and Araz, R and Scaini, D},
  journal={arXiv preprint arXiv:2206.03065},
  year=2022
}

@misc{resembleenhance,
  title        = {{Resemble Enhance}},
  author={Zhe Niu},
  year         = 2024,
  journal      = {GitHub repository},
  publisher    = {GitHub},
  howpublished = {\url{https://github.com/resemble-ai/resemble-enhance}},
  commit       = {acfc52f4ba24a492d4ad8209d43df34a78e91381}
}

@inproceedings{hiresldm,
  title={High-Resolution Speech Restoration with Latent Diffusion Model},
  author={Dhyani, T and Lux, F and Mancusi, M and Fabbro, G and Hohl, F and Vu, N},
  booktitle={ICASSP 2025-2025 IEEE International Conference on Acoustics, Speech and Signal Processing (ICASSP)},
  pages={1--5},
  year={2025},
  organization={IEEE}
}

@article{su2021hifigan2,
    author =  "Su, Jiaqi and Jin, Zeyu and Finkelstein, Adam",
    title = "HiFi-GAN-2: Studio-quality Speech Enhancement via Generative Adversarial Networks Conditioned on Acoustic Features",
    journal = {2021 IEEE Workshop on Applications of Signal Processing to Audio and Acoustics (WASPAA)},
    year = {2021}
}

@article{finally,
  title={FINALLY: fast and universal speech enhancement with studio-like quality},
  author={Babaev, N and Tamogashev, K and Saginbaev, A and Shchekotov, I and Bae, H and Sung, H and Lee, W and Cho, H and Andreev, P},
  journal={Advances in Neural Information Processing Systems},
  volume={37},
  pages={934--965},
  year={2024}
}

@inproceedings{mandel2023aero,
  title={Aero: Audio super resolution in the spectral domain},
  author={Mandel, Moshe and Tal, Or and Adi, Yossi},
  booktitle={ICASSP 2023-2023 IEEE International Conference on Acoustics, Speech and Signal Processing (ICASSP)},
  pages={1--5},
  year={2023},
  organization={IEEE}
}

@article{kouzelis2025eq,
  title={EQ-VAE: Equivariance Regularized Latent Space for Improved Generative Image Modeling},
  author={Kouzelis, T and Kakogeorgiou, I and Gidaris, S and Komodakis, N},
  journal={arXiv preprint arXiv:2502.09509},
  year={2025}
}

@article{yao2025reconstruction,
  title={Reconstruction vs. generation: Taming optimization dilemma in latent diffusion models},
  author={Yao, Jingfeng and Yang, Bin and Wang, Xinggang},
  journal={arXiv preprint arXiv:2501.01423},
  year={2025}
}

@article{skorokhodov2025improving,
  title={Improving the diffusability of autoencoders},
  author={Skorokhodov, Ivan and Girish, Sharath and Hu, Benran and Menapace, Willi and Li, Yanyu and Abdal, Rameen and Tulyakov, Sergey and Siarohin, Aliaksandr},
  journal={arXiv preprint arXiv:2502.14831},
  year={2025}
}

@article{yu2024representation,
  title={Representation alignment for generation: Training diffusion transformers is easier than you think},
  author={Yu, Sihyun and Kwak, Sangkyung and Jang, Huiwon and Jeong, Jongheon and Huang, Jonathan and Shin, Jinwoo and Xie, Saining},
  journal={arXiv preprint arXiv:2410.06940},
  year={2024}
}

@misc{gan,
      title={Generative Adversarial Networks}, 
  author={Ian J. Goodfellow and Jean Pouget-Abadie and Mehdi Mirza and Bing Xu and David Warde-Farley and Sherjil Ozair and Aaron Courville and Yoshua Bengio},
      year={2014},
      eprint={1406.2661},
      archivePrefix={arXiv},
      primaryClass={stat.ML},
      url={https://arxiv.org/abs/1406.2661}, 
}

@inproceedings{dit,
  title        = {{Scalable Diffusion Models with Transformers}},
  author={William Peebles and
                  Saining Xie},
  year         = 2023,
  booktitle    = {ICCV}
}

@inproceedings{audiolbm,
  title        = {{Audio Super-Resolution with Latent Bridge Models}},
  author={Chang Li and
          Zehua Chen and
          Liyuan Wang and
          Jun Zhu},
  year         = 2025,
  booktitle    = {NeurIPS}
}

@inproceedings{metis,
  title        = {{Metis: A Foundation Speech Generation Model with Masked Generative Pre-training}},
  author={Yuancheng Wang and
          Jiachen Zheng and 
          Junan Zhang and
          Xueyao Zhang and
          Huan Liao and
          Zhizheng Wu},
  year         = 2025,
  booktitle    = {NeurIPS}
}

@inproceedings{u-vit,
  title        = {{All are Worth Words: {A} ViT Backbone for Diffusion Models}},
  author={Fan Bao and
                  Shen Nie and
                  Kaiwen Xue and
                  Yue Cao and
                  Chongxuan Li and
                  Hang Su and
                  Jun Zhu},
  year         = 2023,
  booktitle    = {CVPR}
}

@inproceedings{song2020score,
  title        = {{Score-Based Generative Modeling through Stochastic Differential Equations}},
  author={Song, Y and Sohl-Dickstein, J and Kingma, D and others},
  year         = 2020,
  booktitle    = {ICLR}
}

@article{stableaudio,
  title        = {{Fast Timing-Conditioned Latent Audio Diffusion}},
  author={Evans, Z and Carr, C and Taylor, J and Hawley, S and Pons, J},
  year         = 2024,
  journal      = {arXiv:2402.04825}
}

@article{evans2024long,
  title={Long-form music generation with latent diffusion},
  author={Evans, Zach and Parker, Julian D and Carr, CJ and Zukowski, Zack and Taylor, Josiah and Pons, Jordi},
  journal={arXiv preprint arXiv:2404.10301},
  year={2024}
}

@article{kong2020diffwave,
  title={Diffwave: A versatile diffusion model for audio synthesis},
  author={Kong, Zhifeng and Ping, Wei and Huang, Jiaji and Zhao, Kexin and Catanzaro, Bryan},
  journal={arXiv preprint arXiv:2009.09761},
  year={2020}
}

@software{Harper_NeMo_a_toolkit,author = {Harper, Eric and Majumdar, Somshubra and Kuchaiev, Oleksii and Jason, Li and Zhang, Yang and Bakhturina, Evelina and Noroozi, Vahid and Subramanian, Sandeep and Nithin, Koluguri and Jocelyn, Huang and Jia, Fei and Balam, Jagadeesh and Yang, Xuesong and Livne, Micha and Dong, Yi and Naren, Sean and Ginsburg, Boris},title = {{NeMo: a toolkit for Conversational AI and Large Language Models}},url = {https://github.com/NVIDIA/NeMo}}

@inproceedings{lemercier2023analysing,
  author={Lemercier, J and Richter, J and Welker, S and Gerkmann, T},
  booktitle={ICASSP}, 
  title={Analysing Diffusion-based Generative Approaches Versus Discriminative Approaches for Speech Restoration}, 
  year={2023},
  volume={},
  number={},
  pages={1-5},
  doi={10.1109/ICASSP49357.2023.10095258}}

@article{vae,
  title={An introduction to variational autoencoders},
  author={Kingma, D and Welling, M and others},
  journal={Foundations and Trends{\textregistered} in Machine Learning},
  volume={12},
  number={4},
  pages={307--392},
  year={2019},
  publisher={Now Publishers, Inc.}
}

@inproceedings{lim2018time,
  title={Time-frequency networks for audio super-resolution},
  author={Lim, Teck Yian and Yeh, Raymond A and Xu, Yijia and Do, Minh N and Hasegawa-Johnson, Mark},
  booktitle={2018 IEEE International Conference on Acoustics, Speech and Signal Processing (ICASSP)},
  pages={646--650},
  year={2018},
  organization={IEEE}
}

@article{baevski2020wav2vec,
  title={wav2vec 2.0: A framework for self-supervised learning of speech representations},
  author={Baevski, Alexei and Zhou, Yuhao and Mohamed, Abdelrahman and Auli, Michael},
  journal={Advances in neural information processing systems},
  volume={33},
  pages={12449--12460},
  year={2020}
}

@article{kong2025a2sb,
  title={A2SB: Audio-to-Audio Schrodinger Bridges},
  author={Kong, Zhifeng and Shih, Kevin J and Nie, Weili and Vahdat, Arash and Lee, Sang-gil and Santos, Joao Felipe and Jukic, Ante and Valle, Rafael and Catanzaro, Bryan},
  journal={arXiv preprint arXiv:2501.11311},
  year={2025}
}

@article{bridgesr,
  title={Bridge-SR: Schr$\backslash$" odinger Bridge for Efficient SR},
  author={Li, C and Chen, Z and Bao, F and Zhu, J},
  journal={arXiv preprint arXiv:2501.07897},
  year={2025}
}

@inproceedings{bunne2023schrodinger,
  title={The schrödinger bridge between gaussian measures has a closed form},
  author={Bunne, C and Hsieh, Y and Cuturi, M and Krause, A},
  booktitle={International Conference on Artificial Intelligence and Statistics},
  pages={5802--5833},
  year={2023},
  organization={PMLR}
}

@article{sbsen,
  title={Schrödinger Bridge for Generative Speech Enhancement},
  author={Juki{\'c}, A and Korostik, R and Balam, J and others},
  journal={arXiv preprint arXiv:2407.16074},
  year={2024}
}

@article{sbsel,
  title={Diffusion-based Speech Enhancement with Schrödinger Bridge and Symmetric Noise Schedule},
  author={Wang, S and Liu, S and Harper, A and Kendrick, P and Salzmann, M and Cernak, M},
  journal={arXiv preprint arXiv:2409.05116},
  year={2024}
}

@inproceedings{richter2025investigating,
  title={Investigating training objectives for generative speech enhancement},
  author={Richter, J and De Oliveira, D and Gerkmann, T},
  booktitle={ICASSP 2025-2025 IEEE International Conference on Acoustics, Speech and Signal Processing (ICASSP)},
  pages={1--5},
  year={2025},
  organization={IEEE}
}

@article{zhang2025sb,
  title={SB-SENet: Diffusion model based on Schr{\"o}dinger bridge for speech enhancement},
  author={Zhang, Huaifeng and Li, Guigeng and Wu, Pengfei and Gao, Yong and Zhang, Hao},
  journal={Applied Acoustics},
  volume={236},
  pages={110742},
  year={2025},
  publisher={Elsevier}
}

@article{han2025few,
  title={Few-step Adversarial Schr$\backslash$"$\{$o$\}$ dinger Bridge for Generative Speech Enhancement},
  author={Han, Seungu and Lee, Sungho and Lee, Juheon and Lee, Kyogu},
  journal={arXiv preprint arXiv:2506.01460},
  year={2025}
}

@article{bridgetts,
  title={Schrodinger bridges beat diffusion models on text-to-speech synthesis},
  author={Chen, Z and He, G and Zheng, K and Tan, X and Zhu, J},
  journal={arXiv preprint arXiv:2312.03491},
  year={2023}
}

@inproceedings{sbtheory,
  title={Sur la th{\'e}orie relativiste de l'{\'e}lectron et l'interpr{\'e}tation de la m{\'e}canique quantique},
  author={Schr{\"o}dinger, E},
  booktitle={Annales de l'institut Henri Poincar{\'e}},
  volume={2},
  number={4},
  pages={269--310},
  year={1932}
}

@inproceedings{wang2021deep,
  title={Deep generative learning via Schrödinger bridge},
  author={Wang, G and Jiao, Y and Xu, Q and others},
  booktitle={International conference on machine learning},
  pages={10794--10804},
  year={2021},
  organization={PMLR}
}

@article{chen2021likelihood,
  title={Likelihood training of schrödinger bridge using forward-backward sdes theory},
  author={Chen, T and Liu, G and Theodorou, E},
  journal={arXiv preprint arXiv:2110.11291},
  year={2021}
}

@inproceedings{bibletts,
    title={BibleTTS: a large, high-fidelity, multilingual, and uniquely African speech corpus},
  author={J Meyer and D Adelani and E Casanova and others},
    booktitle={Interspeech},
    publisher = {{ISCA}},
    year={2022},
    url={https://arxiv.org/pdf/2207.03546.pdf}
  }

@misc{vctk,
  title        = {{96kHz version of the CSTR VCTK Corpus}},
  author={Veaux, C and Yamagishi, J},
  year         = 2017,
  publisher    = {University of Edinburgh. CSTR},
}

@misc{valentini2017noisy,
  title={Noisy speech database for training speech enhancement algorithms and TTS models},
  author={Valentini-Botinhao, C},
  year={2017},
  publisher={University of Edinburgh. School of Informatics. Centre for Speech Technology Research (CSTR)},
  doi={10.7488/ds/2117},
}

@article{hifitts,
  title={{Hi-Fi Multi-Speaker English TTS Dataset}},
  author={Bakhturina, E and Lavrukhin, V and Ginsburg, B and Zhang, Y},
  journal={arXiv preprint arXiv:2104.01497},
  year={2021}
}

@article{aishell4,
  title={Aishell-4: An open source dataset for speech enhancement, separation, recognition and speaker diarization in conference scenario},
  author={Fu, Y and Cheng, L and Lv, S and Jv, Y and Kong, Y and Chen, Z and Hu, Y and Xie, L and Wu, J and Bu, H and others},
  journal={arXiv preprint arXiv:2104.03603},
  year={2021}
}

@article{expresso,
  title={Expresso: A benchmark and analysis of discrete expressive speech resynthesis},
  author={Nguyen, T and Hsu, W and d'Avirro, A and others},
  journal={arXiv preprint arXiv:2308.05725},
  year={2023}
}

@article{ears,
  title={EARS: An anechoic fullband speech dataset benchmarked for speech enhancement and dereverberation},
  author={Richter, J and Wu, Y and Krenn, S and others},
  journal={arXiv preprint arXiv:2406.06185},
  year={2024}
}

@inproceedings{aishell1,
  title={Aishell-1: An open-source mandarin speech corpus and a speech recognition baseline},
  author={Bu, H and Du, J and Na, X and Wu, B and Zheng, H},
  booktitle={2017 20th conference of the oriental chapter of the international coordinating committee on speech databases and speech I/O systems and assessment (O-COCOSDA)},
  pages={1--5},
  year={2017},
  organization={IEEE}
}

@inproceedings{chime2,
  title={The second ‘CHiME’speech separation and recognition challenge: An overview of challenge systems and outcomes},
  author={Vincent, E and Barker, J and Watanabe, S and others},
  booktitle={2013 IEEE Workshop on Automatic Speech Recognition and Understanding},
  pages={162--167},
  year={2013},
  organization={IEEE}
}

@article{chime3,
  author={Barker, J and Marxer, R and Vincent, E and others},
    title = {The third ‘CHiME’speech separation and recognition challenge: Dataset, task and baselines},
    journal = {IEEE Workshop on Automatic Speech Recognition and Understanding (ASRU)},
    year = 2015,
    pages = {504-511}
}

@inproceedings{ko2017study,
  title={A study on data augmentation of reverberant speech for robust speech recognition},
  author={Ko, T and Peddinti, V and Povey, D and Seltzer, M and Khudanpur, S},
  booktitle={2017 IEEE international conference on acoustics, speech and signal processing (ICASSP)},
  pages={5220--5224},
  year={2017},
  organization={IEEE}
}

@inproceedings{turpault2019sound,
  title={Sound event detection in domestic environments with weakly labeled data and soundscape synthesis},
  author={Turpault, N and Serizel, R and Shah, A and Salamon, J},
  booktitle={Workshop on Detection and Classification of Acoustic Scenes and Events},
  year={2019}
}

@article{mesaros2018multi,
  title={A multi-device dataset for urban acoustic scene classification},
  author={Mesaros, A and Heittola, T and Virtanen, T},
  journal={arXiv preprint arXiv:1807.09840},
  year={2018}
}

@article{reddy2021interspeech,
  title={Interspeech 2021 deep noise suppression challenge},
  author={Reddy, C and Dubey, H and Koishida, K and Nair, A and Gopal, V and Cutler, R and Braun, S and Gamper, H and Aichner, R and Srinivasan, S},
  journal={arXiv preprint arXiv:2101.01902},
  year={2021}
}

@inproceedings{pesq,
  title={Perceptual evaluation of speech quality (PESQ)-a new method for speech quality assessment of telephone networks and codecs},
  author={Rix, A and Beerends, J and Hollier, M and Hekstra, A},
  booktitle={2001 IEEE international conference on acoustics, speech, and signal processing. Proceedings (Cat. No. 01CH37221)},
  volume={2},
  pages={749--752},
  year={2001},
  organization={IEEE}
}

@inproceedings{stoi,
  title={A short-time objective intelligibility measure for time-frequency weighted noisy speech},
  author={Taal, Cees H and Hendriks, Richard C and Heusdens, Richard and Jensen, Jesper},
  booktitle={2010 IEEE international conference on acoustics, speech and signal processing},
  pages={4214--4217},
  year={2010},
  organization={IEEE}
}

@article{srmr,
  title={A non-intrusive quality and intelligibility measure of reverberant and dereverberated speech},
  author={Falk, Tiago H and Zheng, Chenxi and Chan, Wai-Yip},
  journal={IEEE Transactions on Audio, Speech, and Language Processing},
  volume={18},
  number={7},
  pages={1766--1774},
  year={2010},
  publisher={IEEE}
}

@inproceedings{wvmos,
  title={Can we use Common Voice to train a Multi-Speaker TTS system?},
  author={Ogun, S and Colotte, V and Vincent, E},
  booktitle={2022 IEEE Spoken Language Technology Workshop (SLT)},
  pages={900--905},
  year={2023},
  organization={IEEE}
}

@inproceedings{nisqa,
  title        = {{NISQA: A Deep CNN-Self-Attention Model for Multidimensional Speech Quality Prediction with Crowdsourced Datasets}},
  author={Mittag, G and Naderi, B and Chehadi, A and M{\"o}ller, S},
  year         = 2021,
  booktitle    = {Interspeech}
}

@INPROCEEDINGS{dnsmos,
  author={Reddy, C and Gopal, V and Cutler, R},
  booktitle={ICASSP}, 
  title={Dnsmos: A Non-Intrusive Perceptual Objective Speech Quality Metric to Evaluate Noise Suppressors}, 
  year={2021},
  volume={},
  number={},
  pages={6493-6497},
  doi={10.1109/ICASSP39728.2021.9414878}}

@inproceedings{utmos,
  title     = {UTMOS: UTokyo-SaruLab System for VoiceMOS Challenge 2022},
  author={Takaaki Saeki and Detai Xin and Wataru Nakata and Tomoki Koriyama and Shinnosuke Takamichi and Hiroshi Saruwatari},
  year      = {2022},
  booktitle = {Interspeech 2022},
  pages     = {4521--4525},
  doi       = {10.21437/Interspeech.2022-439},
  issn      = {2958-1796},
}

@article{pekmezci2024evaluation, 
  title={Evaluation of SSIM loss function in RIR generator GANs},
  author={Pekmezci, M and Genc, Y},
  journal={Digital Signal Processing},
  volume={154},
  pages={104685},
  year={2024},
  publisher={Elsevier}
}

@article{maaten2008visualizing,
  title={Visualizing data using t-SNE},
  author={Maaten, Laurens van der and Hinton, Geoffrey},
  journal={Journal of machine learning research},
  volume={9},
  number={Nov},
  pages={2579--2605},
  year={2008}
}

@article{defossez2022high,
  title={High fidelity neural audio compression},
  author={D{\'e}fossez, Alexandre and Copet, Jade and Synnaeve, Gabriel and Adi, Yossi},
  journal={arXiv preprint arXiv:2210.13438},
  year={2022}
}

@article{anastassiou2024seed,
  title={Seed-tts: A family of high-quality versatile speech generation models},
  author={Anastassiou, Philip and Chen, Jiawei and Chen, Jitong and Chen, Yuanzhe and Chen, Zhuo and Chen, Ziyi and Cong, Jian and Deng, Lelai and Ding, Chuang and Gao, Lu and others},
  journal={arXiv preprint arXiv:2406.02430},
  year={2024}
}

@article{wang2024maskgct,
  title={Maskgct: Zero-shot text-to-speech with masked generative codec transformer},
  author={Wang, Yuancheng and Zhan, Haoyue and Liu, Liwei and Zeng, Ruihong and Guo, Haotian and Zheng, Jiachen and Zhang, Qiang and Zhang, Xueyao and Zhang, Shunsi and Wu, Zhizheng},
  journal={arXiv preprint arXiv:2409.00750},
  year={2024}
}

@article{ju2025mooncast,
  title={MoonCast: High-quality zero-shot podcast generation},
  author={Ju, Zeqian and Yang, Dongchao and Yu, Jianwei and Shen, Kai and Leng, Yichong and Wang, Zhengtao and Tan, Xu and Zhou, Xinyu and Qin, Tao and Li, Xiangyang},
  journal={arXiv preprint arXiv:2503.14345},
  year={2025}
}

@article{thiemann2013demand,
  title={DEMAND: a collection of multi-channel recordings of acoustic noise in diverse environments},
  author={Thiemann, Joachim and Ito, Nobutaka and Vincent, Emmanuel},
  journal={(No Title)},
  year={2013},
  publisher={Zenodo}
}

@misc{higgsaudio2025,
  author       = {{Boson AI}},
  title        = {{Higgs Audio V2: Redefining Expressiveness in Audio Generation}},
  year         = {2025},
  howpublished = {\url{https://github.com/boson-ai/higgs-audio}},
  note         = {GitHub repository. Release blog available at \url{https://www.boson.ai/blog/higgs-audio-v2}},
}

@article{peng2025vibevoice,
  title={Vibevoice technical report},
  author={Peng, Zhiliang and Yu, Jianwei and Wang, Wenhui and Chang, Yaoyao and Sun, Yutao and Dong, Li and Zhu, Yi and Xu, Weijiang and Bao, Hangbo and Wang, Zehua and others},
  journal={arXiv preprint arXiv:2508.19205},
  year={2025}
}

@article{panaretos2019statistical,
  title={Statistical aspects of Wasserstein distances},
  author={Panaretos, Victor M and Zemel, Yoav},
  journal={Annual review of statistics and its application},
  volume={6},
  number={1},
  pages={405--431},
  year={2019},
  publisher={Annual Reviews}
}
\bibliographystyle{icml2026}

%%%%%%%%%%%%%%%%%%%%%%%%%%%%%%%%%%%%%%%%%%%%%%%%%%%%%%%%%%%%%%%%%%%%%%%%%%%%%%%
%%%%%%%%%%%%%%%%%%%%%%%%%%%%%%%%%%%%%%%%%%%%%%%%%%%%%%%%%%%%%%%%%%%%%%%%%%%%%%%
% APPENDIX
%%%%%%%%%%%%%%%%%%%%%%%%%%%%%%%%%%%%%%%%%%%%%%%%%%%%%%%%%%%%%%%%%%%%%%%%%%%%%%%
%%%%%%%%%%%%%%%%%%%%%%%%%%%%%%%%%%%%%%%%%%%%%%%%%%%%%%%%%%%%%%%%%%%%%%%%%%%%%%%
\newpage
\appendix
\onecolumn

% \begingroup
% \small  
% \setstretch{0.7} 
\tableofcontents
% \endgroup

\section{Related Work}
\label{appendix:related}

\subsection{Bridge-based Speech Enhancement}
Schr\"odinger Bridge (SB) models~\citep{sbtheory, chen2021likelihood, wang2021deep, bunne2023schrodinger} explore learning an optimal stochastic trajectory between two boundary distributions with iterative fitting, allowing for efficient and high-quality generation.
Built upon SB models, tractable bridge models have been recently developed~\citep{bunne2023schrodinger}, which achieves~\textit{data-to-data} generation process effectively exploiting the instructive information contained in the observed prior distribution. 
These bridge models have shown advantages over diffusion models in tasks where indicative prior information has already been provided, such as image-to-image translation~\citep{liu20232}, text-to-speech synthesis~\citep{bridgetts}, image-to-video generation~\citep{wang2024framebridge}, and speech restoration (SE)~\citep{sbsen}.

In the field of SE, to exploit the advantages of the~\textit{data-to-data} generative process on SE tasks, recent works have designed bridge-based speech denoising, super-resolution, and dereveration models.
In task-specific benchmark datasets, these works have proposed different innovations, such as designing the noise schedule~\citep{sbsen}, examining the model parameterization~\citep{bridgetts}, incorporating an adversarial training objective~\citep{han2025few}, and introducing a two-stage generation framework~\citep{sbsel}.
Although these attempts have improved the generation quality and the inference speed of bridge-based SE systems, their applicability may still be restricted because of the narrow-band generation target, the specific degradation method, or the model training on a task-specific benchmark dataset. 

In VoiceBridge, we propose a bridge-based general speech restoration system (GSR), restoring different low-quality (LQ) signals to the high-quality (HQ) target with a latent bridge model (LBM). By compressing the speech waveform into continuous latent representations, we model~\textit{diverse LQ-to-HQ restoration tasks} with~\textit{a unified latent-to-latent generative framework}, and propose three techniques to holistically strengthen the system, achieving GSR with high perceptual quality.

\subsection{General Speech Restoration}
GSR, shown by~\citep{voicefixer}, focuses on the task of retrieving clean speech out of lossy recordings, such as noise addition, bandwidth limitation, dereverberation, or other acoustic degradations. Unlike the traditional SE task which restores speech from only noise addition or another single kind of degradation, GSR pays extra focus on generating audio with high perceptual quality from acoustic scenarios with mixed degradations~\citep{voicefixer, finally}. 

Prior works explored different methods to solve the GSR task. As the initial proposer of GSR, VoiceFixer~\citep{voicefixer} introduced a mapping-based method with two stages: firstly synthesizes the mel-spectrogram of clean audio from the distorted waveform, and then synthesizes the clean audio waveform through a vocoder. UniverSE~\citep{universe} and UniverSE++~\citep{scheibler_universalpp_2024} formulate the restoration as a diffusion generative process, where the degraded signal and the mel-spectrogram are taken as the condition information of diffusion models. 
Hi-ResLDM~\citep{hiresldm} is a two-stage model, including a mapping-based recovery stage and a diffusion-based restoration stage. 
%However, diffusion models are gradually outperformed by bridge models in the data-to-data generation scenario~\citep{bridgetts, sbsen, bridgesr}, where bridge models accomplish better results with the same neural network settings.
FINALLY~\citep{finally} is an adversarial network, with additional alignment to self-supervised speech features and perceptual quality loss terms. 
%It achieves strong performance on different test settings, while requiring a lengthy and complex training strategy. 
MaskSR~\citep{li2024masksr} and AnyEnhance~\citep{anyenhance} are masked generative models, which generate the speech signal from discrete tokens. 
In VoiceBridge, we propose the first LBM-based GSR system, exploiting the two advantages of bridge models for GSR: full exploitation of the indicative LQ prior information, and iterative refinement nature along the entire sampling trajectory. 
%causing extra errors in the codec translation process. As far as we are concerned, VoiceBridge is the first to utilize latent bridge modeling in the Speech Restoration job, setting up a milestone for the new generative approach.

\section{GSR Task Formulation and Data Construction}
\label{appendix:augmentation}

In the formulation of GSR, the LQ speech sample is a degradation of the HQ target. The degradation operator $\mathcal{T}$ is a mixture of random degrading functions, i.e. $\mathcal{T} = \mathcal{T}_1 \circ\mathcal{T}_2\circ\cdots\circ\mathcal{T}_N$, where each $\mathcal{T}_n$ is a specific degradation method including noise addition, bandwidth limitation, reverberation, clipping, etc. 
At the training stage of VoiceBridge, we follow previous works and construct the HQ and LQ speech data pairs with 
\begin{align}
\label{eq:aug}
    \mathcal{T} = \mathcal{T}_\text{bw}\circ\mathcal{T}_\text{clip}\circ\mathcal{T}_\text{rev}\circ\mathcal{T}_\text{noise}\circ\mathcal{T}_\text{rev}\circ\mathcal{T}_\text{eq},
\end{align}
where each degradation operator is applied with a certain probability, otherwise leaving the speech unchanged. 
These degradation operators are configured as follows.
\begin{itemize}
    \item $\mathcal{T}_\text{bw}$: down-sampling the HQ speech sample to one of $\{2,4,8,12,16,24,32\}$ kHz (by uniformly random choice), with 0.5 probability. The down-sampling filter is chosen from the Bessel Filter, Chebyshev Filter, and Butterworth Filter with uniform randomness.
    \item $\mathcal{T}_\text{clip}$: clipping the HQ amplitude to the range $[0.06, 0.9]$ times of its original maximum absolute value, with 0.25 probability.
    \item $\mathcal{T}_\text{rev}$: applying room reverberation, with 0.5 probability. The reverberation is applied by convolving the speech signal with a room impulse response from a mixture of real-world and simulated RIR datasets.
    \item $\mathcal{T}_\text{noise}$: Adding noise with a random SNR in $[-5, 20]$ dB with 0.9 probability. Noise is randomly selected from the employed noise datasets
    \item $\mathcal{T}_\text{eq}$: Applying 1–3 random bell filters (frequency $\in [10, 12000]$ Hz, gain $\in [-5, 5]$ dB, $Q \in [0.5, 2]$, all sampled from a uniform distribution across the range) to each time window independently, with 0.5 probability.
\end{itemize}
Note that $\mathcal{T}_\text{rev}$ appears twice in Equation~\ref{eq:aug}. This design follows~\citep{anyenhance}, where the reverberation is applied twice, both inside and outside of noise addition, to ensure that additional noise involves audio pieces in both reverberant or non-reverberant environments. The parameters are chosen such that the degradation is noticeable by listeners.

\section{Schr\"odinger Bridge Training and Inference}
\label{appendix:sb}

 The SB problem seeks the stochastic trajectory connecting two distributions $p_0$ and $p_1$ that minimizes the KL divergence to a reference diffusion process $p_{\text{ref}}$~\citep{sbtheory, chen2021likelihood}:
\begin{align}
\min_{p\in\mathcal{P}{[0,1]}} \mathcal{D}_{\mathrm{KL}}(p \| p_{\text{ref}}),\, \text{subject to } p_0 = p_{\text{data}}, p_1 = p_{\text{prior}}.
\end{align}
where $\mathcal{P}{[0,1]}$ is the space of path measures on the interval [0,1]. Here, we define $p_{\text{ref}}$ as the marginal distribution accompanying the forward SDE~\citep{song2020score}:
\begin{align}
\dl{\mathbf{z}_t} = f(\mathbf{z}_t, t)\dl{t} + g(t)\dl{\mathbf{w}_t},
\end{align}
where $w_t$ is a standard Wiener process. Under this framework, the optimal SB dynamics are given by a pair of forward-backward SDEs with data-driven drift corrections:
\begin{align}
\dl{\mathbf{z}_t} &= \left[ f(\mathbf{z}_t, t) + g^2(t) \nabla \log \Psi_t(\mathbf{z}_t) \right] \dl{t} + g(t) \dl{\mathbf{w}_t}, \mathbf{z}_0\sim p_0\\
\label{eq:reverse_SDE}\dl{\mathbf{z}_t} &= \left[ f(\mathbf{z}_t, t) - g^2(t) \nabla \log \bar{\Psi}_t(\mathbf{z}_t) \right] \dl{t} + g(t) \dl{\bar{\mathbf{w}}_t}, \mathbf{z}_1\sim p_1
\end{align}
where $\Psi_t$ and $\bar{\Psi}_t$ are solutions to the corresponding Schrödinger system PDEs, and their product defines the marginal density $p_t = \Psi_t \bar{\Psi}_t$ of the SB process. The drift term is typically chosen linearly as $f(\mathbf{z}_t, t)=f(t)\mathbf{z}_t$ with a predefined schedule $f(t)$. Although generally intractable, the SB problem has an analytical solution assuming that $p_0$ and $p_1$ are Gaussians centered on $\mathbf{z}_0$ and $\mathbf{z}_1$~\citep{bridgetts,bunne2023schrodinger}. Under this setting, the interpolated distribution $p_t$ at time $t$ is itself Gaussian:

\begin{align}
\label{eq:interpolation}
p_t= \mathcal{N}(
    \frac{\alpha_t\bar{\sigma}_t^2}{\sigma_1^2}\mathbf{z}_0 + \frac{\bar{\alpha_t}\sigma_t^2}{\sigma_1^2} \mathbf{z}_1,
    \frac{\alpha_t^2\bar{\sigma}_t^2\sigma_t^2}{\sigma_1^2}\mathbf{I}
)
\end{align}
where
\begin{align}
\alpha_t=e^{\int_0^tf(\tau)\dl\tau},\quad\bar\alpha_t=e^{-\int_t^1f(\tau)\dl\tau},\quad
\sigma_t^2 = \int_0^t \frac{g^2(\tau)}{\alpha_\tau^2} \dl\tau,\quad \bar\sigma_t^2 = \int_t^1 \frac{g^2(\tau)}{\alpha_\tau^2} \dl\tau.
\end{align}
are noise schedules analytic from the reference SDE's $f,g$ functions~\citep{bridgetts}. 

Given the condition $\mathbf{z}_1$, the model is trained to predict unknown terms related to $(\mathbf{z}_t,t)$ in the reverse SB process (Equation~\ref{eq:reverse_SDE}) to generate $\mathbf{z}_0$ from $t=1$ to $t=0$. The model can be parameterized to predict either $\nabla \log \bar{\Psi}_t(\mathbf{z}_t)$ or other equivalent forms, similar to diffusion models. The most common and effective choice is to predict $\mathbf{z}_0$ directly and to minimize the mean square error (MSE) loss:
\begin{align}
\label{eq:appendix:bridgeloss}
\mathcal{L}(\varphi)
=
\mathbb{E}_{(\mathbf{z}_0,\mathbf{z_1})}
\bigl\|\hat{\mathbf{z}}_{0,\varphi}(\mathbf{z}_t,t,\mathbf{z}_1) - \mathbf{z}_0 \bigr\|_2^2,
\end{align}
where $\mathbf{z}_t$ is the noisy interpolation of $\mathbf{z}_0,\mathbf{z}_1$ on the bridge according to Equation~\ref{eq:interpolation}.

After training, samples can be generated by simulating Equation~\ref{eq:reverse_SDE} in reverse time, starting from condition $\mathbf{z}_1$. The sampling process can be principled accelerated by exponential integrators~\citep{bridgetts}. Specifically, for the transition from $s$ to $t<s$, the first-order discretization takes the form
\begin{align}
      \label{eq:first-order-SDE}
    \mathbf{z}_t&=\frac{\alpha_t\sigma_t^2}{\alpha_s\sigma_s^2}\mathbf{z}_s+\alpha_t\left(1-\frac{\sigma_t^2}{\sigma_s^2}\right)\mathbf{z}_\theta(\mathbf{z}_s,s)
    +\alpha_t\sigma_t\sqrt{1-\frac{\sigma_t^2}{\sigma_s^2}}\boldsymbol{\epsilon},\quad \boldsymbol{\epsilon}\sim\mathcal{N}(0,\mathbf{I})\\
     \label{eq:first-order-ODE}
    \mathbf{z}_t&=\frac{\alpha_t\sigma_t\bar\sigma_t}{\alpha_s\sigma_s\bar\sigma_s}\mathbf{z}_s+\frac{\alpha_t}{\sigma_1^2}\left[\left(\bar\sigma_t^2-\frac{\bar\sigma_s\sigma_t\bar\sigma_t}{\sigma_s}\right)\mathbf{z}_\theta(\mathbf{z}_s,s) \right. 
     \left. +\left(\sigma_t^2-\frac{\sigma_s\sigma_t\bar\sigma_t}{\bar\sigma_s}\right)\frac{\mathbf{z}_1}{\alpha_1}\right]
\end{align}
for SDE and ODE, respectively.

\section{Model Architecture and Training Setup}
\label{appendix:training}
\subsection{Model Architecture}
VoiceBridge consists of a VAE for prior latent encoding and target latent decoding, and a transformer backbone for solving the bridge model trajectory. For the autoencoder, we utilize the Oobleck VAE~\citep{evans2024long} architecture. The encoder and decoder are symmetric and contain 156M parameters each. 
The VAE compresses 48 kHz audio into 64-channel latent representations at 23.4 Hz, yielding a temporal downsampling factor of 2048. 
The discriminator architecture follows~\citep{stableaudio}, where we employ a multi-scale STFT discriminator with 5 separate models modeling STFT spectrograms of different window lengths.
For the SB model, we adopt the transformer backbone from Stable Audio 2~\citep{stableaudio, evans2024long}, removing the text-conditioning cross-attention layers. We set the transformer backbone to have 24 transformer layers with a hidden dimension of 1152, resulting in 544~M parameters in total. 

As Stable Audio 2 itself is a text-to-music generation model for a cross-modal audio generation task, we change its model architecture to GSR, which is a speech processing task. 
Specifically, we remove the text encoder and the cross-attention layers, reducing the model parameters and computation complexity. 
To include the low-quality audio as a condition, we concatenate the conditional LQ latent with the input latent on the channel dimension. Classifier-free guidance is also removed, as cross-modal generation is not involved.

\subsection{Model Training Details}

The training pipeline of VoiceBridge consists of four stages: (1) pre-training a VAE with the Energy Preserving training objective $\mathcal{L}_\text{ep-vae}$ on clean speech; (2) fine-tuning the VAE encoder on distorted inputs with the joint neural prior loss $\mathcal{L}_\text{np-enc}$, while keeping the decoder frozen; (3) training a transformer-based bridge model to solve the SB in the latent space by $\mathcal{L}_\text{bridge}$, with both the encoder and decoder frozen; and (4) jointly fine-tuning the Bridge Transformer and the VAE decoder with $\mathcal{L}_\text{pt}$, improving perceptual quality of generation process. 

In the EP-VAE training section, the training objective has the form:
\begin{align}
\label{eq:appendix:epvae}
    \mathcal{L}_\text{ep-vae}
    %&=
    %\mathcal{L}_\text{data}^\text{ep} + \mathcal{L}_\text{latent} \nonumber\\
    &=\mathcal{L}^{\text{ep}}_\text{data}(\mathcal{D}_\theta(s\cdot \mathcal{E}_\theta(\mathbf{x})), s\cdot \mathbf{x}) + \mathcal{L}_\text{latent}(\mathcal{E}_\theta(\mathbf{x}),\mathbf{z}_{\text{ref}}),
\end{align}
where the $\mathcal{L}_\text{data}^\text{ep}$ and $\mathcal{L}_\text{latent}$ are the data-space and latent-space losses, respectively, consisting of the following loss terms:
\begin{align}
\label{eq:appendix:epdataloss}
&\mathcal{L}_\text{data}^\text{ep} = \lambda_\text{rec}\mathcal{L}_\text{rec} + \lambda_\text{adv}\mathcal{L}_\text{adv} + \lambda_\text{fm}\mathcal{L}_\text{fm},\\
\label{eq:appendix:eplatentloss}
&\mathcal{L}_\text{latent} =\lambda_\text{kl}\mathcal{L}_\text{kl},
\end{align}
in which $\mathcal{L}_\text{rec}, \mathcal{L}_\text{adv}, \mathcal{L}_\text{fm}$ and $\mathcal{L}_\text{kl}$ stand for the Multi-Resolution STFT reconstruction loss, adversarial loss, feature matching loss, and KL regularization loss, respectively. The loss weights $\lambda$ are used to balance the training objectives. 
In practice, we use the weights ensuring that each loss term has a similar order of magnitude, resulting in $\lambda_\text{rec} = 1, \lambda_\text{adv} = 0.1, \lambda_\text{fm}=5,\lambda_\text{kl} = 1\text{e}-4$. 
In this stage, the VAE is pre-trained on 8 A800 GPUs with a batch size of 16 for 800k steps, on the combined dataset for clean speech signals.

At the second stage, namely fine-tuning the encoder for the joint neural prior, we use the training objective
\begin{align}
\label{eq:appendix:pc-encoder}
\mathcal{L}_\text{np-enc} &= \mathcal{L}^\text{ep}_\text{data}(\mathcal{D}(s\cdot \mathcal{E}^\text{np}_{\theta}(\mathbf{x}_1)), s\cdot \hat{\mathbf{x}}_0)+\mathcal{L}_{\text{latent}}(\mathcal{E}^\text{np}_{\theta}(\mathbf{x}_1), \mathbf{z}_0),
\end{align}
for optimization. The $\mathcal{L}^\text{ep}_\text{data}$ term holds the same meaning as in Equation~\ref{eq:appendix:epdataloss}, while the $\mathcal{L}_\text{latent}$ now stands for the latent convergence objective that aligns different LQ prior latent with the ground-truth HQ target latent. 
In this stage, we hope to converge LQ prior latent representations in both scale and direction, thus designing both the MSE term and the cosine similarity term in the $\mathcal{L}_\text{latent}$ objective:
\begin{align}
\label{eq:appendix:pclatentloss}
    \mathcal{L}_\text{latent} = \lambda_\text{mse}\mathcal{L}_\text{mse} + \lambda_\text{cos}\mathcal{L}_\text{cos},
\end{align}
where $\lambda_\text{mse}=\lambda_\text{cos} = 2.5$.
We fine-tune the encoder from the initial state of the EP-VAE encoder, using paired data created with our construction pipeline introduced above. 
The fine-tuning process is conducted on 8 A800 GPUs with a batch size of 16 for 500k iterations.

At the third stage, we train the bridge transformer model to learn the bridge trajectory using Equation~\ref{eq:appendix:bridgeloss}. 
This training stage is implemented on 32 A800 GPUs with a batch size of 256 for 1300k iterations. 
%We found that this stage's training is determinant to the final performance of VoiceBridge, especially the batch size causes a significant impact. For this reason, we exceptionally used 4 times amount of GPUs to open up a batch size that is 16 times larger than the VAE training stage.

At the last stage, we jointly fine-tune the Bridge Transformer and the VAE decoder to achieve higher perceptual quality. The training objective
\begin{align}
\label{eq:appendix:posttraining}
    \mathcal{L}_\text{pt} 
    &= \mathcal{L}_\text{bridge}(\varphi) + 
    \mathcal{L}_\text{data}(\mathcal{D}_{\theta}(\hat{\mathbf{z}}_{0,\varphi}(\mathbf{z}_t,t,\mathbf{z}^{\text{np}}_1)), \mathbf{x}_0) \nonumber \\
    &+ 
    \mathcal{L}_\text{GAN}(\mathcal{D}_{\theta}(\hat{\mathbf{z}}_{0,\varphi}(\mathbf{z}_t,t,\mathbf{z}^{\text{np}}_1)), \mathbf{x}_0)
    +
    \mathcal{L}_\text{perc}(\mathcal{D}_{\theta}(\hat{\mathbf{z}}_{0,\varphi}(\mathbf{z}_t,t,\mathbf{z}^{\text{np}}_1)), \mathbf{x}_0),
\end{align}
includes the bridge loss shown in Equation~\ref{eq:appendix:bridgeloss}, VAE data-space loss shown in Equation~\ref{eq:appendix:epdataloss}, and two extra loss terms to align our generation with human perceptual quality:
\begin{align}
    \mathcal{L}_\text{perc} = \lambda_\text{pesq}\mathcal{L}_\text{pesq} + \lambda_\text{utmos}\mathcal{L}_\text{utmos},
\end{align}
where $\mathcal{L}_\text{pesq}$ and $\mathcal{L}_\text{utmos}$ are two loss functions based on the PESQ and UTMOS scores, with $\lambda_\text{pesq}=1$ and $\lambda_\text{utmos}=10$. The joint finetuning is achieved on 8 A800 GPUs with a batch size of 16 for 200k iterations.

\section{Derivations on Denoiser to Generator Post-training Objective}
\label{appendix:math}

In this appendix, we provide a concise derivation of the optimization target of our denoiser-to-generator post-training, and explain why it changes the learning objective from \emph{conditional expectation prediction} (typical of MSE-trained bridge models) to \emph{conditional distribution matching}.

\paragraph{Problem setup.}
Let the conditioning signal be
\[
c := (x_t, t, x_1),
\]
where $x_t$ is the intermediate state used by the bridge model, $t$ is the normalized time index, and $x_1$ is the LQ prior (or its latent counterpart), consistent with the main text.
Let $x:=x_0$ denote the paired HQ target waveform.
The training pairs are sampled from a joint data distribution
\[
p_{\text{data}}(x,c)=p(c)\,p_{\text{data}}(x\mid c).
\]
Our post-training refines a generator that produces an HQ estimate under condition $c$.
To model a \emph{conditional distribution} rather than a deterministic point estimate, we assume the generator is stochastic:
\[
\hat{x} = G_\varphi(c,\epsilon),\qquad \epsilon\sim p(\epsilon),
\]
which induces a conditional model distribution $p_G(x\mid c)$.

\paragraph{(A) Denoiser-to-generator post-training objective.}
We use the vanilla (logistic) conditional GAN with a discriminator $D_\omega(x,c)\in(0,1)$ and augment it with a perceptual loss.
The min--max objective is
\begin{align}
\label{eq:app_minmax}
\min_{G}\ \max_{D}\ 
\Big\{
&\mathbb{E}_{(x,c)\sim p_{\text{data}}}\big[\log D(x,c)\big]
+
\mathbb{E}_{c\sim p(c),\,\epsilon\sim p(\epsilon)}\big[\log(1-D(G(c,\epsilon),c))\big]
\nonumber\\
&\quad+\ \lambda_{\text{perc}}\,
\mathbb{E}_{(x,c)\sim p_{\text{data}},\,\epsilon\sim p(\epsilon)}
\big[\ell_{\text{perc}}(G(c,\epsilon),x)\big]
\Big\},
\end{align}
where $\ell_{\text{perc}}(\hat{x},x)$ is the perceptual loss used in our method (e.g., a weighted combination of PESQ and UTMOS losses, or their differentiable surrogates).
For clarity, we omit other auxiliary terms in the main text (e.g., $\mathcal{L}_{\text{data}}$ or $\mathcal{L}_{\text{bridge}}$); adding them does not change the conclusion below, since they simply introduce additional regularization on top of the distribution-matching objective.

\paragraph{(B) What the adversarial term optimizes (JS divergence).}
Fix a generator $G$ (hence $p_G(\cdot\mid c)$). Under standard assumptions that
$p_{\text{data}}(\cdot\mid c)$ and $p_G(\cdot\mid c)$ admit densities with respect to a common base measure for $p(c)$-a.e.\ $c$,
the discriminator that maximizes the GAN value function (the first two terms in Eq.~\ref{eq:app_minmax}) is given by
\begin{equation}
\label{eq:app_Dstar}
D^*(x,c)=\frac{p_{\text{data}}(x\mid c)}{p_{\text{data}}(x\mid c)+p_G(x\mid c)}.
\end{equation}
Substituting $D^*$ into the GAN value function yields the classical Jensen--Shannon divergence form:
\begin{equation}
\label{eq:app_vstar}
V(D^*,G)
=
-\log 4
+
2\,\mathbb{E}_{c\sim p(c)}\Big[
\mathrm{JSD}\big(p_{\text{data}}(\cdot\mid c)\ \|\ p_G(\cdot\mid c)\big)
\Big],
\end{equation}
where $\mathrm{JSD}(p\|q)$ is the Jensen--Shannon divergence.
Therefore, the min--max training in Eq.~\ref{eq:app_minmax} is equivalent (up to the constant $-\log 4$) to minimizing
\begin{equation}
\label{eq:app_reduced_objective}
\min_{G}\ 
2\,\mathbb{E}_{c}\Big[
\mathrm{JSD}\big(p_{\text{data}}(\cdot\mid c)\ \|\ p_G(\cdot\mid c)\big)
\Big]
+
\lambda_{\text{perc}}\,
\mathbb{E}_{(x,c),\,\epsilon}
\big[\ell_{\text{perc}}(G(c,\epsilon),x)\big].
\end{equation}
Eq.~\ref{eq:app_reduced_objective} makes explicit that the adversarial component performs \emph{conditional distribution matching} (via JS divergence to the data conditional), while the perceptual term encourages high subjective quality.

\paragraph{(C) Why this differs from expectation prediction.}
Many bridge models are trained as $x_0$-predictors with a squared error objective (in latent or waveform space), which yields conditional expectation estimation.
Concretely, consider a deterministic predictor $f(c)$ trained with
\begin{equation}
\label{eq:app_mse}
\min_{f}\ \mathbb{E}\big[\|x-f(c)\|_2^2\big].
\end{equation}
The population minimizer of Eq.~\ref{eq:app_mse} satisfies
\begin{equation}
\label{eq:app_cond_mean}
f^*(c)=\mathbb{E}[x\mid c],
\end{equation}
i.e., the model learns the \emph{conditional mean} (``expectation modeling'').
This explains why MSE-trained $x_0$ predictors often produce over-smoothed outputs when the conditional distribution is multi-modal.

In contrast, the adversarial term in Eq.~\ref{eq:app_reduced_objective} is minimized when
\begin{equation}
\label{eq:app_js_zero}
p_G(\cdot\mid c)=p_{\text{data}}(\cdot\mid c)\qquad \text{for }p(c)\text{-a.e. }c,
\end{equation}
since $\mathrm{JSD}(p\|q)\ge 0$ with equality iff $p=q$.
Thus, the GAN component explicitly optimizes a \emph{distributional} objective rather than a point-estimation objective.

\paragraph{(D) Why initializing from a bridge model helps.}
Our post-training starts from a bridge model that already provides a strong conditional estimate (often close to $\mathbb{E}[x\mid c]$ under MSE-style training), and then refines it using Eq.~\ref{eq:app_minmax}.
Intuitively, the original bridge training provides a stable initialization that preserves the learned transport structure, while adversarial fine-tuning reduces the discrepancy between the model conditional $p_G(\cdot\mid c)$ and the data conditional $p_{\text{data}}(\cdot\mid c)$, and the perceptual term further steers the solutions toward human-aligned quality.
Importantly, to represent a non-degenerate conditional distribution in Eq.~\ref{eq:app_js_zero}, the generator must include stochasticity (e.g., the noise $\epsilon$); otherwise $p_G(\cdot\mid c)$ collapses to a point mass and cannot match a multi-modal $p_{\text{data}}(\cdot\mid c)$.

\section{Dataset Details}
\label{appendix:dataset}
We construct our HQ speech dataset by combining publicly available datasets: 
VCTK~\citep{vctk} is a multi-speaker English dataset with 109 speakers of diverse accents each reading about 400 sentences from newspaper excerpts and phonetic-rich passages, at a sampling rate of 48~kHz. 
HiFi-TTS~\citep{hifitts} contains speech from 10 English speakers reading LibriVox audiobooks and Project Gutenberg texts, specially filtered by a threshold of bandwidth and signal-to-noise ratio.
HQ-TTS~\citep{voicefixer} is a collection of publicly available 44.1kHz or 48kHz clean speech data on the OpenSLR website, including multilingual speech from a wide range of speakers.
AiShell-1~\citep{aishell1} and AiShell-4~\citep{aishell4} are two multi-speaker Mandarin datasets collected from noiseless indoor environments through different recording devices, providing non-English speech context for VoiceBridge to learn multilingual speech patterns.
Bible-TTS~\citep{bibletts} is also a multilingual dataset, composed of speech for six languages spoken in Sub-Saharan Africa.
Expresso~\citep{expresso} is a high-quality expressive speech dataset, which includes both expressively rendered read speech and improvised dialogues, offering speech in various tones and styles.
EARS~\citep{ears} contains anechoic speech recordings from over 100 English speakers with high demographic diversity, spanning the full range of human speech.
Some of these datasets include speech samples recorded in different acoustic backgrounds, from which we select the clean audio in noiseless environments. 
For the datasets used for later evaluation,~\textit{e.g.}, VCTK, the corresponding testset is removed from the training data. 
The combination of training data, after the filtering, totals in around 1138 hours. All recordings are resampled to a sampling rate of 48 kHz. 

For the noise addition and reverberation application processes of data construction, extra noise data and room impulse response (RIR) signals are required. 
For noise samples, we utilize the following datasets. 
DEMAND~\citep{valentini2017noisy} includes multi-channel recordings of acoustic noise in diverse environments. 
CHiME-2/3~\citep{chime2, chime3} have noise recordings taken from the two ChiME ASR challenges.
TUT Urban Acoustic Scenes 2018~\citep{mesaros2018multi} is a dataset of multi-device recordings, originally used for urban acoustic scene classification.
DESED~\citep{turpault2019sound} is an audio dataset designed for recognizing sound event classes in domestic environments. 
These datasets have covered common noise in different acoustic scenarios, supporting our denoising-related tasks. 
Similar to the speech datasets, the testing portion is also removed for DEMAND and CHiME3, as they are later used for evaluation. 
Room Impulse Responses are drawn from the following datasets.
SLR26/28~\citep{ko2017study} contains simulated RIRs in different room conditions. 
GTU-RIR~\citep{pekmezci2024evaluation} is a real RIR dataset, including the recordings of the impulse signal with different devices in different rooms.
We also include other simulated RIR signals created by ourselves, following previous works.

\begin{table*}[t]
\centering
\small
\caption{\small Bandwidth extension results on VCTK-BWE~\citep{andreev2023hifi++}. VoiceBridge beats other GSR models and audio super-resolution models on both subjective and objective metrics.}
\label{tab:sr}
% \begin{adjustbox}{width=\linewidth, center}
\begin{tabular}{llccccc}
  \toprule
  Dataset & Model & WVMOS($\uparrow$) & NISQA($\uparrow$) & PESQ($\uparrow$) & STOI($\uparrow$) & LSD($\downarrow$) \\
  \midrule
  % ---------- 1K ----------
  \multirow{7}{*}{BW=1K}
    & TFiLM             & 1.650*            & --                & --                  & 0.810*               & --      \\
    & HiFi++            & 3.710*            & --                & --                  & \textbf{0.860}*      & --      \\
    & AudioSR           & --                & --                & --                  & --                   & --      \\
    & NuWave2           & 1.895             & 2.958             & 1.671               & 0.751                & 1.762   \\
    & VoiceFixer                & \underline{3.822} & \underline{3.691} & 1.567               & 0.818                & \underline{1.510} \\
    & ResembleEnhance                & 3.033             & 3.348             & \underline{2.021}   & 0.745                & 2.223   \\
    & \textbf{VoiceBridge} & \textbf{4.154}    & \textbf{4.58}     & \textbf{2.11}       & \underline{0.824}    & \textbf{1.433} \\
  \midrule
  % ---------- 2K ----------
  \multirow{7}{*}{BW=2K}
    & TFiLM             & 2.270*            & --                & --                  & \underline{0.910}*   & --      \\
    & HiFi++            & 3.950*            & --                & --                  & \textbf{0.940}*      & --      \\
    & AudioSR           & 3.435             & 4.176             & 1.752               & 0.907                & 1.539   \\
    & NuWave2           & 3.208             & 3.755             & \underline{2.098}   & 0.885                & 1.580   \\
    & VoiceFixer                & 4.031             & 4.033             & 1.997               & 0.868                & \underline{1.499} \\
    & ResembleEnhance                & \underline{4.141} & \underline{4.642} & 2.038               & 0.873                & 1.913   \\
    & \textbf{VoiceBridge} & \textbf{4.306}    & \textbf{4.656}    & \textbf{2.696}      & 0.897                & \textbf{1.392} \\
  \midrule
  % ---------- 4K ----------
  \multirow{7}{*}{BW=4K}
    & TFiLM             & 3.490*            & --                & --                  & \textbf{1.000}*      & --      \\
    & HiFi++            & 4.160*            & --                & --                  & \textbf{1.000}*      & --      \\
    & AudioSR           & 4.138             & 4.373             & 2.678               & 0.983                & 1.479   \\
    & NuWave2           & 4.169             & 3.870             & \underline{3.114}   & \underline{0.992}    & \underline{1.348} \\
    & VoiceFixer                & 4.144             & 4.290             & 2.471               & 0.909                & 1.507   \\
    & ResembleEnhance                & \textbf{4.439}    & \underline{4.684} & 2.685               & 0.941                & 1.845   \\
    & \textbf{VoiceBridge} & \underline{4.404} & \textbf{4.703}    & \textbf{3.318}      & 0.940                & \textbf{1.323} \\
  \bottomrule
\end{tabular}

% \end{adjustbox}
\end{table*}

\begin{table*}[!h]
\centering
\small
\caption{\small Dereverberation Results on WSJ-Reverb~\citep{richter2023speech}}
\label{tab:derev}
% \begin{adjustbox}{width=\linewidth,
% center}
\begin{tabular}{lcccc}
      \toprule
      \multicolumn{5}{c}{WSJ0–Reverb~\cite{sgmse}} \\
      \cmidrule(lr){1-5}
      Model      & WVMOS ($\uparrow$) & DNSMOS ($\uparrow$) & NISQA ($\uparrow$) & SRMR ($\uparrow$) \\
      \midrule
      SGMSE+     & 3.285               & 2.815               & 3.210              & 8.768             \\
      StoRM      & \underline{3.583}   & 2.969               & 3.823              & 9.615             \\
      RE         & 3.474               & \underline{3.263}   & \underline{4.610}  & 8.556             \\
      VF         & 3.093               & 3.056               & 3.849              & \textbf{10.278}   \\
      UniverSE++ & 2.572               & 2.313               & 3.131              & 8.472             \\
      \textbf{VB(Ours)}       & \textbf{4.403}      & \textbf{3.286}      & \textbf{4.617}     & \underline{9.929} \\
      \bottomrule
    \end{tabular}
% \end{adjustbox}
\end{table*}

\section{Evaluation Details}
\label{appendix:evaluation}
In this section, we provide a detailed introduction to the benchmark datasets, baseline methods, and evaluation metrics that we used for evaluating VoiceBridge's performance. We also list further evaluation settings and results on broken down degradation types in detail, and make comparison to other models. Note that results in appendix are achieved by 4 inference steps.

\begin{table*}[t]
\centering
\small
\caption{\small Dereverberation results on VCTK~\citep{vctk} convolved with simulated RIR signals with different RT60 time.}
\label{tab:derev_breakdown}

\begin{tabular}{llccccccc}
  \toprule
  Dataset & Method & PESQ($\uparrow$) & WVMOS($\uparrow$) & UTMOS($\uparrow$) & DNSMOS($\uparrow$)& NISQA($\uparrow$) & SRMR($\uparrow$) \\
  \midrule
  \multirow{4}{*}{RT60=0.3s}
    & Voicefixer         & 2.321          & 4.124          & 3.533          & \underline{3.138} & 4.369 & 9.044\\
    & Resemble-Enhance   & 2.333          & \textbf{4.376} & \underline{3.628} & 3.101          & \underline{4.381} & 9.208\\
    & UniverSE++         & \textbf{2.599} & 4.102          & 3.296          & 3.010             & 4.024 & \underline{9.744} \\
    & \textbf{VoiceBridge}        & \underline{2.393} & \underline{4.286} & \textbf{3.864} & \textbf{3.185} & \textbf{4.488} & \textbf{9.767}\\
  \midrule
  \multirow{4}{*}{RT60=0.6s}
    & Voicefixer         & \underline{2.060} & 4.068          & 3.394          & \textbf{3.111} & 4.294 & 9.398\\
    & Resemble-Enhance   & 1.653          & \textbf{4.244} & \underline{3.419} & 3.057        & \underline{4.317} & 9.208\\
    & UniverSE++         & 1.652          & 3.553          & 2.639          & 2.830          & 3.695 &\textbf{9.794} \\
    & \textbf{VoiceBridge}        & \textbf{2.127} & \underline{4.242} & \textbf{3.821} & \underline{3.081} & \textbf{4.448} & \underline{9.762}\\
  \midrule
  \multirow{4}{*}{RT60=0.9s}
    & Voicefixer         & \underline{1.915} & 4.024          & \underline{3.342} & \textbf{3.108} & 4.292 & \underline{9.641}\\
    & Resemble-Enhance   & 1.394          & \underline{4.139} & 3.232          & \underline{3.067} & \underline{4.353} & 9.207\\
    & UniverSE++         & 1.407          & 3.251          & 2.329          & 2.755          & 3.602 & 9.604\\
    & \textbf{VoiceBridge}        & \textbf{1.970} & \textbf{4.195} & \textbf{3.757} & 3.058          & \textbf{4.358} & \textbf{9.919}\\
  \midrule
  \multirow{4}{*}{RT60=1.2s}
    & Voicefixer         & \underline{1.837} & 3.995          & \underline{3.292} & \textbf{3.105} & 4.261 & \underline{9.658}\\
    & Resemble-Enhance   & 1.323          & \underline{4.100} & 3.175          & 3.051            & \underline{4.271} & 9.334\\
    & UniverSE++         & 1.351          & 3.170          & 2.193          & 2.706             & 3.540 & 8.999\\
    & \textbf{VoiceBridge}        & \textbf{1.873} & \textbf{4.177} & \textbf{3.732} & \underline{3.058} & \textbf{4.338} & \textbf{9.968} \\
  \bottomrule
\end{tabular}

\end{table*}

\subsection{Evaluation datasets}

To comprehensively evaluate VoiceBridge's GSR performance at scale, we utilized three test datasets as follows. 
VoiceFixer-GSR~\citep{voicefixer} is a simulated testset, which produces LQ speech samples with a random degradation process combining noise addition, low-pass filtering, clipping, and reverberation. 
We utilize this benchmark to evaluate VoiceBridge's multi-degradation restoration ability and to enable direct comparison with VoiceFixer. 

DNS-with-Reverb~\citep{reddy2021interspeech} augments DNS Challenge samples with reverberation and noise, generating a set of noisy audio in strong reverberant conditions. 
This testset evaluates VoiceBridge's performance under challenging degradation scenarios. Meanwhile,~\textit{all training data samples from the DNS Challenge are excluded from the training dataset of VoiceBridge, causing an extra training-inference gap, and testing VoiceBridge's robustness in unseen conditions}. 

DNS-Real-Data~\citep{reddy2021interspeech} is an in-the-wild dataset, composed of real recordings from the DNS Challenge. 
It contains lossy audio samples recorded in a real acoustic environment, demonstrating the real-world restoration abilities of VoiceBridge.

As GSR is defined as general restoration, including different restoration processes, the restoration process from a single degradation naturally becomes its subtask. 
To evaluate VoiceBridge's zero-shot ability on these restoration subtasks, we choose the traditional Speech Enhancement (SE) task as a typical example.
We utilize VoiceBank-Demand~\citep{valentini2017noisy} and WSJ0-CHiME3~\citep{chime3}. The VoiceBank-Demand dataset mixes utterances from the Voicebank speech corpus with real-world noise from the DEMAND dataset, serving as a benchmark testset for the SE. WSJ0-CHiME3, similarly, mixes clean speech signals from the WSJ0 dataset with noise from CHiME3, but with a smaller signal-to-noise ratio, making the denoising task harder. The two benchmark datasets are widely used in prior SE works~\citep{lu2021study, lu2022conditional, richter2023speech, storm, sbsen}.

To further verify the zero-shot performance of bridge models, we report VoiceBridge's performance on \textit{Out-of-Domain} (OOD) tasks. We include test results for codec artifact removal (CAR), and text-to-speech (TTS) refinement, which are especially significant for contemporary TTS applications. 

For the CAR test, we utilize the VCTK~\citep{vctk} testset, using Encodec~\citep{defossez2022high} to compress the 48kHz audio to a codec at a 3kbps frame rate, and then reconstruct it. 

For the TTS task, we choose the Seed-TTS~\citep{anastassiou2024seed} benchmark, which is widely used in various TTS works~\citep{wang2024maskgct, ju2025mooncast, higgsaudio2025, peng2025vibevoice}, and generate speech and podcast using MaskGCT~\citep{wang2024maskgct} and MoonCast~\citep{ju2025mooncast} respectively. Note that the MaskGCT is used with a prompt speech for the speaker condition, and MoonCast is called without a prompt audio.

\begin{table*}[t]
\centering
\caption{\small Denoising results on VCTK~\citep{vctk} with additive noise from DEMAND~\citep{thiemann2013demand} at different SNR levels.}
\label{tab:se_breakdown}

\begin{tabular}{llccccccc}
  \toprule
  Dataset & Method & PESQ($\uparrow$) & WVMOS($\uparrow$) & UTMOS($\uparrow$) & DNSMOS($\uparrow$) & NISQA($\uparrow$) \\
  \midrule
  \multirow{5}{*}{SNR=-15dB}
    & SBSE & 1.119 & -0.949 & 1.480 & 1.513 & 1.869 \\
    & Voicefixer         & \underline{1.225} & 2.423 & 2.278 & 2.313 & 2.848 \\
    & Resemble-Enhance   & 1.168          & \underline{3.394} & \underline{2.440}  & \underline{2.640} & \underline{3.130} \\
    & UniverSE++         & 1.211          & 2.772 & 2.271 & 2.374 & 2.950 \\
    & \textbf{VoiceBrigde}        & \textbf{1.352} & \textbf{3.479} & \textbf{2.755} & \textbf{2.698} & \textbf{3.270} \\
  \midrule
  \multirow{5}{*}{SNR=-10dB}
    & SBSE & 1.224 & 0.347 & 1.829 & 1.822 & 2.389 \\
    & Voicefixer         & \textbf{1.588} & 3.263 & 2.845 & 2.791 & 3.576 \\
    & Resemble-Enhance   & 1.345          & \underline{3.783} & 2.929 & 2.876 & 3.693 \\
    & UniverSE++         & \underline{1.527} & 3.704 & \underline{3.075} & \underline{2.933} & \textbf{3.971} \\
    & \textbf{VoiceBrigde}        & 1.453          & \textbf{3.908} & \textbf{3.339}& \textbf{2.944} & \underline{3.725} \\
  \midrule
  \multirow{5}{*}{SNR=-5dB}
    & SBSE & 1.410 & 1.698 & 2.301 & 2.275 & 2.933\\
    & Voicefixer         & \textbf{1.871} & 3.901 & \underline{3.346}& \textbf{3.106} & \underline{4.133} \\
    & Resemble-Enhance   & 1.586          & \underline{4.001} & 3.266 & 3.009 & 4.075 \\
    & UniverSE++         & \underline{1.790} & 3.820 & 3.320& 2.913 & \textbf{4.177} \\
    & \textbf{VoiceBrigde}        & 1.730          & \textbf{4.093} & \textbf{3.641} & \underline{3.022} & 3.969 \\
  \midrule
  \multirow{5}{*}{SNR=0dB}
    & SBSE & 1.663 & 2.938 & 2.826 & 2.665 & 3.468 \\
    & Voicefixer         & 2.143 & 4.074 & 3.548 & \textbf{3.170} & \textbf{4.350} \\
    & Resemble-Enhance   & 1.891 & \underline{4.135} & 3.491 & 3.070 & \underline{4.322} \\
    & UniverSE++         & \textbf{2.240} & 4.100 & \underline{3.674} & 2.974 & 4.291 \\
    & \textbf{VoiceBrigde}        & \underline{2.200} & \textbf{4.200} & \textbf{3.765}& \underline{3.098} & 4.201 \\
  \bottomrule
\end{tabular}

\end{table*}

\subsection{Baseline Methods}
As a GSR model, we compare VoiceBridge against other strong GSR baseline methods for both GSR tasks and OOD subtasks. 
VoiceFixer~\citep{voicefixer} is a mapping-based model with two stages: synthesizes the mel-spectrogram of the clean speech from the distorted one at the first stage, and then generates the clean audio waveform through a vocoder. 
As the first proposer of the GSR task, it stands as a classic baseline to compare with. 
Resemble-Enhance~\citep{resembleenhance} is an open-sourced GSR tool based on the flow-matching generative process, and a popular baseline model used in several GSR works. UniverSE++~\citep{scheibler_universalpp_2024} is an improved version of UniverSE~\citep{universe}, which formulates the restoration as a conditional diffusion generation process, where the degraded observation is taken as condition information. 
We use the pre-trained checkpoints of these models for inference on the benchmark test sets mentioned above. 
AnyEnhance~\citep{anyenhance} is a masked generative model that demonstrates strong performance on GSR. We use their reported values in their paper due to a lack of publicly available implementation.

To further demonstrate the performance of VoiceBridge on SE, we include not only the GSR models listed above, but also strong task-specific baseline methods for comparison. 
For SE, we include the following baseline models:
SGMSE+~\citep{richter2023speech, sgmse} is a simple diffusion model used in both denoising and dereverberation scenarios, but needs to be trained separately. 
StoRM~\citep{storm} makes further improvement based on SGMSE+, adding an extra predictive stage before diffusion generation, reducing the gap between diffusion target and prior conditions. SBSE~\citep{sbsen} is an SB model trained for the SE task, demonstrating data-space SB performance. 
Noticeably, we use the pre-trained checkpoint of these models on VoiceBank-Demand, WSJ0-CHiME3, or WSJ0-Reverb training set, and report the inference result on the corresponding test sets, whereas VoiceBridge is used in a cross-validation way where the training set is composed of other audios.

\subsection{Metrics}
To make a comprehensive and fair comparison, we use both the intrusive and the non-intrusive metrics to evaluate our GSR performance. 
As to intrusive metric, we use PESQ~\citep{pesq} which is a widely adopted ITU-T standard. It was developed to model the subjective test and works at a 16~kHz sampling rate. 
Nowadays, non-intrusive metrics measure perceptual quality without reference signals, and are believed to be more relevant to generative SE~\citep{manjunath2009limitations, manocha2022audio, finally}. 
We follow previous GSR works and adopt several non-intrusive metrics for comprehensive evaluation.
DNSMOS~\citep{dnsmos} is based on a multi-stage CNN trained on human MOS scores. It takes the log Mel spectrogram as the input feature and predicts the estimated signal quality, background noise quality, and overall quality (SIG, BAK, OVRL) scores. WV-MOS is~\citep{wvmos} is a non-intrusive metric based on self-supervised wav2vec2~\citep{baevski2020wav2vec} audio features.
UTMOS~\citep{utmos} is an evaluation system based on ensemble learning of strong and weak learners, achieving high overlap with human opinion on both in-domain and out-of-domain data. 
All the above metrics operate at 16~kHz, where the inference results are first down-sampled before metric calculation. 
To evaluate the perceptual quality of the high-frequency components in the generated full-band audio, we include NISQA~\citep{nisqa}, a quality prediction model that takes 48~kHz audio as input.

\subsection{Degradation Breakdown}
To evaluate the zero-shot capability of VoiceBridge and baseline models across various types and degrees of degradations, we further break down several degradation type into different levels by severity, either following already proposed benchmarks or curating new testsets, and report the performance. 
For the Speech Bandwidth Extension (BWE) task, we follow previous works and utilize the VCTK-BWE~\citep{andreev2023hifi++} benchmark, which is composed of VCTK speech samples with bandwidth limited to 1,2,4 kHz Nyquist frequencies, where models are required to generate full-band (\textit{i.e.}, 48kHz) target from the severely down-sampled signals. 
Such BWE task settings are aligned with the setting of bandwidth extension in other speech restoration works ~\citep{andreev2023hifi++, lemercier2023analysing, ku2025generative}, where the lowpass filtering causes severe information loss in the speech signals, challenging the models to restore high-quality speech using limited prior knowledge. 
For the Dereverberation (Derev) task, we report experiment results on the WSJ-Reverb~\citep{richter2023speech} benchmark, which constructs reverberant signals from speech in the WSJ0 corpora and the simulated room-impulse-responses. This benchmark is utilized in prior speech dereverberation works~\citep{sbsen}. 
Meanwhile, to test model performance under different severity of reverberations, we create another test set by adding simulated RIR signals to speech from the VCTK dataset, with the RT60 time (time needed for sound energy to decay by 60 dB) ranging in 0.3, 0.6, 0.9, and 1.2 seconds.
For the SE task, we still utilize the VoiceBank corpus and the DEMAND noise dataset, but this time adding noise at a fixed signal-to-noise ratio (SNR). We curated 6 testsets, with SNR at -15,-10,-5,0dB respectively, covering a wide range of additive noise from slight disturbance to severe degradation.

To make comprehensive evaluation, we include task-specific baseline models and evaluation metrics. For the BWE task, we include audio super-resolution models and other speech BWE model, including AudioSR~\citep{liu2024audiosr}, NuWave2~\citep{han2022nu},
TFiLM~\citep{birnbaum2019temporal}, and HiFi++~\citep{andreev2023hifi++}. For dereverberation on WSJ-Reverb, we include SGMSE+~\citep{richter2023speech, sgmse} and StoRM~\citep{storm}. For SE, we include SBSE~\citep{sbsen}.
Additional metrics involves the following: STOI~\citep{stoi} is an objective metric that measures the intelligibility of degraded speech. LSD~\citep{voicefixer} directly calculates the log-spectrogram-distance between the enhanced and ground truth signals, which is a commonly used metric in BWE scenarios~\citep{liu2024audiosr, bridgesr}. SRMR~\citep{srmr}, a non-intrusive metric specially designed for measuring signal to reverberation energy ratio, is utilized for measuring dereverberation performance in our experiments.

The results are are listed in  Table~\ref{tab:sr},Table~\ref{tab:derev}, Table~\ref{tab:derev_breakdown} and Table~\ref{tab:se_breakdown}. For the bandwidth extension results shown in Table~\ref{tab:sr}, VoiceBridge consistently outperforms baseline models on the NISQA, PESQ, and LSD metrics, outperforming the specifically trained bandwidth extension models and other generalist models. For the dereverberation results shown in Table~\ref{tab:derev}, VoiceBridge achieves the highest WVMOS, DNSMOS, and NISQA score, and second-best SRMR score, showing strong ability despite the cross-dataset setting. In the detailed experiments of reverberation and additional noise at different severities, shown in Table~\ref{tab:derev_breakdown} and Table~\ref{tab:se_breakdown}, VoiceBridge exhibits strong performance in all settings, including those with degradation severity well beyond the training data. These results demonstrate VoiceBridge's strong zero-shot performance across various Speech Restoration scenario.

\begin{table*}[!h]
\centering
\caption{\small Comparison with Metis~\citep{metis} on the two DNS testsets. Note that Metis used 300 times more data than hours for pretraining, and meanwhile generating only 24kHz audio, thus gaining better results in DNSMOS metrics (which operate at 16kHz). NISQA provides more comprehensive evaluation by taking full-band audio as input.}
\label{tab:metis}
\begin{tabular}{llcccc}
  \toprule
  Dataset & Method & SIG($\uparrow$) & BAK($\uparrow$) & OVRL($\uparrow$) & NISQA($\uparrow$) \\
  \midrule
  \multirow{2}{*}{DNS with Reverb}
    & Metis & 3.68 & 4.14 & 3.44 & 4.56 \\
    & \textbf{VoiceBridge}  & 3.58 & 4.13 & 3.32  & 4.59 \\ 
  \midrule
  \multirow{2}{*}{DNS Real}
    & Metis & 3.59 & 4.01 & 3.27 & 3.95 \\
    & \textbf{VoiceBridge} & 3.47   & 4.03 & 3.19  & 4.50 \\
  \bottomrule
\end{tabular}

\end{table*}

\subsection{Comparison with Pretrained Models}
We further compare VoiceBridge to contemporary generative pretrained models. Metis~\citep{metis}, is a foundational speech model pretrained on 300~K hours of speech data using a masked-generative strategy. It can then be fine-tuned to multiple generative speech tasks, including Speech Restoration. 
We report the comparison of Metis and VoiceBridge on the two DNS testsets~\citep{reddy2021interspeech} (DNS-with-Reverb and DNS-Real) in Table~\ref{tab:metis}. VoiceBridge exhibits slightly lower performance than Metis on DNSMOS metrics but achieves comparable results on NISQA under simulated test conditions and substantially surpasses Metis in real-world NISQA evaluations. This discrepancy can be attributed in part to the characteristics of the evaluation metrics: DNSMOS operates only at 16 kHz and therefore fails to capture fine-grained high-frequency details. Metis, which generates speech at 24 kHz, benefits from modeling a narrower frequency range and thus has an inherent advantage under DNSMOS evaluation. By contrast, NISQA provides a more comprehensive assessment, as it considers full-band audio. Furthermore, Metis was trained with approximately 300 times more pre-training data, much of which is closed-source~\citep{metis}. In comparison, VoiceBridge achieves competitive performance using only public datasets, highlighting its superior data efficiency.

\subsection{Comparison with other Closed-Source Models}

\begin{table}[!h]
\small
\centering
\caption{\small  Comparison with FINALLY~\citep{finally} and HiFi-GAN-2~\cite{su2021hifigan2}, with samples from FINALLY's demo page.}
\label{tab:finally}

\begin{tabular}{lccc}
  \toprule
  \multicolumn{4}{c}{FINALLY~\citep{finally} Demonstration Samples} \\
  \cmidrule(lr){1-4}
  Model & DNSMOS ($\uparrow$) & UTMOS ($\uparrow$) & WV-MOS ($\uparrow$) \\
  \midrule
  HiFi-GAN-2   & 3.283 & 3.947         & 3.470         \\
  FINALLY      & \textbf{3.313} & 3.847         & 3.989         \\
  VoiceBridge  & 3.093 & \textbf{4.122} & \textbf{4.061} \\
  \bottomrule
\end{tabular}

\end{table}

Moreover, we compare VoiceBridge to other closed-source models such as FINALLY~\cite{finally}, HiFi-GAN-2~\cite{su2021hifigan2}, SpeechFlow~\cite{speechflow}, and Ku et al.~\cite{ku2025generative}. For comparison with FINALLY and HIFI-GAN-2, we use the open samples on the demonstration page of FINALLY at \url{https://mmacosha.github.io/finally-demo/}. The results are shown in Table~\ref{tab:finally}. We can see that, VoiceBridge beat FINALLY and HiFi-GAN-2 on the WV-MOS and UTMOS metric, while reaching on-par performance on the DNSMOS metric, demonstrating competitive performance against strong closed-source baselines. We need to note that, FINALLY is \textbf{pretrained on large internal} datasets, where as VoiceBridge is only trained on publicly available data, which highlights the advantage of our modeling. Meanwhile, the testing samples are from the FINALLY's demo page, which might be cherry-picked results for the baseline model, yet completely random for VoiceBridge.

\begin{table}[!h]
\small
\centering
\caption{\small  Comparison with SpeechFlow~\citep{speechflow} and Ku et al.~\cite{ku2025generative}, on corresponding benchmarks.}
\label{tab:speechflow}
\begin{tabular}{lccccc}
  \toprule
  \multicolumn{5}{c}{Downstream Task Evaluation on Curated Testsets} \\
  \cmidrule(lr){1-5}
  
  \multicolumn{5}{c}{VB-Demand} \\
  \cmidrule(lr){1-5}
  Model & PESQ ($\uparrow$) & WVMOS ($\uparrow$) & ESTOI ($\uparrow$) & SISDR ($\uparrow$) \\
  \midrule
  SpeechFlow   & 3.13              & -                        & 0.87              & -                \\
  Ku et al.    & \textbf{3.27}     & \textbf{4.41}            & \textbf{0.88}     & \textbf{19.1}    \\
  VoiceBridge  & 2.83              & 4.39                     & 0.79              & 4.8              \\
  \midrule
  \multicolumn{5}{c}{WSJ0-CHiME3} \\
  \cmidrule(lr){1-5}
  Model & PESQ ($\uparrow$) & WVMOS ($\uparrow$) & ESTOI ($\uparrow$) & SISDR ($\uparrow$) \\
  \midrule
  Ku et al.    & \textbf{2.85}     & 4.26                     & \textbf{0.92}     & \textbf{16.2}    \\
  VoiceBridge  & 1.74              & \textbf{4.37}            & 0.83              & 3.4              \\
  \midrule
  \multicolumn{5}{c}{WSJ0-BWE} \\
  \cmidrule(lr){1-5}
  Model & WVMOS ($\uparrow$) & SQUIM-MOS ($\uparrow$) & ESTOI ($\uparrow$) & SISDR ($\uparrow$) \\
  \midrule
  Ku et al.                       & 3.93             & 4.41   & \textbf{0.92}     & \textbf{8.5}     \\
  VoiceBridge                     & \textbf{4.29}    & \textbf{4.49} & 0.86        & 3.5              \\
  \bottomrule
\end{tabular}

\end{table}

To compare against SpeechFlow and Ku et al.'s model, we evaluated VoiceBridge on the benchmark and metrics used in the corresponding papers. We reported the evaluation results on two denoising benchmarks VB-Demand~\citep{valentini2017noisy} and WSJ0-CHiME3~\citep{chime3}, and the bandwidth extension benchmark WSJ0-BWE~\citep{ku2025generative}. The results are shown in Table~\ref{tab:speechflow}. VoiceBridge achievew comparable performance with the two baseline models on non-intrusive metrics, such as WVMOS and SQUIM-MOS, but showcase lower performance on other metrics. For this, we must state that:
SpeechFlow and Ku et al.'s methods need specific finetuning for each task, while VoiceBridge is a general model without any task-specific adaptation. Meanwhile, both SpeechFlow and Ku et al.'s models work on 16kHz audio, while VoiceBridge generates 48kHz full-band audio, which brings significantly more challenges for the increased data dimension and high-frequency modeling. The used metrics only evaluate the lower 16kHz band, bringing inherent disadvantages to VoiceBridge.
On waveform-level metrics like SISDR, latent-space methods like VoiceBridge typically underperforms data-level methods due to the latent reconstrucion error (Especially, VAE reconstruction is trained mainly on MR-STFT maginitude, not on the waveform dimension, and no phase preservation is guaranteed) and the cascading error from latent modeling. However, these metrics do not necessarily reflect the generated audio quality, as pointed out in~\citep{manocha2022audio, manjunath2009limitations, andreev2023hifi++, finally}.

\section{Further Studies on Latent Modeling}
\label{appendix:latents}

\subsection{Quantitative Results on the Joint Neural Prior}
\begin{figure}[t]
    \centering
    \includegraphics[width=1\linewidth]{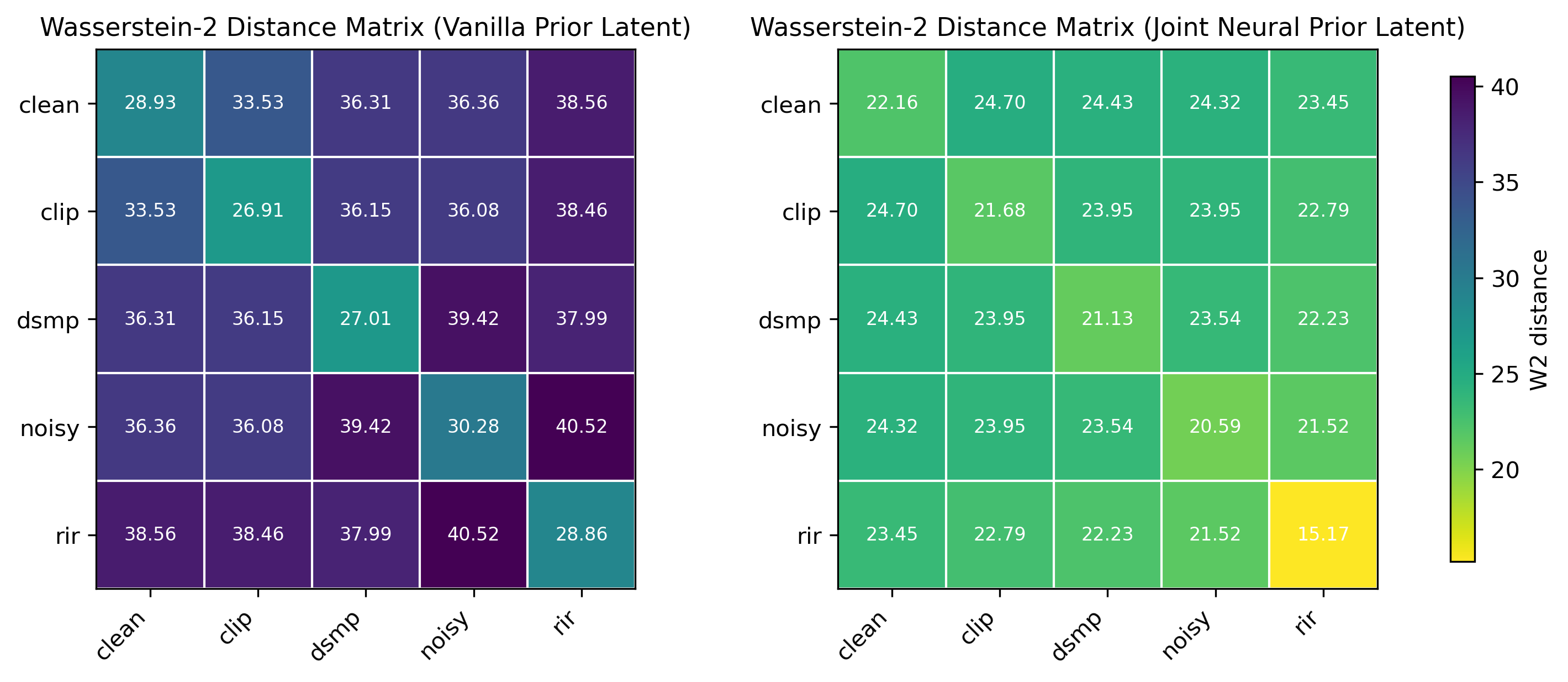}
    \caption{Wasserstein Distance Matrix of latents with different degradation types for the vanilla prior and the joint neural prior.}
    \label{fig:was}
\end{figure}

To quantitatively analyze the effect of prior convergence beyond tSNE visualization, we calculate the 2-Wasserstein distance~\citep{panaretos2019statistical} matrix over the latent distribution for different priors. We model the latent distribution for each degradation type as a Gaussian distribution with the sample mean and variance.
The 2-Wasserstein (Fréchet) distance between two Gaussian-fitted distributions
$\mathcal{N}(\mu_i,\Sigma_i)$ and $\mathcal{N}(\mu_j,\Sigma_j)$ is defined as
\begin{align}
W_2^2\big(\mathcal{N}(\mu_i,\Sigma_i),\,\mathcal{N}(\mu_j,\Sigma_j)\big)
= \|\mu_i-\mu_j\|_2^2 \;+\;
\mathrm{Tr}\Big(\Sigma_i+\Sigma_j - 2\big(\Sigma_i^{1/2}\Sigma_j\Sigma_i^{1/2}\big)^{1/2}\Big).
\end{align}
The resulting matrix represents all pairwise distances between the latent distributions of different degradation types.
For the diagonal entries, we use a within-label variance proxy by measuring the Wasserstein distance to a Dirac at the mean,
which equals the square root of the total variance:
\begin{align}
W_2\big(\mathcal{N}(\mu_c,\Sigma_c),\,\delta_{\mu_c}\big) = \sqrt{\mathrm{Tr}(\Sigma_c)}.
\end{align}
As shown in the Figure~\ref{fig:was}, The joint neural prior yields closer distance between priors of different distributions than the vanilla priors. Moreover, the Wasserstein distance between different degradation types converges toward the diagonal terms, indicating that all priors converge to a single distribution.

\subsection{Exceeding the Perceptual Quality Limits of Latent Models}

Without our proposed post-training process with perceptual awareness, the LBM is trained to capture the distribution of the latent of the HQ target, which may not be fully aligned with human perceptual quality. 
Moreover, without joint post-training, the perceptual generation quality of VoiceBridge will be limited by VAE as well. 
As the perceptual objectives have not been explicitly considered in the training of VAE, the perceptual quality of VoiceBridge will be inevitably restricted. 
By the perceptual joint post-training process, we can improve the perceptual generation quality of both LBMs and the VAE decoder, meanwhile reducing their cascading error in generation.

\begin{table}[!h]
\centering
\small
\caption{\small Comparison of perceptual quality between VAE reconstruction, VoiceBridge restoration, VoiceBridge's reconstruction results on clean audio input, and Ground Truth clean audio.}
\label{tab:recon}

\begin{tabular}{lccc}
\toprule
  % \multicolumn{4}{c}{VB-Demand~\cite{valentini2017noisy} }\\
  % \cmidrule(lr){1-4}
 & DNSMOS($\uparrow$) & WVMOS($\uparrow$) & NISQA($\uparrow$) \\
 \midrule
VAE-Rec. & 3.11 & 4.05 & 4.5 \\
VB-Gen. & 3.20 & 4.39 & 4.51 \\
VB-Rec. & \textbf{3.25} & \underline{4.43} & \textbf{4.71} \\
GT & \underline{3.22} & \textbf{4.5} & \underline{4.62} \\
\bottomrule
\end{tabular}
\end{table}

To verify that VoiceBridge surpasses the upper-bound reconstruction performance of the original VAE, we compare the restoration results of noisy inputs with EP-VAE reconstructions of clean ones. Specifically, we report four groups of results: (1) the EP-VAE reconstruction of clean audio (VAE-Rec.); (2) the restoration results of corresponding degraded audio of VoiceBridge (VB-Gen.); (3) the restoration results of VoiceBridge when passed the clean audio as input (VB-Rec.), which can be seen as a version of reconstruction with enhanced perceptual quality by VoiceBridge; (4) the ground-truth clean target (GT). Clean and degraded data are taken from the VoiceBank-Demand~\citep{valentini2017noisy} dataset. 
As shown in Table~\ref{tab:recon}, the post-training stage enables VoiceBridge's restoration to surpass the VAE reconstruction limits in perceptual quality. 
Moreover, when GT is passed as input for VoiceBridge instead of its noisy version, the perceptual quality further improves, achieving near-GT or even better performance on the reconstruction setting. 
This indicates that jointly training the LBM and the decoder together breaks through the VAE reconstruction upper-bound.
%and enables VoiceBridge to further enhance audio quality even in studio environments.

\begin{table*}[!h]
\centering
\small
\caption{\small VAE-only restoration results on different degradations}
\label{tab:vae}
\begin{tabular}{lccccc}
  \toprule
  \multicolumn{6}{c}{Curated Validation Sets for Different Degradations} \\
  \cmidrule(lr){1-6}
  Model & PESQ ($\uparrow$) & DNSMOS ($\uparrow$) & WVMOS ($\uparrow$) & UTMOS ($\uparrow$) & NISQA ($\uparrow$) \\
  \midrule
  \multicolumn{6}{c}{Noise} \\
  \cmidrule(lr){1-6}
  VoiceBridge & \textbf{2.34} & \textbf{3.17} & \textbf{4.30} & \textbf{4.26} & \textbf{4.18} \\
  VAE         & 1.71          & 2.58          & 3.51          & 3.37          & 3.23          \\
  \midrule
  \multicolumn{6}{c}{Reverb} \\
  \cmidrule(lr){1-6}
  VoiceBridge & \textbf{2.67} & \textbf{3.24} & \textbf{4.41} & \textbf{4.31} & \textbf{4.63} \\
  VAE         & 1.52          & 2.72          & 3.85          & 3.69          & 4.20          \\
  \midrule
  \multicolumn{6}{c}{BWE} \\
  \cmidrule(lr){1-6}
  VoiceBridge & \textbf{3.32} & \textbf{3.23} & \textbf{4.41} & \textbf{4.31} & \textbf{4.64} \\
  VAE         & 2.63          & 2.83          & 3.75          & 3.79          & 4.41          \\
  \midrule
  \multicolumn{6}{c}{CLIP} \\
  \cmidrule(lr){1-6}
  VoiceBridge & \textbf{3.57} & \textbf{3.23} & \textbf{4.43} & \textbf{4.31} & \textbf{4.69} \\
  VAE         & 2.94          & 2.82          & 3.96          & 3.90          & 4.44          \\
  \bottomrule
\end{tabular}

\end{table*}

\subsection{VAE-only Restoration}
In the joint neural prior training stage, the VAE is trained to align the degraded prior to clean prior in multiple measures, including those measuring reconstruction distances, which means the reconstruction of degraded speech will move forward the direction of the restoration target. After the post-training stage, the decoder is further adapted to generate speech with high perceptual quality. This indicate that the VAE itself have certain amount of restoration abilities. To further evaluate the ability of the VAE and the role of the latent bridge model, we test the performance of VAE-only restoration compared with VoiceBridge on different degradation types. The results are shown in Table~\ref{tab:vae}. Basically, we can find that VAE-only enhancement works well on Bandwidth extension task and de-clipping tasks. These two tasks are considered easier, since they preserve the main structure of the speech signal. For the other harder tasks, generative modeling gives more effort. The bridge model successfully refines all degradations to a similar HQ leel beyond the VAE reconstruction.

\begin{table}[!h]
\centering
\small
\caption{\small Inference efficiency, evaluated by Real-Time-Factor (RTF) and Number of Function Evaluation (NFE)}
\label{tab:rtf}
\begin{tabular}{lcccc}
      \toprule
      Model      & VF & RE & UPP & \textbf{VB} \\
      \midrule
      RTF ($\downarrow$)    & 0.015 & 0.833 & 0.075 & 0.025 \\
      NFE & 1 & 64 & 8 & 1 \\
      \bottomrule
\end{tabular}
\end{table}

%unlike the usual "the more the better" sampling logic in diffusion models. 
%where also approximately 4-5 sampling steps reaches the best quality, demonstrating the efficient nature of bridge models compared with diffusion models.

\section{Details on Ablation Studies}
\label{appendix:ablation}

\subsection{Experimental Setup}
We conduct ablation studies on each innovation of VoiceBridge: EP-VAE, joint neural prior, and denoiser-to-generator post-training. 
For the first part of the ablation study, which mainly focuses on the variation of VAE networks, the VAE networks (with or without EP) are pre-trained for 500k steps on 8 A800 GPUs with a batch size of 16. The fine-tuning for joint neural prior costs 150k steps with the same settings. 
For fair comparison, the groups without joint neural prior are trained using the original objective for the same amount of steps.
For the second part, which studies the different loss terms in the post-training stage, all variants are fine-tuned from the pre-trained bridge model in the main experiment for 50k steps on 8 A800 GPUs with a batch size of 16.

\begin{table*}[!h]
\centering
\small
\caption{\small Evaluation results on VoiceFixer-GSR~\citep{voicefixer} and DNS-Real~\citep{reddy2021interspeech} for bridge modeling in different spaces.}
\label{tab:abla-space}
\begin{tabular}{lccccccc}
  \toprule
  \multicolumn{8}{c}{Voicefixer-GSR~\citep{voicefixer}} \\
  \cmidrule(lr){1-8}
  Model & PESQ ($\uparrow$) & SIG ($\uparrow$) & BAK ($\uparrow$) & OVRL ($\uparrow$) & UTMOS ($\uparrow$) & WV-MOS ($\uparrow$) & NISQA ($\uparrow$) \\
  \midrule
  Waveform SB & 1.56 & 2.65 & 3.83 & 2.41 & 2.21 & 2.87 & 1.73 \\
  STFT SB     & 2.06 & 3.07 & 3.34 & 2.58 & 2.69 & 2.44 & 2.73 \\
  Latent SB   & \textbf{2.07} & \textbf{3.43} & \textbf{4.09} & \textbf{3.20} & \textbf{3.58} & \textbf{3.86} & \textbf{3.76} \\
  \midrule
  \multicolumn{8}{c}{DNS-Real~\citep{reddy2021interspeech}} \\
  \cmidrule(lr){1-8}
  Model & PESQ ($\uparrow$) & SIG ($\uparrow$) & BAK ($\uparrow$) & OVRL ($\uparrow$) & UTMOS ($\uparrow$) & WV-MOS ($\uparrow$) & NISQA ($\uparrow$) \\
  \midrule
  Waveform SB & - & 2.64 & 2.55 & 1.94 & 1.61 & 1.55 & 1.56 \\
  STFT SB     & - & 2.96 & 3.57 & 2.55 & 2.27 & 2.59 & 3.39 \\
  Latent SB   & - & \textbf{3.39} & \textbf{4.03} & \textbf{3.12} & \textbf{3.15} & \textbf{3.69} & \textbf{4.34} \\
  \bottomrule
\end{tabular}

\end{table*}

\subsection{Ablation on Bridge Modeling Space}

To validate the necessity of latent modeling for the GSR task, we further conduct an ablation study on the modeling space for SB models. We train three bridge models on VAE latent, waveform, and complex STFT data space respectively. For the waveform-space bridge, we adopt the diffwave~\citep{kong2020diffwave} architecture. For the STFT space, the NeMo~\citep{Harper_NeMo_a_toolkit} (adopted in ~\citep{speechflow} and ~\cite{ku2025generative}) architecture is used. All networks are scaled to around 540M parameters by increasing the number of layers and hidden channels. For the VAE latent, we use VoiceBridge's configuration without the denoiser-to-generator post-training, for equal comparison on the modeling spaces. All models are trained for 500k steps on the same data mixture as VoiceBridge. We evaluate the three models on the VoiceFixer-GSR and the DNS-Real benchmarks. The results are shown in Table~\ref{tab:abla-space}
We can clearly observe that latent bridge models far outperform waveform and STFT domain bridge models, given the high information density of the compressed latent space. Meanwhile, the latent SB models take much less computation than the other two models, as the sequence modeled is much shorter.

\begin{table*}[!h]
\centering
\small
\caption{\small Evaluation results on VoiceFixer-GSR~\citep{voicefixer} and DNS-Real~\citep{reddy2021interspeech} for different joint neural prior training objectives}
\label{tab:abla-jnp}
\begin{tabular}{lccccccc}
  \toprule
  \multicolumn{8}{c}{Voicefixer-GSR~\citep{voicefixer}} \\
  \cmidrule(lr){1-8}
  Model & PESQ ($\uparrow$) & SIG ($\uparrow$) & BAK ($\uparrow$) & OVRL ($\uparrow$) & UTMOS ($\uparrow$) & WV-MOS ($\uparrow$) & NISQA ($\uparrow$) \\
  \midrule
  KL-only & 2.08 & 3.15 & 3.91 & 3.13 & 3.12 & 3.37 & 3.68 \\
  JNP     & \textbf{2.15} & \textbf{3.43} & \textbf{4.09} & \textbf{3.20} & \textbf{3.58} & \textbf{3.86} & \textbf{3.76} \\
  \midrule
  \multicolumn{8}{c}{DNS-Real~\citep{reddy2021interspeech}} \\
  \cmidrule(lr){1-8}
  Model &       & SIG ($\uparrow$) & BAK ($\uparrow$) & OVRL ($\uparrow$) & UTMOS ($\uparrow$) & WV-MOS ($\uparrow$) & NISQA ($\uparrow$) \\
  \midrule
  KL-only &       & 3.03 & 3.88 & 2.97 & 2.75 & 3.31 & 4.16 \\
  JNP     &       & \textbf{3.39} & \textbf{4.03} & \textbf{3.12} & \textbf{3.15} & \textbf{3.69} & \textbf{4.34} \\
  \bottomrule
\end{tabular}

\end{table*}

\begin{table}[!h]
\centering
\small
\caption{\small Evaluation results on curated testsets with different degradation types for bridge models with and without joint neural prior.}
\label{tab:abla-jnp-type}
\begin{tabular}{lccccc}
  \toprule
  \multicolumn{6}{c}{Effect of JNP on Perceptual Metrics} \\
  \cmidrule(lr){1-6}
  Model & PESQ ($\uparrow$) & DNSMOS ($\uparrow$) & WVMOS ($\uparrow$) & UTMOS ($\uparrow$) & NISQA ($\uparrow$) \\
  \midrule
  \multicolumn{6}{c}{Noise} \\
  \cmidrule(lr){1-6}
  w/ JNP   & \textbf{1.88} & \textbf{3.17} & \textbf{4.05} & \textbf{3.69} & \textbf{4.36} \\
  w/o JNP  & 1.75          & 2.95          & 4.01          & 3.65          & 4.10          \\
  \midrule
  \multicolumn{6}{c}{Reverb} \\
  \cmidrule(lr){1-6}
  w/ JNP   & \textbf{2.12} & \textbf{3.24} & \textbf{4.11} & \textbf{3.77} & \textbf{4.60} \\
  w/o JNP  & 2.01          & 3.13          & 4.10          & 3.72          & 4.43          \\
  \midrule
  \multicolumn{6}{c}{BWE} \\
  \cmidrule(lr){1-6}
  w/ JNP   & \textbf{2.70} & \textbf{3.30} & \textbf{4.32} & \textbf{3.97} & \textbf{4.69} \\
  w/o JNP  & 2.51          & 3.19          & 4.26          & 3.95          & 4.50          \\
  \midrule
  \multicolumn{6}{c}{CLIP} \\
  \cmidrule(lr){1-6}
  w/ JNP   & \textbf{3.26} & \textbf{3.25} & \textbf{4.31} & \textbf{4.01} & \textbf{4.66} \\
  w/o JNP  & 2.95          & 3.16          & 4.20          & 3.91          & 4.63          \\
  \bottomrule
\end{tabular}
\end{table}

\subsection{Ablation on joint neural prior training objective design}

A core difference of our joint neural prior with~\citep{fang2021variational} is that we align the LQ and HQ priors not only by the Euclidean measure in latent space (wich is basically equivalent with the KL alignment done in~\citep{fang2021variational} if regardless of the variance).
In contrast, we add objectives including cosine similarity in latent space and multiple supervision signals in the reconstructed data space, in order to reduce the distance between the latents in all geometrical measures. We conduct an ablation study with two bridge models trained with the multi-measure aligned joint neural prior and KL-aligned-only prior respectively, to prove that our multi-measure alignment is necessary. The results are shown in Table~\ref{tab:abla-jnp}. Our joint neural prior clearly outperforms the KL-only alignment proposed in~\cite{fang2021variational}, proving our method's effectiveness.

\subsection{Ablation on joint neural prior's effectiveness for different degradation types}

To further discuss whether the joint neural prior can converge the prior of diverse degradation types to a unified clean distribution, rather than only working on certain types, we conduct an ablation study on the effectiveness of joint neural prior for different degradation types. We train two bridge models with and without joint neural prior respectively, and evaluate their performance on different degradation types on curated testsets with additional noise, reverberation, bandwidth limitation, and clipping on the VCTK corpus. The models are trained for 500k steps. Results are shown in Table~\ref{tab:abla-jnp-type}. The results demonstrate consistent improvement across all dereverberation types, indicating that JNP benefit all distortions simultaneously.

%%%%%%%%%%%%%%%%%%%%%%%%%%%%%%%%%%%%%%%%%%%%%%%%%%%%%%%%%%%%%%%%%%%%%%%%%%%%%%%
%%%%%%%%%%%%%%%%%%%%%%%%%%%%%%%%%%%%%%%%%%%%%%%%%%%%%%%%%%%%%%%%%%%%%%%%%%%%%%%

\end{document}